\documentclass[12pt,letterpaper,a4paper]{article}
\usepackage[includeheadfoot,
            marginratio={1:1,2:3},
            width=475pt,
            height=786pt,]{geometry}
\usepackage{amsmath}
\usepackage{amsfonts}
\usepackage{amssymb}
\usepackage{graphicx}
\usepackage{cancel}

\usepackage{empheq}
\usepackage{color}
\usepackage{hyperref}

\usepackage{float}
\restylefloat{table}
\restylefloat{figure}

\newcommand{\nc}{\newcommand}
\nc{\beq}{\begin{equation}}
\nc{\eeq}{\end{equation}}
\nc{\bea}{\begin{eqnarray}}
\nc{\eea}{\end{eqnarray}}
\def\IZ{\mathbb{Z}}

\def\ov{\overline}

\allowdisplaybreaks[2]
\numberwithin{equation}{section}

\begin{document}

\vspace{1.5cm}
\begin{center}
{\LARGE 
%A modular invariant symplectic formulation of non-geometric scalar potentials
%Modular invariant non-geometric scalar potentials of type IIB on CY orientifolds: A symplectic formulation
%Modular invariant non-geometric scalar potentials via the intersection numbers of compactifying CYs
Reading off the non-geometric scalar potentials via \vskip0.1cm topological data of the Calabi Yau manifolds}
\vspace{0.4cm}
\end{center}

\vspace{0.35cm}
\begin{center}
Pramod Shukla \footnote{Email: shukla.pramod@ictp.it}
\end{center}

\vspace{0.1cm}
\begin{center}
{ICTP, Strada Costiera 11, Trieste 34151, Italy}
\end{center}

\vspace{1cm}

%%%%%%%%%%%%%%%%%%%%%%%%%%%%%%%%%%%%%%%%%%%%%%%
%%%%%%%%%%%%%%%%%%%%%%%%%%%%%%%%%%%%%%%%%%%%%%%
%%%%%%%%%%%%%%%%%%%%%%%%%%%%%%%%%%%%%%%%%%%%%%%
%%%%%%%%%%%%%%%%%%%%%%%%%%%%%%%%%%%%%%%%%%%%%%%
%%%%%%%%%%%%%%%%%%%%%%%%%%%%%%%%%%%%%%%%%%%%%%%
%%%%%%%%%%%%%%%%%%%%%%%%%%%%%%%%%%%%%%%%%%%%%%%
%%%%%%%%%%%%%%%%%%%%%%%%%%%%%%%%%%%%%%%%%%%%%%%
%%%%%%%%%%%%%%%%%%%%%%%%%%%%%%%%%%%%%%%%%%%%%%%

\begin{abstract}
In the context of studying the 4D effective potentials of type IIB non-geometric flux compactifications, this article has a twofold goal. First, we present a modular invariant symplectic rearrangement of the tree level non-geometric scalar potential arising from a flux superpotential which includes the S-dual pairs of non-geometric fluxes $(Q, P)$, the standard NS-NS and RR three-form fluxes $(F_3, H_3)$ and the geometric flux ($\omega$). This `symplectic formulation' is valid for arbitrary numbers of K\"ahler moduli, and the complex structure moduli which are implicitly encoded in a set of symplectic matrices. 

In the second part, we further explicitly rewrite all the symplectic ingredients in terms of saxionic and axionic components of the complex structure moduli. The same leads to a compact form of the generic scalar potential being explicitly written out in terms of all the real moduli/axions. Moreover, the final form of the scalar potential needs only the knowledge of some topological data (such as hodge numbers and the triple intersection numbers) of the compactifying (CY) threefolds and their respective mirrors. Finally, we demonstrate how the same is equivalent to say that, for a given concrete example, various pieces of the scalar potential can be directly read off from our generic proposal, without the need of starting from the K\"ahler- and super-potentials. %The applicability of these two proposals have been demonstrated in a couple of concrete type IIB toroidal compactification examples with a ${\mathbb T}^6/({\mathbb Z}_2 \times {\mathbb Z}_2)$ orientifold and  a ${\mathbb T}^6/{\mathbb Z}_4$ orientifold. 

\end{abstract}

\clearpage

\tableofcontents

%%%%%%%%%%%%%%%%%%%%%%%%%%%%%%%%%%%%%%%%%%%%%%%
%%%%%%%%%%%%%%%%%%%%%%%%%%%%%%%%%%%%%%%%%%%%%%%
%%%%%%%%%%%%%%%%%%%%%%%%%%%%%%%%%%%%%%%%%%%%%%%
%%%%%%%%%%%%%%%%%%%%%%%%%%%%%%%%%%%%%%%%%%%%%%%
%%%%%%%%%%%%%%%%%%%%%%%%%%%%%%%%%%%%%%%%%%%%%%%
%%%%%%%%%%%%%%%%%%%%%%%%%%%%%%%%%%%%%%%%%%%%%%%
%%%%%%%%%%%%%%%%%%%%%%%%%%%%%%%%%%%%%%%%%%%%%%%
%%%%%%%%%%%%%%%%%%%%%%%%%%%%%%%%%%%%%%%%%%%%%%%

%\newpage

\section{Introduction}
\label{sec_intro}
In the context of moduli stabilization, the study of four dimensional effective potentials arising from type IIB superstring compactifications have continuously absorbed a lot of attention since long \cite{Kachru:2003aw, Balasubramanian:2005zx, Grana:2005jc, Blumenhagen:2006ci, Douglas:2006es, Denef:2005mm, Blumenhagen:2007sm}. In this regard, non-geometric flux compactification scenario has emerged as an interesting playground for model builders \cite{Derendinger:2004jn,Grana:2012rr,Dibitetto:2012rk, Danielsson:2012by, Blaback:2013ht, Damian:2013dq, Damian:2013dwa, Hassler:2014mla, Ihl:2007ah, deCarlos:2009qm, Danielsson:2009ff, Blaback:2015zra, Dibitetto:2011qs}. The existence of non-geometric fluxes are rooted through a successive application of T-duality on the three form $H$-flux of the type II orientifold theories, where a chain with geometric and non-geometric fluxes appears as \cite{Shelton:2005cf},
\bea
\label{eq:Tdual}
& & H_{ijk} \longrightarrow \omega_{ij}{}^k  \longrightarrow Q^{jk}{}_i  \longrightarrow R^{ijk} \, .
\eea
Moreover, S-duality invariance of type IIB superstring compactification demands for including additional $P$-fluxes, which are S-dual to the non-geometric $Q$-fluxes \cite{Aldazabal:2008zza,Font:2008vd,Guarino:2008ik, Hull:2004in, Aldazabal:2008zza, Kumar:1996zx, Hull:2003kr}. If one could consistently turn-on these fluxes on the compactification background, one induces a 4D-effective potential which generically depends on all of such flux parameters, creating the possibility of such fluxes being utilized for moduli stabilization and in search of string vacua \cite{Derendinger:2004jn,Grana:2012rr,Dibitetto:2012rk, Danielsson:2012by, Blaback:2013ht, Damian:2013dq, Damian:2013dwa, Hassler:2014mla, Ihl:2007ah, deCarlos:2009qm, Danielsson:2009ff, Blaback:2015zra, Dibitetto:2011qs}. Moreover, the inclusion of various kinds of flux parameters with a diverse superpotential coupling makes the compactification background richer and more flexible for model building.

The magnitude of their importance can be illustrated by the fact that generically speaking, one can stabilize all moduli  at the tree level, including the K\"ahler moduli which, in conventional flux compactification, are protected by the No-scale structure. However, the complexity with introducing many flux parameters not only facilitates a possibility for the easier samplings to fit the values, but also backreacts on the strategy itself via putting some inevitably hard challenges, which sometimes can make the situation worse. For example the resulting 4D scalar potential are very often so huge in concrete examples (say in Type IIB on ${\mathbb T}^6/({\mathbb Z}_2\times{\mathbb Z}_2)$ orientifold) that even it gets hard to analytically solve the extremization conditions, and one has to look either for simplified ansatz by switching off certain flux components at a time, or else one has to opt for an involved numerical analysis \cite{Aldazabal:2008zza,Font:2008vd,Guarino:2008ik,Danielsson:2012by,Damian:2013dq, Damian:2013dwa}. On top of solving the extremization conditions, another obstacle comes with imposing a huge amount of quadratic flux constraints coming from a  set of Bianchi identities and tadpole cancellation conditions. Nevertheless, the possibility of stabilizing all moduli at tree level still put the non-geometric flux compactification scenarios quite attractive as well as relevant framework for future investigations.

Apart from the direct  model building motivations, the interesting relations among the ingredients of superstring flux-compactifications and those of the gauged supergravities have significant relevance in understanding both sectors as fluxes in one setting are related to the gauging in the other one  \cite{ Derendinger:2004jn, Derendinger:2005ph, Shelton:2005cf, Aldazabal:2006up, Dall'Agata:2009gv, Aldazabal:2011yz, Aldazabal:2011nj,Geissbuhler:2011mx,Grana:2012rr,Dibitetto:2012rk, Villadoro:2005cu}. In the conventional approach of studying 4D type II effective theories in a non-geometric flux compactification framwork, most of the studies have been centered around toroidal examples; or in particular with a ${\mathbb T}^6/({\mathbb Z}_2 \times {\mathbb Z}_2)$ orientifold. A simple justification for the same lies in their relatively simpler structure to perform explicit computations, which led toroidal setups to serve as promising toolkits in studying concrete examples. However, the very recent developments in \cite{Hassler:2014mla, Blumenhagen:2015qda, Blumenhagen:2015kja,Blumenhagen:2015jva, Blumenhagen:2015xpa,  Li:2015taa, Blumenhagen:2015xpa} regarding the formal developments along with applications towards moduli stabilization, searching de-Sitter vacua as well as  building inflationary models have boosted the interests in setups beyond toroidal examples, say Calabi Yaus. As the explicit form of the metric for a Calabi Yau threefold is not known, while understanding the ten dimensional origin of the 4D effective scalar potential, one should preferably represent the same in a framework where one could bypass the need of knowing the Calabi Yau metric. This can be done via trading the lack of knowledge about the Calabi Yau metric against knowing the period matrices along with the moduli space matrices for the K\"ahler moduli \cite{Taylor:1999ii, Blumenhagen:2003vr,Shukla:2015hpa}.

Moreover, there have been close connections between the symplectic geometry and effective potentials of type II supergravity theories \cite{Ceresole:1995ca, D'Auria:2007ay, Taylor:1999ii}, and as motivated above, the role of symplectic geometry gets crucially important while dealing with Calabi Yau orientifolds. For example, in the context of type IIB orientifolds with the presence of standard NS-NS three-form flux ($H_3$) and RR three-form flux ($F_3$), the two scalar potentials, one arising from the  $F/D$-term contributions while the other being derived from the dimensional reduction of the ten-dimensional kinetic pieces, could be matched via merely using the period matrices and without the need of knowing CY metric \cite{Taylor:1999ii, Blumenhagen:2003vr}. This strategy was also recently extended with the inclusion of non-geometric $Q$-flux in \cite{Shukla:2015hpa} which we plan to further generalize into a modular invariant non-geometric framework in this article.

\subsubsection*{Motivation and goals} 
In the context of non-geometric flux compactifications, there have been great amount of studies via considering the 4D effective potential merely derived by knowing the K\"ahler and super-potentials \cite{Danielsson:2012by, Blaback:2013ht, Damian:2013dq, Damian:2013dwa, Blumenhagen:2013hva, Villadoro:2005cu, Robbins:2007yv, Ihl:2007ah, Gao:2015nra}, and without having a complete understanding of their ten-dimensional origin. Some significant steps have been taken towards exploring the form of non-geometric 10D action via Double Field Theory (DFT) \cite{Andriot:2013xca, Andriot:2011uh, Blumenhagen:2015lta} as well as supergravity \cite{Villadoro:2005cu, Blumenhagen:2013hva, Gao:2015nra, Shukla:2015rua, Shukla:2015bca} \footnote{See \cite{Andriot:2012wx, Andriot:2012an,Andriot:2014qla,Blair:2014zba} also.}. In this article, our main aim is to provide a modular completion of the symplectic formulation proposed for expressing the type IIB non-geometric scalar potentials in \cite{Shukla:2015hpa}. The same may be considered as another iterative step taken on the lines of attempts for finding a compact and `suitable' rearrangement \footnote{On the lines of \cite{Blumenhagen:2013hva, Gao:2015nra, Shukla:2015rua, Shukla:2015bca, Shukla:2015hpa}, by a `suitable' rearrangement of the four dimensional scalar potential, we mean an equivalent  collection of its terms which gives some insights about their 10D origin.} of the scalar potential as enumerated below,
\begin{itemize}
 \item{In the context of standard type IIB flux compactification with the usual NS-NS and RR fluxes, $H_3$, and $F_3$, the four dimensional scalar potential has been compactly rewritten in a symplectic manner in \cite{Taylor:1999ii, Blumenhagen:2003vr}}.
 \item{Motivated by the type IIA flux compactification with geometric flux in \cite{Villadoro:2005cu}, a rearrangement of the scalar potential induced via the inclusion of $Q$-fluxes on top of the standard $H_3/F_3$ within a type IIB non-geometric framework, has been proposed in \cite{Blumenhagen:2013hva}.}
 \item{The proposal of \cite{Blumenhagen:2013hva} was further extended with the inclusion of $P$-flux as a modular completion counterpart of non-geometric $Q$-flux \cite{Gao:2015nra}. However the analysis in \cite{Blumenhagen:2013hva, Gao:2015nra} were limited to a toroidal (${\mathbb T}^6/{\left({\mathbb Z}_2 \times {\mathbb Z}_2\right)}$-orientifold) setup without odd axions. Though an extension of the scalar potential rearrangement with the odd axions has been proposed later on in \cite{Shukla:2015rua} within a type IIB compactification on ${\mathbb T}^6/{{\mathbb Z}_4}$-orientifold, however by now all the attempts made in \cite{Blumenhagen:2013hva, Gao:2015nra, Shukla:2015rua} needed the explicit knowledge of the internal background metric.}
 \item{With the motivation to promote the aforementioned studies towards the non-toroidal non-geometric backgrounds, a `symplectic formulation' of the non-geometric scalar potential (induced from a generic tree level generalized flux superpotential and the $D$-term contributions,) having all the NS-NS (non-)geometric fluxes, has been proposed recently  in  \cite{Shukla:2015hpa}. This formulation is valid for arbitrary number of complex structure moduli as well as K\"ahler moduli, and on top of it, this does not need the knowledge of internal (CY) metric.}
 \item{Meanwhile, some crucial attempts have also been made from DFT framework  to understand the ten-dimensional origin of the 4D effective non-geometric potentials involving all the NS-NS fluxes and also incorporating the odd axions $B_2/C_2$  \cite{Blumenhagen:2015lta}. }
 \item{As a next step in this iterative chain we enumerated as above, in this article, we provide a modular completion of the symplectic formulation proposed in \cite{Shukla:2015hpa} by including the non-geometric RR flux, namely $P$-flux which is S-dual to the NS-NS $Q$-flux.}
 \item{Finally, after rewriting the symplectic quantities in terms of saxion/axion parts of the complex structure moduli, we present the most general form of the tree level scalar potential which can be explicitly written for a particular compactification by merely knowing (some of) the topological data (such as hodge numbers and intersection numbers) of the compactifying (CY) threefolds and their mirrors.}
\end{itemize}
The article is organized as follows: In section \ref{sec_setup} we provide some limited (but self-contained) preliminaries regarding type IIB non-geometric flux compactification in a modular invariant framework. Subsequently, in section \ref{sec_symplectic_formulation}, using a set of new generalized flux orbits and some non-trivial symplectic identities, we rewrite the generic tree level scalar potential using symplectic quantities such as period matrices of the compactifying (CY) threefolds. We demonstrate the applicability of the symplectic proposal for two explicit toroidal examples in section \ref{sec_2examples}. In section \ref{sec_Intersections1}, we further extend the analysis to another representation of the scalar potential by expanding out the symplectic ingredients in terms of the complex structure moduli, and subsequently we provide a final and very compact version. Subsequently, in section \ref{sec_Intersections},  we demonstrate how one can directly read-off the various scalar potential pieces (for a given particular compactification) via simply knowing some of the topological data of the compactifying CYs and their respective mirrors. Finally, in section \ref{sec_conclusion} we present an overall conclusion. In addition, we provide four appendices. In appendix \ref{sec_symbolsList} we provide a list of various distinct symbols utilized in writing down our two proposals for non-geometric scalar potential. Further, we provide some lengthy intermediate steps for arriving at the `symplectic formulation' proposals of the scalar potential in appendix \ref{sec_ExpandedVersion1}. In the next step, in appendix \ref{sec_proof}, we provide the proof of some relevant symplectic identities (conjectured in \cite{Shukla:2015hpa}) for generic prepotentials in the absence of non-perturbative effects. Finally, we provide expanded version of the final proposal along with some additional details on the intermediate stages in the appendix \ref{sec_csVExpanded}.

\section{A short review of the relevant preliminaries}
\label{sec_setup}
%The dynamics of low energy effective supergravity action is encoded in three building blocks; namely a K\"{a}hler potential ($K$), a holomorphic superpotential ($W$) and a holomorphic gauge kinetic function ($\hat{\cal G}$) written in terms of appropriate chiral variables. Subsequently, t
The $F$-term contributions to ${N}=1$ scalar potential governing the dynamics of low energy effective supergravity are computed from the K\"ahler and flux induced super-potential via the following well known relation,
\bea
\label{eq:Vtot}
& & V=e^{K}\Big(K^{I\bar J}D_I W\, D_{\bar J} \ov W-3\, |W|^2\Big)  \,,\,
\eea
where the covariant derivatives are defined with respect to all the chiral variables on which the K\"{a}hler potential ($K$) and the holomorphic superpotential ($W$) generically depend on. The massless states in the four dimensional effective theory are in one-to-one correspondence
with harmonic forms which are either  even or odd
under the action of an isometric, holomorphic involution ($\sigma$) acting on the internal compactifying Calabi Yau threefolds, and these do generate the equivariant  cohomology groups $H^{p,q}_\pm (X)$. Let us fix our conventions, and denote the bases  of even/odd two-forms as $(\mu_\alpha, \, \nu_a)$ while four-forms as $(\tilde{\mu}_\alpha, \, \tilde{\nu}_a)$ where $\alpha\in h^{1,1}_+(X), \, a\in h^{1,1}_-(X)$. Considering setups with $h^{1,1}_-(X) \neq 0$ are usually less studied as compared to the simpler case of $h^{1,1}_-(X) = 0$, and explicit construction of such orientifold odd two-cycles can be found in \cite{Lust:2006zg,Lust:2006zh,Blumenhagen:2008zz,Cicoli:2012vw,Gao:2013rra,Gao:2013pra}. Also, we denote the zero- and six- even forms as ${\bf 1}$ and $\Phi_6$ respectively. In addition, the bases for the even and odd cohomologies  of three-forms $H^3_\pm(X)$ are denoted as the symplectic pairs $(a_K, b^J)$ and $({\cal A}_\Lambda, {\cal B}^\Delta)$ respectively. Here we fix the normalization in the various cohomology bases as under,
\bea
\label{eq:intersection}
& & \hskip-0.7cm \int_X \Phi_6 = f, \, \, \int_X \, \mu_\alpha \wedge \tilde{\mu}^\beta = \hat{d}_\alpha^{\, \, \, \beta} , \, \, \int_X \, \nu_a \wedge \tilde{\nu}^b = {d}_a^{\, \, \,b}, \quad \int_X \, \mu_\alpha \wedge \mu_\beta \wedge \mu_\gamma = k_{\alpha \beta \gamma},\\
& & \hskip0.7cm  \int_X \, \mu_\alpha \wedge \nu_a \wedge \nu_b = \hat{k}_{\alpha a b}, \quad \int_X a_K \wedge b^J = \delta_K{}^J, \, \, \, \, \, \int_X {\cal A}_\Lambda \wedge {\cal B}^\Delta = \delta_\Lambda{}^\Delta\nonumber
\eea
Here, for the orientifold choice with $O3/O7$-planes, $K\in \{1, ..., h^{2,1}_+\}$ and $\Lambda\in \{0, ..., h^{2,1}_-\}$ while for $O5/O9$-planes, one has $K\in \{0, ..., h^{2,1}_+\}$ and $\Lambda\in \{1, ..., h^{2,1}_-\}$. Also, let us note that if four-form basis is appropriately chosen to be dual of the two-form basis, one will of course have $\hat{d}_\alpha^{\, \, \, \beta} = {\delta}_\alpha^{\, \, \, \beta}$ and ${d}_a^{\, \, \,b} = {\delta}_a^{\, \, \,b}$. However for the present work, we follow the conventions of \cite{Robbins:2007yv}, and take a bit more generic case. 

Now, the various field ingredients can be expanded in appropriate bases of the equivariant cohomologies. For example, the K\"{a}hler form $J$, the
two-forms $B_2$,  $C_2$ and the RR four-form $C_4$ can be expanded as \cite{Grimm:2004uq}
\bea
\label{eq:fieldExpansions}
& & \hskip-1.7cm J = t^\alpha\, \mu_\alpha, \quad  B_2= b^a\, \nu_a , \quad C_2 =c^a\, \nu_a, \quad  C_4 = D_2^{\alpha}\wedge \mu_\alpha + V^{K}\wedge a_K + U_{K}\wedge b^K + {\rho}_{\alpha} \, \tilde\mu^\alpha\,,
\eea
where $t^\alpha$ denotes the two-cycle volume moduli, while $b^a, \, c^a$ and $\rho_\alpha$ are various axions. Further, ($V^K$, $U_K$) forms a dual pair of space-time one-forms
and $D_2^{\alpha}$ is a space-time two-form dual to the scalar field $\rho_\alpha$. %Due to the self-duality of RR four-form, half of the degrees of freedom of $C_4$ are removed. Note that the even component of the Kalb-Ramond field $B_{+} = b^\alpha \, \omega_\alpha$, though  not a continuous modulus, can take the two  discrete values $b^\alpha\in\{0, 1/2\}$. 
Also, since $\sigma^*$ reflects the holomorphic three-form $\Omega_3$, we have $h^{2,1}_-(X)$ complex structure moduli $z^{i}$ appearing as complex scalars. Let us also mention that under the full orientifold action, we can only have the following flux-components of the standard fluxes $(F_3, H_3)$, the geometric flux $(\omega)$ and the non-geometric fluxes ($Q$, $P$ and $R$) \cite{Robbins:2007yv, Blumenhagen:2015kja},
\bea
\label{eq:allowedFluxes}
& & \hskip-0.7cm  H\equiv \left(H_\Lambda, H^\Lambda\right), \, \, \omega\equiv \left({\omega}_a{}^\Lambda, {\omega}_{a \Lambda} , \hat{\omega}_\alpha{}^K, \hat{\omega}_{\alpha K}\right), \, \, Q\equiv \left({Q}^{a{}K}, \, {Q}^{a}{}_{K}, \, \hat{Q}^{\alpha{}\Lambda} , \, \hat{Q}^{\alpha}{}_{\Lambda}\right), \, \, R\equiv \left(R_K, R^K \right)\, \nonumber\\
& & \hskip3cm F\equiv \left(F_\Lambda, F^\Lambda\right), \quad P\equiv \left({P}^{a{}K}, \, {P}^{a}{}_{K}, \, \hat{P}^{\alpha{}\Lambda} , \, \hat{P}^{\alpha}{}_{\Lambda}\right).
\eea
We hereby state that for the present analysis we mainly focus on the cases where the orientifold involution is such that $h^{2,1}_+(X) =0$, and so we do not have non-geometric $R$-flux and the subsequent possibly generated D-terms \cite{Robbins:2007yv, Shukla:2015bca, Shukla:2015rua, Shukla:2015hpa, Blumenhagen:2015lta}.

\subsection{The K\"ahler potential ($K$)}
Using appropriate chiral variables, a generic form of the tree level K\"{a}hler potential can be written as a sum of two pieces motivated from their underlying ${N}=2$ special K\"ahler and quaternionic structure, and the same is give as under,
\bea
\label{eq:K}
& & \hskip-2.0cm K = -\ln\left(i\int_{X}\Omega_3\wedge{\bar\Omega_3}\right) - \ln\left(-i(\tau-\ov\tau)\right) -2\ln\left({\cal V}_E\,(\tau, G^a, T_\alpha; \ov \tau, \ov G^a, \ov T_\alpha)\right) \,.
\eea
where one has to supplement the following ingredients,
\begin{itemize}
\item{Here, the involutively-odd holomorphic three-form $\Omega_3$ generically depends on the complex structure moduli ($z^k$) and can be written out in terms of period vectors, 
\bea
\label{eq:Omega3}
& &  \Omega_3\, \equiv  {\cal X}^\Lambda \, {\cal A}_\Lambda - \, {\cal F}_{\Lambda} \, {\cal B}^\Lambda \,  
\eea
via using a genetic tree level pre-potential given as under,
\bea
\label{eq:prepotential}
& & {\cal F} = ({\cal X}^0)^2 \, \, f({z^i}) \,, \quad \quad  f({z^i}) = \frac{1}{6}\,{\hat{l}_{ijk} \, z^i\, z^j \, z^k} +  \frac{1}{2} \,{a_{ij} \, z^i\, z^j} +  \,{b_{i} \, z^i} +  \frac{1}{2} \,{i\, \gamma} \,.
\eea 
Here, the special coordinates $z^i =\frac{\delta^i_\Lambda \, {\cal X}^\Lambda}{{\cal X}^0}$ are used, and {\it $ \hat{l}_{ijk}$ are triple intersection numbers on the mirror (Calabi Yau) threefold.} Further,  the quantities $a_{ij}, b_i$ and $\gamma$ are real numbers where $\gamma$ is related to the perturbative $(\alpha^\prime)^3$-corrections on the mirror side, and so is proportional to the Euler characteristic of the mirror Calabi Yau \cite{Hosono:1994av,Arends:2014qca, Blumenhagen:2014nba}.  In general, $f({z^i})$ will also have an infinite series of non-perturbative contributions (say ${\cal F}_{\rm inst.}(z^i)$), however  for the current purpose, we are assuming the large complex structure limit to suppress the same.}
\item{The overall internal volume ${\cal V}_E$ in the Einstein frame can be generically written in terms of two-cycle volume moduli as below,
\begin{eqnarray}
& & {\cal V}_E = \frac{1}{6} \,{k_{\alpha \beta \gamma} \, t^\alpha\, t^\beta \, t^{\gamma}} 
\end{eqnarray}
One should represent ${\cal V}_E$ in terms of chiral variables ($\tau, G^a, T_\alpha$) given as under \cite{Benmachiche:2006df},
\bea
\label{eq:N=1_coords}
& & \hskip-2cm \tau \equiv C_0 + \, i \, e^{-\phi} = C_0 + i\, s\, , \qquad \qquad G^a= c^a + \tau \, b^a \, ,\\
& & \hskip-1cm T_\alpha= \left({\rho}_\alpha +  \hat{\kappa}_{\alpha a b} c^a b^b + \frac{1}{2} \, \tau \, \hat{\kappa}_{\alpha a b} b^a \, b^b \right)  -\frac{i}{2} \, \kappa_{\alpha\beta\gamma} t^\beta t^\gamma\, ,\nonumber
\eea
where $\kappa_{\alpha\beta\gamma}=(\hat{d^{-1}})_\alpha^{ \, \,\delta} \, k_{\delta\beta\gamma}$ and $\hat{\kappa}_{\alpha a b} = (\hat{d^{-1}})_\alpha^{ \, \,\delta} \, \hat{k}_{\delta a b}$ are some rescaled triple intersection numbers surviving under the orientifold action \cite{Shukla:2015hpa}.}
\end{itemize}
Let us mention that the chiral variable $T_\alpha$ is simplified to be,
\bea
\label{eq:N=1_coords2}
& & T_\alpha= \left({\rho}_\alpha +  \hat{\kappa}_{\alpha a b} c^a b^b + \frac{1}{2} \, C_0 \, \hat{\kappa}_{\alpha a b} b^a \, b^b \right) -i\, \biggl(\frac{1}{2}\,\kappa_{\alpha\beta\gamma} t^\beta t^\gamma -\frac{s}{2} \, \hat{\kappa}_{\alpha a b} \, b^a \, b^b \biggr)\\
& & \hskip0.5cm = \left(\tilde{\rho}_\alpha + \frac{1}{2}\, \hat{\kappa}_{\alpha a b} c^a b^b + \frac{1}{2} \, C_0 \, \hat{\kappa}_{\alpha a b} b^a \, b^b \right) -i\, \biggl(\sigma_\alpha - \frac{s}{2} \, \hat{\kappa}_{\alpha a b} \, b^a \, b^b \biggr)\,,\nonumber
\eea 
where in the second line we have introduced $\sigma_\alpha = \frac{1}{2} \, \kappa_{\alpha\beta\gamma} t^\beta t^\gamma$ which being purely the (Einstein-frame) four-cycle volume is invariant under S-duality. Moreover, the new axionic combination utilized as $\tilde{\rho}_\alpha = {\rho}_\alpha + \frac{1}{2} \hat{\kappa}_{\alpha a b} c^a b^b$, is an S-duality invariant collection which will be later observed to be useful for the manifestations of the S-dulality transformations of the new flux orbits. %The detailed expressions of moduli space K\"ahler metrices as well as the K\"ahler derivative to be heavily utilized for the scalar potential simplification have been collected in the appendix \ref{sec_detailedV}.

\subsection{Modular completed non-geometric superpotential ($W$)}
The effective four dimensional scalar potential generically have an S-duality invariance following from the underlying ten-dimensional type IIB supergravity, and this corresponds to the following $SL(2, \mathbb{Z})$ transformation,
\bea
\label{eq:SL2Za}
& & \hskip-1.5cm \tau\to \frac{a \tau+ b}{c \tau + d}\, \quad \quad {\rm where} \quad a d- b c = 1\,;\quad a,\ b,\ c,\ d\in \mathbb{Z}
\eea
Under this $SL(2, \mathbb{Z})$ transformation, complex structure moduli and the Einstein frame Calabi Yau volume (${\cal V}_E$) are invariant, and so the tree level K\"ahler potential given in eqn. (\ref{eq:K}) transforms as:
\bea
\label{eq:S-duality1}
e^K \longrightarrow |c \, \tau + d|^2 \, e^K\, , 
\eea
and subsequently the S-duality invariance of physical quantities (e.g. gravitino mass-square which involves a factor of $e^K |W|^2$-type) suggests that the holomorphic superpotential, $W$ should have a modularity of weight $-1$, and so one has \cite{Font:1990gx, Cvetic:1991qm, Grimm:2007xm}
\bea
\label{eq:modularW}
& & W \to \frac{W}{c \, \tau + d}
\eea
Now, in a given flux compactification scenario, the various fluxes have to readjust among themselves to respect this modularity condition (\ref{eq:modularW}). Such $SL(2, \IZ)$ transformations involved in the flux adjustments have two physically different cases which, in our conventions \cite{Shukla:2015bca}, are described as under 
\begin{itemize}
\item{{\bf First transformation ($\tau \to -\frac{1}{\tau}$):} This is popularly known as strong-weak duality, and in addition, results in the following transformations,
\bea
\label{eq:S-duality}
& &  \hskip-1.5cm B_2 \to C_2, \quad C_2 \to - \, B_2, \quad \quad \quad G^a \to - \, \frac{G^a}{\tau},  \quad \quad \quad T_{\alpha} \to {T_{\alpha}}- \frac{1}{2} \frac{\hat{\kappa}_{\alpha a b} G^a G^b}{\tau}\\
%& & C_4 \to C_4, \,\,C_8 \to \tilde C_8,\,\, \tilde C_8 \to C_8,\,\, C'_8 \to - C'_8,\\
& & \hskip-1.0cm {H} \to {F}, \quad {F} \to -{H}, \quad \quad  \quad {Q} \to - {P}, \quad {P}\to {Q} \,. \nonumber
\eea}
\item{{\bf Second transformation ($\tau \to \tau + 1$):} This corresponds to a shift in the universal axion $C_0 \to C_0 + 1$ along with the following additional transformations,
\bea
\label{eq:S-duality1}
& &  \hskip-2.5cm B_2 \to B_2, \quad C_2 \to C_2 - \, B_2, \quad \quad \quad G^a \to \, {G^a},  \quad \quad \quad T_{\alpha} \to {T_{\alpha}}\\
%& & C_4 \to C_4, \,\,C_8 \to \tilde C_8,\,\, \tilde C_8 \to C_8,\,\, C'_8 \to - C'_8,\\
& & \hskip-2.5cm {H} \to {H}, \quad {F} \to {F}-{H}, \hskip2cm {Q} \to {Q} - {P}, \quad {P}\to {P} \,. \nonumber
\eea}
\end{itemize}
From the point of view of K\"ahler potential and the superpotential, the later case amounts to have some constant rescalings $e^K \to |d|^2\, e^K$ and $W \to W/d$ as $\tau \to \tau +1$ simply implies $c=0$. Hence, from now onwards our main focus will be only on the first case, i.e. on strong/weak duality. 

With these ingredients in hand, a generic modular completed form of the flux superpotential can be given as under \cite{Aldazabal:2006up,Aldazabal:2008zza, Guarino:2008ik,Blumenhagen:2015kja},
\bea
\label{eq:W1}
& & \hskip-0.5cm W  = - \int_{X} \biggl[\left({F} +\tau \, {H} \right)+ \, \omega_a {G}^a + \,\left( {\hat Q}^{\alpha} + \tau  {\hat P}^{\alpha}\right)  \,{T}_\alpha - \frac{1}{2} {\hat P}^{\alpha}\, \hat{\kappa}_{\alpha a b} \, G^a\, G^b \biggr]_3 \wedge \Omega_3. 
\eea
Now using eqn. (\ref{eq:Omega3}), this generic flux superpotential $W$ can be equivalently written as,
\begin{eqnarray}
\label{eq:W_gen}
& & W = e_\Lambda \, {\cal X}^\Lambda + m^\Lambda \, {\cal F}_\Lambda,
\end{eqnarray} 
where the components of symplectic vector $e_\Lambda$ and $m^\Lambda$ are given as,
\begin{eqnarray}
\label{eq:eANDm}
& &  e_\Lambda = \left({F}_\Lambda + \tau \, H_\Lambda \right) + \omega_{a\, \Lambda}\, G^a+ \left( \hat{Q}^\alpha{}_{\Lambda} +\tau \, \hat{P}^\alpha{}_{\Lambda} \right) \, T_\alpha - \frac{1}{2} {\hat P}^{\alpha}{}_{\Lambda}\, \hat{\kappa}_{\alpha a b} \, G^a\, G^b , \, \\
& &  m^\Lambda = \left({F}^\Lambda + \tau \, {H}^\Lambda \right) + {\omega}_a{}^{\Lambda}\, G^a + \left( \hat{Q}^{\alpha \,\Lambda} + \tau \,  \hat{P}^{\alpha \,\Lambda}\right) \, T_\alpha  \,- \frac{1}{2} {\hat P}^{\alpha \Lambda}\, \hat{\kappa}_{\alpha a b} \, G^a\, G^b .\nonumber
\end{eqnarray} 
Using the superpotential (\ref{eq:W_gen}), one can compute the various derivatives with respect to chiral variables, $\tau, G^a$ and $T_\alpha$ to be given as followings,
\begin{eqnarray}
\label{eq:DerW}
&  W_\tau = \left(H_\Lambda + \hat{P}^\alpha{}_\Lambda \, T_\alpha\right) \, {\cal X}^\Lambda + \left( {H}^\Lambda + \hat{P}^{\alpha \Lambda} \, T_\alpha\right) \, {\cal F}_\Lambda,  \quad \nonumber\\
&  W_{G^a} = \left(\omega_{a \Lambda} - \hat{P}^\alpha{}_\Lambda \, \hat{\kappa}_{\alpha a b} \, G^b\right) \, {\cal X}^\Lambda + \left( {\omega}_a{}^\Lambda - \hat{P}^{\alpha \Lambda} \, \hat{\kappa}_{\alpha a b} \, G^b \right)\, {\cal F}_\Lambda, \quad \nonumber\\
& W_{T_\alpha} = \left(\hat{Q}^\alpha{}_\Lambda + \tau \hat{P}^\alpha{}_\Lambda \right)\, {\cal X}^\Lambda + \left( \hat{Q}^{\alpha \Lambda} + \tau \, \hat{P}^{\alpha \Lambda} \right)\, {\cal F}_\Lambda \quad.
\end{eqnarray}
Note that, apart from the standard $H_3$ and $F_3$ fluxes, only $\omega_a$, $\hat{Q}^\alpha$ and $\hat{P}^\alpha$ components are allowed by the choice of involution to contribute into the superpotential.

\subsection{Some useful modular-completed generalized flux-orbits}
Let us consider the components of the three-form flux factor appearing in eqns. (\ref{eq:W_gen}) and (\ref{eq:eANDm}), which after using the definitions of chiral variables in eqn. (\ref{eq:N=1_coords}) simplifies as under,
\bea
\label{eq:orbitsA0}
& & e_\Lambda = \left({\mathbb F}_\Lambda + s\, \hat{\mathbb P}^{\alpha}{}_{\Lambda} \sigma_\alpha\right)  + i \, \left(s \, {\mathbb H}_\Lambda -\hat{{\mathbb Q}}^{\alpha}{}_{\Lambda} \sigma_\alpha \right) \nonumber\\
& & m^\Lambda = \left({\mathbb F}^\Lambda + s\, \hat{\mathbb P}^{\alpha\,\Lambda} \sigma_\alpha\right)  + i \, \left(s \, {\mathbb H}^\Lambda -\hat{{\mathbb Q}}^{\alpha\,\Lambda} \sigma_\alpha \right)\,,
\eea
where recall that $\sigma_\alpha = \frac{1}{2}\, {\kappa}_{\alpha \beta \gamma} t^\beta t^\gamma$ has been used along with the following new flux-orbits \cite{Shukla:2015bca}, 
\bea
\label{eq:orbitsB1}
& \hskip-1.5cm {\mathbb H}_\Lambda =  {\bf h}_\Lambda~, \, \, \, \, \, &\hat{\mathbb Q}^{\alpha}{}_{\Lambda} ={ \bf{\hat{q}^{\alpha}{}_{\Lambda} }} + C_0 \, {\bf \hat{p}^{\alpha}{}_{\Lambda}} ~, \nonumber\\
& {\mathbb F}_\Lambda=   {\bf f}_\Lambda + C_0 \, ~{\bf h}_\Lambda~, \, \, \, \, \, &\hat{\mathbb P}^{\alpha}{}_{\Lambda} = {\bf \hat{p}^{\alpha}{}_{\Lambda}}~; \nonumber\\
& & \\
& \hskip-1.5cm {\mathbb H}^\Lambda =  {\bf h}^\Lambda~, \, \, \, \, \, &{\hat{\mathbb Q}}^{\alpha \Lambda} = {\bf \hat{q}^{\alpha \Lambda}} + C_0 \, {\bf \hat{p}^{\alpha \Lambda}} ~, \nonumber\\
& {\mathbb F}^\Lambda=   {\bf f}^\Lambda + C_0 \, ~{\bf h}^\Lambda~, \, \, \, \, \, &{\hat{\mathbb P}}^{\alpha \Lambda} = {\bf {\hat{p}}^{\alpha \Lambda}}~, \nonumber
\eea
where  
\bea
\label{eq:orbits11A}
& &  {\bf h}_\Lambda = H_\Lambda + (\omega_{a\Lambda} \, {b}^a) + \hat{Q}^\alpha{}_\Lambda \, \left(\frac{1}{2}\, \hat{\kappa}_{\alpha a b} b^a b^b\right)  + \hat{P}^\alpha{}_\Lambda \, \left(\tilde{\rho}_\alpha -\frac{1}{2} \hat{\kappa}_{\alpha a b} c^a b^b\right) \nonumber\\
& &  {\bf f}_\Lambda = F_\Lambda + (\omega_{a\Lambda} \, {c}^a) - \hat{P}^\alpha{}_\Lambda \, \left(\frac{1}{2}\, \hat{\kappa}_{\alpha a b} c^a c^b\right) + \hat{Q}^\alpha{}_\Lambda \, \left(\tilde{\rho}_\alpha +\frac{1}{2} \hat{\kappa}_{\alpha a b} c^a b^b\right)\,, \\
& &  {\bf h}^\Lambda = H^\Lambda + (\omega_{a}{}^{\Lambda} \, {b}^a) + \hat{Q}^{\alpha \Lambda} \, \left(\frac{1}{2}\, \hat{\kappa}_{\alpha a b} b^a b^b\right)  + \hat{P}^{\alpha \Lambda} \, \left(\tilde{\rho}_\alpha -\frac{1}{2} \hat{\kappa}_{\alpha a b} c^a b^b\right) \nonumber\\
& &  {\bf f}^\Lambda = F^\Lambda + (\omega_{a}{}^{\Lambda} \, {c}^a) - \hat{P}^{\alpha \Lambda} \, \left(\frac{1}{2}\, \hat{\kappa}_{\alpha a b} c^a c^b\right) + 
\hat{Q}^{\alpha \Lambda} \, \left(\tilde{\rho}_\alpha +\frac{1}{2} \hat{\kappa}_{\alpha a b} c^a b^b\right), \nonumber\\
& & {\bf \hat{q}^{\alpha}{}_{\Lambda}} = \hat{Q}^{\alpha}{}_{\Lambda} , \quad {\bf \hat{q}^{\alpha \Lambda}} = \hat{Q}^{\alpha \Lambda} , \qquad  {\bf \hat{p}^{\alpha}{}_{\Lambda}} = \hat{P}^{\alpha}{}_{\Lambda}, \quad {\bf {\hat{p}}^{\alpha \Lambda}} = {\hat{P}}^{\alpha \Lambda} \, .\nonumber
\eea
Here note that $\tilde{\rho}_\alpha = {\rho}_\alpha + \frac{1}{2} \hat{\kappa}_{\alpha a b} c^a b^b$ is the S-duality invariant axionic combination. Moreover, we have modular completion of generalized geometric flux orbits as under,
\bea
\label{eq:orbits11B}
& & {\bf \mho_{a\Lambda}}\equiv {\bf  \omega_{a\Lambda}} = \omega_{a\Lambda} + \hat{Q}^\alpha{}_\Lambda \, \left(\hat{\kappa}_{\alpha a b}\, b^b\right) -\, \hat{P}^\alpha{}_\Lambda \, \left(\hat{\kappa}_{\alpha a b}\, c^b\right) \\
& & {\bf \mho_{a}{}^{\Lambda}} \equiv {\bf \omega_{a}{}^{\Lambda}}= \omega_{a}{}^{\Lambda} + \hat{Q}^{\alpha \Lambda} \, \left(\hat{\kappa}_{\alpha a b}\, b^b\right)-\hat{P}^{\alpha \Lambda} \, \left(\hat{\kappa}_{\alpha a b}\, c^b\right)\nonumber
\eea
Using these crucial flux combinations will help us to club many terms of the scalar potential leading to a very compact representation as we will see later. Moreover, as a supplementary remark, let us recall (from \cite{Shukla:2015bca}) that the new generalized flux combinations in eqns. (\ref{eq:orbits11A})-(\ref{eq:orbits11B}) transform under S-duality: $\tau \to -\frac{1}{\tau}$ as,
\bea
\label{eq:orbits11C}
& & \hskip-1.5cm  {\bf f^\Lambda} \to  -{\bf h^\Lambda}, \quad {\bf h^\Lambda} \to {\bf f^\Lambda}, \quad  {\bf {\cal \omega}_{a}{}^{\Lambda}} \to {\bf {\cal \omega}_{a}{}^{\Lambda}},  \quad {\bf \hat{q}^{\alpha \Lambda}} \to - {\bf \hat{p}^{\alpha \Lambda}},   \quad  {\bf \hat{p}^{\alpha \Lambda}} \to  {\bf \hat{q}^{\alpha \Lambda}} \\
& & \hskip-1.5cm {\bf f_\Lambda} \to  -{\bf h_\Lambda}, \quad {\bf h_\Lambda} \to {\bf f_\Lambda}, \quad  {\bf {\cal \omega}_{a\Lambda}} \to {\bf {\cal \omega}_{a \Lambda}},  \quad {\bf \hat{q}^{\alpha}{}_{\Lambda}} \to - {\bf \hat{p}^{\alpha}{}_{\Lambda}},   \quad  {\bf \hat{p}^{\alpha}{}_{ \Lambda}} \to  {\bf \hat{q}^{\alpha}{}_{\Lambda}} \nonumber
\eea
while under $\tau \to \tau +1$ one finds the same transforming as,
\bea
\label{eq:orbits11D}
& & \hskip-1.9cm  {\bf f^\Lambda} \to {\bf f^\Lambda} -{\bf h^\Lambda}, \quad {\bf h^\Lambda} \to {\bf h^\Lambda}, \quad  {\bf {\cal \omega}_{a}{}^{\Lambda}} \to {\bf {\cal \omega}_{a}{}^{\Lambda}},  \quad {\bf \hat{q}^{\alpha \Lambda}} \to {\bf \hat{q}^{\alpha \Lambda}} - {\bf \hat{p}^{\alpha \Lambda}},   \quad  {\bf \hat{p}^{\alpha \Lambda}} \to  {\bf \hat{p}^{\alpha \Lambda}} \\
& & \hskip-1.9cm {\bf f_\Lambda} \to {\bf f_\Lambda} -{\bf h_\Lambda}, \quad {\bf h_\Lambda} \to {\bf h_\Lambda}, \quad  {\bf {\cal \omega}_{a\Lambda}} \to {\bf {\cal \omega}_{a \Lambda}},  \quad {\bf \hat{q}^{\alpha}{}_{\Lambda}} \to {\bf \hat{q}^{\alpha}{}_{\Lambda}} - {\bf \hat{p}^{\alpha}{}_{\Lambda}},   \quad  {\bf \hat{p}^{\alpha}{}_{ \Lambda}} \to  {\bf \hat{p}^{\alpha}{}_{\Lambda}} \nonumber
\eea
which simply implies that all orbits with fluxes ${\mathbb H}, {\mathbb F}, \hat{\mathbb Q}$ and $\hat{\mathbb P}$ as defined in eqn. (\ref{eq:orbitsB1})
are invariant under $\tau \to \tau +1$. These will be the ones directly appearing in our symplectic formulation as we will see later in the upcoming sections.

\subsection{Some important symplectic identities}
Let us also recollect some relevant ingredients for rewriting the $F$-terms scalar potential into a symplectic formalism. The strategy we follow is an extension of the previous proposal made in \cite{Shukla:2015hpa}. To be specific, for simplifying the complex structure moduli dependent piece of the scalar potential, we will use the following symplectic ingredients,
\begin{itemize}
\item{A symplectic identity \cite{Ceresole:1995ca},
\begin{eqnarray}
\label{eq:Identity1}
&& K^{i \ov j} \, (D_i {\cal X}^{\Lambda}) \, (\ov D_{\ov j} \ov{{\cal X}^{\Delta}}) = - \ov{{\cal X}^{\Lambda}} \,  {\cal X}^{\Delta} - \frac{1}{2} \, e^{-K_{cs}} \, {\rm Im{\cal N}}^{\Lambda \Delta}
\end{eqnarray}
where the period matrix ${\cal N}$ for the involutively odd (2,1)-cohomology sector is given as,
\bea
\label{eq:periodN}
& & {\cal N}_{\Lambda\Delta} = \ov{\cal F}_{\Lambda\Delta} + 2 \, i \, \frac{Im({\cal F}_{\Lambda\Gamma}) \, {\cal X}^\Gamma X^\Sigma \, (Im{\cal F}_{\Sigma \Delta}) }{Im({\cal F}_{\Gamma\Sigma}) {\cal X}^\Gamma X^\Sigma}
\eea
}
\item{Using period matrix components, one can introduce the following definitions of the new-matrices (${\cal M}$) for computing the hodge star of various odd three-forms \cite{Ceresole:1995ca}, 
\begin{eqnarray}
\label{stardef}
&& \star \, {\cal A}_\Lambda =  {\cal M}_{\Lambda}^{\, \, \, \, \, \Sigma} \, \, {\cal A}_\Sigma + {\cal M}_{\Lambda \Sigma}\, \, {\cal B}^\Sigma, \, \, \, {\rm and} \, \, \, \, \star\, {\cal B}^\Lambda = {\cal M}^{\Lambda \Sigma} \, \, {\cal A}_\Sigma + {\cal M}^\Lambda_{\, \, \, \, \Sigma} \, \, {\cal B}^\Sigma
\end{eqnarray}
where we also define the following useful components to be utilized later on,
\begin{eqnarray}
\label{coff}
&& {\cal M}^{\Lambda \Delta} = {\rm Im{\cal N}}^{\Lambda \Delta}, \qquad \qquad \qquad {\cal M}_{\Lambda}^{\, \, \, \, \, \Delta}  = {\rm Re{\cal N}}_{\Lambda \Gamma} \, \, {\rm Im{\cal N}}^{\Gamma \Delta} \\
&& {\cal M}^\Lambda_{\, \, \, \, \Delta} =- \left({\cal M}_{\Lambda}^{\, \, \, \, \, \Delta}\right)^{T} , \qquad \qquad {\cal M}_{\Lambda \Delta}\, =  -{\rm Im{\cal N}}_{\Lambda \Delta} -{\rm Re{\cal N}}_{\Lambda \Sigma} \, \, {\rm Im{\cal N}}^{\Sigma \Gamma}\, \, {\rm Re{\cal N}}_{\Gamma \Delta} \nonumber
\end{eqnarray}
}
\item{It was observed in \cite{Shukla:2015hpa} that an interesting and very analogous relation as compared to the definition of period matrix (\ref{eq:periodN}) holds, 
\bea
\label{eq:periodF}
& & {\cal F}_{\Lambda\Delta} = \ov{\cal N}_{\Lambda\Delta} + 2 \, i \, \frac{Im({\cal N}_{\Lambda\Gamma}) \, {\cal X}^\Gamma X^\Sigma \, (Im{\cal N}_{\Sigma \Delta}) }{Im({\cal N}_{\Gamma\Sigma}) {\cal X}^\Gamma X^\Sigma}
\eea
Moreover, similar to the definition of the period matrices (\ref{coff}), one can also define another set of symplectic quantities given as under,
\begin{eqnarray}
\label{coff2}
&& {\cal L}^{\Lambda \Delta} = {\rm Im{\cal F}}^{\Lambda \Delta},\qquad \qquad \qquad {\cal L}_{\Lambda}^{\, \, \, \, \, \Delta}  = {\rm Re{\cal F}}_{\Lambda \Gamma} \, \, {\rm Im{\cal F}}^{\Gamma \Delta} \\
&& {\cal L}^\Lambda_{\, \, \, \, \Delta} =- \left({\cal L}_{\Lambda}^{\, \, \, \, \, \Delta}\right)^{T}, \qquad \qquad  {\cal L}_{\Lambda \Delta}\, =  -{\rm Im{\cal F}}_{\Lambda \Delta} -{\rm Re{\cal F}}_{\Lambda \Sigma} \, \, {\rm Im{\cal F}}^{\Sigma \Gamma}\, \, {\rm Re{\cal F}}_{\Gamma \Delta} \nonumber
\end{eqnarray}}
\item{These two sets of matrices ${\cal M}$ and ${\cal L}$ provide the two sets of very crucial identities. The same will be important in our scalar potential rearrangement, and assuming the large complex structure limit, we will give proof of these relations for generic prepotential in the appendix \ref{sec_proof}. These identities are given as under,
\bea
\label{eq:symp10}
& & Re({\cal X}^\Lambda \, \ov{\cal X}^\Delta) = -\frac{1}{4} \, e^{-K_{cs}} \, \left( {\cal M}^{\Lambda \Delta} + \, \, {\cal L}^{\Lambda \Delta} \right) \,%:= -\frac{1}{4} \, e^{-K_{cs}} \, \left( {\cal M}_1\right)^{\Lambda \Delta} 
\nonumber\\
& & Re({\cal X}^\Lambda \, \ov{\cal F}_\Delta) = +\frac{1}{4} \, e^{-K_{cs}} \, \left({\cal M}^\Lambda_{\, \, \, \, \Delta}  +  \, \,  {\cal L}^\Lambda_{\, \, \, \, \Delta} \right)\, \,%:= + \frac{1}{4} \, e^{-K_{cs}} \, \left({\cal M}_1\right)^\Lambda_{\, \, \, \, \Delta} 
\nonumber\\
& & Re({\cal F}_\Lambda \, \ov{\cal X}^\Delta) = - \frac{1}{4} \, e^{-K_{cs}} \, \left(  {\cal M}_{\Lambda}^{\, \, \, \, \, \Delta} +  \, \, {\cal L}_{\Lambda}^{\, \, \, \, \, \Delta}\right) \, %:= - \frac{1}{4} \, e^{-K_{cs}} \, \left({\cal M}_1\right)_{\Lambda}^{\, \, \, \, \, \Delta} 
\\
& & Re({\cal F}_\Lambda \, \ov{\cal F}_\Delta) = + \frac{1}{4} \, e^{-K_{cs}} \, \left({\cal M}_{\Lambda \Delta} + \, \, {\cal L}_{\Lambda \Delta}\right) \, \,\,%:= + \frac{1}{4} \, e^{-K_{cs}} \, \left({\cal M}_1\right)_{\Lambda \Delta}  \, ,
\nonumber
\eea
and
\bea
\label{eq:symp11}
& & Im({\cal X}^\Lambda \, \ov{\cal X}^\Delta) = + \frac{1}{4} \, e^{-K_{cs}} \, \biggl[\left({\cal M}^{\Lambda}{}_{ \Sigma} \, {\cal L}^{\Sigma \Delta} + {\cal M}^{\Lambda \Sigma} \, {\cal L}_{\Sigma}{}^{ \Delta} \right) \biggr] %:= \frac{1}{4} \, e^{-K_{cs}} {\cal S}^{\Lambda \Delta} 
\nonumber\\
& & Im({\cal X}^\Lambda \, \ov{\cal F}_\Delta) = - \frac{1}{4} \, e^{-K_{cs}} \, \biggl[\left({\cal M}^{\Lambda}{}_{ \Sigma} \, {\cal L}^{\Sigma}{}_\Delta + {\cal M}^{\Lambda \Sigma} \, {\cal L}_{\Sigma \Delta}\right) - \delta^\Lambda{}_\Delta \biggr] %:= - \frac{1}{4} \, e^{-K_{cs}} {\cal S}^{\Lambda}{}_{\Delta} 
\nonumber\\
& & Im({\cal F}_\Lambda \, \ov{\cal X}^\Delta) = + \frac{1}{4} \, e^{-K_{cs}} \, \biggl[\left({\cal M}_{\Lambda}{}_{ \Sigma} \, {\cal L}^{\Sigma \Delta} + {\cal M}_{\Lambda}{}^{ \Sigma} \, {\cal L}_{\Sigma}{}^{ \Delta} \right) - \delta_\Lambda{}^\Delta\biggr] %:= \frac{1}{4} \, e^{-K_{cs}} {\cal S}_{\Lambda}{}^{\Delta}\, 
\\
& & Im({\cal F}_\Lambda \, \ov{\cal F}_\Delta) = - \frac{1}{4} \, e^{-K_{cs}} \, \biggl[\left({\cal M}_{\Lambda \Sigma} \, {\cal L}^{\Sigma}{}_{ \Delta} + {\cal M}_{\Lambda}{}^{ \Sigma} \, {\cal L}_{\Sigma \Delta} \right) \biggr]\, %:= - \frac{1}{4} \, e^{-K_{cs}} {\cal S}_{\Lambda \Delta}
. \nonumber
\eea
}
\end{itemize}
Now we end this section on review of the relevant preliminaries by mentioning the two main goals. First we represent the total $F$-term scalar potential in a completely symplectic manner via utilizing the modular invariant version of the {\it new} generalized flux orbits proposed in \cite{Shukla:2015rua}.  In this step, by considering the non-geometric $P$-flux, the present analysis will generalize the results of \cite{Shukla:2015hpa}.  Subsequently, in the second step, we would further expand out all the symplectic ingredients in terms of the various saxionic and axionic parts of the complex structure moduli.

\section{A modular completed symplectic formulation of the 4D scalar potential}
\label{sec_symplectic_formulation}
\subsection{Simplifying the scalar potential in three steps}
The block diagonal nature of K\"ahler metric facilitates the following splitting of pieces coming from generic ${N} = 1$ F-term contribution,
\begin{eqnarray}
\label{eq:V_gen}
& & \hskip-2cm e^{- K} \, V_F = K^{{\cal A} \ov {\cal B}} \, (D_{\cal A} W) \, (\ov D_{\ov {\cal B}} \ov{W}) -3 |W|^2 \equiv V_{cs} + V_{k}\, ,
\end{eqnarray}
where
\begin{eqnarray}
\label{eq:VcsVk}
& & \hskip-1.5cm V_{cs} =  K^{{i} \ov {j}} \, (D_{i} W) \, (\ov D_{\ov {j}} \ov{W}), \quad V_{k} =  K^{{A} \ov {B}} \, (D_{A} W) \, (\ov D_{\ov {B}} \ov{W}) -3 |W|^2 
\end{eqnarray}
Here, the indices $(i,j)$ corresponds to complex structure moduli $z^i$'s while the other indices $(A,B)$ are counted in rest of the chiral variables $\{\tau, G^a, T_\alpha\}$. Note that such an splitting of total scalar potential into two pieces is possible because of the block diagonal nature of the total (inverse) K\"ahler matrix in all moduli, in which complex structure moduli sector is decoupled from the rest. While we will use some symplectic ingredients to simplify the first piece $V_{cs}$, let us recollect the following inverse K\"ahler metric components, $K^{{A} \ov {B}}$ which are relevant to simplify $V_k$ \cite{Grimm:2004uq},
\begin{eqnarray}
\label{eq:InvK}
& & K^{\tau \ov{\tau}} =  4 \, s^2, \quad K^{G^a \, \ov{\tau}} =  4 \, s^2 \, b^a , \quad K^{T_\alpha \, \ov{\tau}} = 2 \, s^2 \, \hat{\kappa}_{\alpha a b} b^a b^b, \\
& & K^{G^a \, \ov{G}^b} = s\, {\cal G}^{ab} + 4 \, s^2\, b^a b^b , \quad K^{T_\alpha \, \ov{G}^a} = \,s \, {\cal G}^{ab} \, \hat{\kappa}_{\alpha b c} b^c + 2 \, s^2 \hat{\kappa}_{\alpha b c} b^b b^c \, b^a , \nonumber\\
& & K^{T_\alpha \, \ov{T}_\beta} = \frac{4}{9}\, k_0^2\, \tilde{\cal G}_{\alpha \beta}  + s \, {\cal G}^{ab}\, {\hat{\kappa}_{\alpha a c} b^c} \, {\hat{\kappa}_{\beta b d} b^d} + s^2\, {\hat{\kappa}_{\alpha a b} b^a b^b} \, {\hat{\kappa}_{\beta c d} b^c b^d}, \nonumber
\end{eqnarray}
where $\tilde{\cal G}_{\alpha \beta}=\left(({\hat{d}^{-1}})_{\alpha}{}^{\alpha'}\, {\cal G}_{\alpha' \beta'}\, ({\hat{d}^{-1}})_{\beta}{}^{\beta'}\right)$ and $\hat{\kappa}_{\alpha a b} = (\hat{d}^{-1})_\alpha{}^{\beta}\hat{k}_{\beta a b}$ \cite{Shukla:2015hpa}. Moreover, we have used the following short hand notations for ${\cal G}$ and ${\cal G}^{-1}$ components,
\begin{eqnarray}
\label{eq:genMetrices}
& & {\cal G}_{\alpha \beta} = -\frac{3}{2} \, \left( \frac{k_{\alpha \beta}}{k_0} - \frac{3}{2} \frac{k_\alpha \,k_\beta}{k_0^2}\right), \quad \quad \quad {\cal G}^{ab} = -\frac{2}{3} \, k_0\, \hat{k}^{ab} \, \\
& & {\cal G}^{\alpha \beta} = -\frac{2}{3} \, k_0 \, k^{\alpha \beta} + 2 \, t^\alpha \, t^\beta , \quad \quad \quad \quad \quad {\cal G}_{ab} = -\frac{3}{2} \frac{\hat{k}_{ab}}{k_0} \nonumber 
\end{eqnarray} 
In addition, we have introduced $k_0 = 6 \, {\cal V}_E = k_\alpha\, t^\alpha$, $k_\alpha =k_{\alpha \beta} \, t^\beta$, $k_{\alpha\beta} = k_{\alpha\beta\gamma} \, t^\gamma$ and $\hat{k}_{a b} = \hat{k}_{\alpha a b} \, t^\alpha$. Apart from the inverse K\"ahler metric components, one would need the following K\"ahler derivatives to simplify the covariant derivatives in $V_k$, 
\begin{eqnarray}
\label{eq:derK}
& & K_\tau = \frac{i}{2 \,s }\left(1 + 2 \, s \, {\cal G}_{ab} \, b^a \, b^b\right) = - K_{\ov \tau} \\
& & \hskip-1cm K_{G^a} = -2 \, i \, {\cal G}_{ab} \, b^b = - K_{\ov{G}^a}, \quad K_{T_\alpha} = -\frac{ 3\, i \, \hat{d}_\beta{}^\alpha \, t^\beta}{k_0} = - K_{\ov{T}_\alpha}, \nonumber
\end{eqnarray}
Now, considering the strategy of \cite{Shukla:2015hpa} and using the famous symplectic identity in eqn. (\ref{eq:Identity1}), one can reshuffle the pieces $V_{cs}$ and $V_k$ of the total $F$-term scalar potential (\ref{eq:V_gen}) into the following three pieces,
\begin{eqnarray}
& & e^{-K}\, V_F = V_1 + V_2 + V_3
\end{eqnarray}
where
\begin{eqnarray}
& & \hskip-1cm V_1 := -\frac{1}{2} \, e^{-K_{cs}} \, \left(e_\Lambda + m^\Sigma \ov {\cal N}_{\Sigma \Lambda} \right) \, {\rm Im{\cal N}}^{\Lambda \Delta} \,   \left(\ov e_\Delta + \ov m^\Gamma {\cal N}_{\Gamma \Delta} \right) \\
& & \hskip-1cm V_2:= - \left(e_\Lambda + m^\Sigma \ov {\cal N}_{\Sigma \Lambda} \right) \, \left(\ov {\cal X}^\Lambda {\cal X}^\Delta \right) \,   \left(\ov e_\Delta + \ov m^\Gamma {\cal N}_{\Gamma \Delta} \right) \nonumber\\
& & + \left(K^{{A} \ov {B}} \, K_A \, K_{\ov {B}} |W|^2 -\,3 |W|^2\right) + K^{{A} \ov {B}} \, \left( (K_A \, W) \, \ov{W}_{\ov B} \, + W_A\, (K_{\ov {B}} \ov W) \right) \nonumber\\
& & \hskip-1cm V_3 := K^{{A} \ov {B}} \, W_A \, \ov{W}_{\ov B} \, .\nonumber
\end{eqnarray}
The reason for such a collection can be anticipated via some more investigations and a closer look which show that in the absence of any geometric and non-geometric fluxes (the simple setup with standard $H_3/F_3$ fluxes), one finds that the total scalar potential is contained in $V_1$ (for example, see \cite{Taylor:1999ii, Blumenhagen:2003vr}), and $V_2 + V_3$ gets trivial via the No-scale structure condition ($K^{{A} \ov {B}} \, K_A \, K_{\ov {B}} =\,4$) along with some more internal cancellations. The same will be elaborated later in the upcoming section. 

Now, our goal is to rewrite these three pieces $V_1, V_2$ and $V_3$ of the scalar potential in terms of new generalized flux combinations on top of using some crucial symplectic identities. 
\subsubsection*{Simplifying $V_1$}
Using the modular completed generalized flux orbits mentioned in eqns. (\ref{eq:orbitsB1})-(\ref{eq:orbits11B}), the pieces in $V_1$ can be considered to split into the following two parts,
\bea
\label{eq:Vcs1aAndb}
& & \hskip-0.6cm V_{1} := V_{1}^{(a)} + V_{1}^{(b)} \, .\nonumber
\eea
where
\bea
\label{eq:Vcs1a}
& & \hskip-0.8cm V_{1}^{(a)} = -\frac{1}{2} \, e^{-K_{cs}} \, \left(e_\Lambda \, {\cal M}^{\Lambda \Delta} \, \ov e_\Delta - e_\Lambda \, {\cal M}^{\Lambda}_{\, \, \, \Delta} \, \ov m^\Delta+ \ov e^\Lambda \, {\cal M}_{\Lambda}^{\, \, \, \Delta} \, m_\Delta- m^\Lambda \, {\cal M}_{\Lambda \Delta} \, \ov m^\Delta\right)\nonumber\\
& &  = -\frac{1}{2} \, e^{-K_{cs}} \, \biggl[\left({\mathbb F}_\Lambda \, {\cal M}^{\Lambda \Delta} \,  {\mathbb F}_\Delta - {\mathbb F}_\Lambda \, {\cal M}^{\Lambda}_{\, \, \, \Delta} \, {\mathbb F}^\Delta+  {\mathbb F}^\Lambda \, {\cal M}_{\Lambda}^{\, \, \, \Delta} \, {\mathbb F}_\Delta- {\mathbb F}^\Lambda \, {\cal M}_{\Lambda \Delta} \, {\mathbb F}^\Delta\right) \\
& & \hskip2.0cm + s^2\, \, \left({\mathbb H}_\Lambda \, {\cal M}^{\Lambda \Delta} \,  {\mathbb H}_\Delta - {\mathbb H}_\Lambda \, {\cal M}^{\Lambda}_{\, \, \, \Delta} \, {\mathbb H}^\Delta+  {\mathbb H}^\Lambda \, {\cal M}_{\Lambda}^{\, \, \, \Delta} \, {\mathbb H}_\Delta- {\mathbb H}^\Lambda \, {\cal M}_{\Lambda \Delta} \, {\mathbb H}^\Delta\right) \nonumber\\
& & \hskip2.0cm + \left(\hat{\mathbb Q}_\Lambda \, {\cal M}^{\Lambda \Delta} \,  \hat{\mathbb Q}_\Delta - \hat{\mathbb Q}_\Lambda \, {\cal M}^{\Lambda}_{\, \, \, \Delta} \, \hat{\mathbb Q}^\Delta+  \hat{\mathbb Q}^\Lambda \, {\cal M}_{\Lambda}^{\, \, \, \Delta} \, \hat{\mathbb Q}_\Delta- \hat{\mathbb Q}^\Lambda \, {\cal M}_{\Lambda \Delta} \, \hat{\mathbb Q}^\Delta\right) \nonumber\\
& & \hskip2.0cm + s^2 \, \left(\hat{\mathbb P}_\Lambda \, {\cal M}^{\Lambda \Delta} \,  \hat{\mathbb P}_\Delta - \hat{\mathbb P}_\Lambda \, {\cal M}^{\Lambda}_{\, \, \, \Delta} \, \hat{\mathbb P}^\Delta+  \hat{\mathbb P}^\Lambda \, {\cal M}_{\Lambda}^{\, \, \, \Delta} \, \hat{\mathbb P}_\Delta- \hat{\mathbb P}^\Lambda \, {\cal M}_{\Lambda \Delta} \, \hat{\mathbb P}^\Delta\right) \nonumber\\
& & \hskip2.0cm  -2 \, s \, \, \left({\mathbb H}_\Lambda \, {\cal M}^{\Lambda \Delta} \,  \hat{\mathbb Q}_\Delta - {\mathbb H}_\Lambda \, {\cal M}^{\Lambda}_{\, \, \, \Delta} \, \hat{\mathbb Q}^\Delta+  {\mathbb H}^\Lambda \, {\cal M}_{\Lambda}^{\, \, \, \Delta} \, \hat{\mathbb Q}_\Delta- {\mathbb H}^\Lambda \, {\cal M}_{\Lambda \Delta} \, \hat{\mathbb Q}^\Delta\right) \nonumber\\
& & \hskip2.0cm  +2 \, s \, \, \left({\mathbb F}_\Lambda \, {\cal M}^{\Lambda \Delta} \,  \hat{\mathbb P}_\Delta - {\mathbb F}_\Lambda \, {\cal M}^{\Lambda}_{\, \, \, \Delta} \, \hat{\mathbb P}^\Delta+  {\mathbb F}^\Lambda \, {\cal M}_{\Lambda}^{\, \, \, \Delta} \, \hat{\mathbb P}_\Delta- {\mathbb F}^\Lambda \, {\cal M}_{\Lambda \Delta} \, \hat{\mathbb P}^\Delta\right)\biggr] \nonumber
\eea
and
\bea\label{eq:Vcs1b}
& & \hskip-1cm V_{1}^{(b)} = \frac{i}{2} \, e^{-K_{cs}} \,\left(\ov e_\Lambda  m^\Lambda - e_\Lambda  \, \ov m^\Lambda \right) \\
& & = -\frac{1}{2} \, e^{-K_{cs}} \, \biggl[2\, s\, \,\left({\mathbb F}_\Lambda \, {\mathbb H}^\Lambda - {\mathbb F}^\Lambda \, {\mathbb H}_\Lambda \right) + 2\, s\, \,\left(\hat{\mathbb Q}_\Lambda \, \hat{\mathbb P}^\Lambda - \hat{\mathbb Q}^\Lambda \, \hat{\mathbb P}_\Lambda \right) \nonumber\\
& & \hskip1cm -2\, \left({\mathbb F}_\Lambda \, \hat{\mathbb Q}^{\Lambda} - {\mathbb F}^\Lambda \,\hat{\mathbb Q}_\Lambda \right)  - 2\,s^2\, \left({\mathbb H}_\Lambda \, \hat{\mathbb P}^{\Lambda} - {\mathbb H}^\Lambda \,\hat{\mathbb P}_\Lambda \right)\biggr]\nonumber
\eea
{\it where $\hat{\mathbb Q}_\Lambda:= \hat{\mathbb Q}^\alpha{}_\Lambda \sigma_\alpha$ and $\hat{\mathbb Q}^\Lambda:= \hat{\mathbb Q}^{\alpha{}\Lambda} \sigma_\alpha$ have been utilized in these expressions, and this convention would be used wherever the $Q/P$ fluxes are seen without an $h^{1,1}_+(X)$-index, $\alpha$.} %The piece (\ref{eq:Vcs1b}) combines various NS-NS and RR-Bianchi identities and can be considered as more generalized RR Tadpoles  on the lines of \cite{Gao:2015nra,Shukla:2015bca}, and these have to vanish by adding local sources. 
\subsubsection*{Simplifying $V_2$}
As we see, there are three pieces within $V_2$, and the first one coming from the complex structure sector contributions can be rewritten as under,
\bea
& & \hskip-0.6cm V_{2}^{(a)} := - \left(e_\Lambda + m^\Sigma \ov {\cal N}_{\Sigma \Lambda} \right) \, \left(\ov {\cal X}^\Lambda {\cal X}^\Delta \right) \,   \left(\ov e_\Delta + \ov m^\Gamma {\cal N}_{\Gamma \Delta} \right) \\
&& \hskip0.2cm =  - e_\Lambda \, (\ov {\cal X}^\Lambda {\cal X}^\Delta) \, \ov e_\Delta - m^\Lambda \, (\ov {\cal F}_\Lambda {\cal X}^\Delta) \, \ov e_\Delta - e_\Lambda \, (\ov {\cal X}^\Lambda {\cal F}_\Delta) \, \ov m^\Delta - m^\Lambda \, (\ov {\cal F}_\Lambda {\cal F}_\Delta) \, \ov m^\Delta  \nonumber\\
& & \hskip0.2cm = - \left(e_\Delta \, \ov {\cal X}^\Delta + m^\Delta \,\ov {\cal F}_\Delta \right) \left(\ov e_\Lambda \, {\cal X}^\Lambda + \ov m^\Lambda \, {\cal F}_\Lambda \right)   \nonumber
\eea
The remaining two pieces of $V_2$ can be expressed as,
\bea
& & \hskip-0.6cm V_{2}^{(b)} := \left(K^{{A} \ov {B}} \, K_A \, K_{\ov {B}} |W|^2 -\,3 |W|^2\right) = |W|^2 \equiv \left(e_\Lambda \,  {\cal X}^\Lambda + m^\Lambda \, {\cal F}_\Lambda \right) \left(\ov e_\Delta \, \ov{\cal X}^\Delta + \ov m^\Delta \, \ov{\cal F}_\Delta \right) \nonumber\\
& & \hskip-0.6cm V_{2}^{(c)} := K^{{A} \ov {B}} \, \left( (K_A \, W) \, \ov{W}_{\ov B} \, + W_A\, (K_{\ov {B}} \ov W) \right) = - 2 \, |W|^2 + \biggl\{W \, \left(e_\Delta \, \ov {\cal X}^\Delta + m^\Delta \,\ov {\cal F}_\Delta \right) \nonumber\\
& & + \ov W\, \left(\ov e_\Lambda \, {\cal X}^\Lambda + \ov m^\Lambda \, {\cal F}_\Lambda \right) \biggr\} - 4 \,s \biggl\{W \, \left({\hat{P}}_\Delta \, \ov {\cal X}^\Delta + {\hat{P}}^\Delta \,\ov {\cal F}_\Delta \right) + \ov W\, \left({\hat{P}}_\Lambda \, {\cal X}^\Lambda + {\hat{P}}^\Lambda \, {\cal F}_\Lambda \right) \biggr\} \,, \nonumber
\eea
where in the last step, we have used the following useful relations which can be easily proven by using eqns. (\ref{eq:InvK})-(\ref{eq:derK}),
\begin{eqnarray}
& & {K}_A\, {K}^{{A} \ov {\tau}} \,  = (\tau -\ov \tau) = - {K}^{{\tau} \ov {B}} \, {K}_{\ov B} \nonumber\\
& & {K}_A\, {K}^{{A} \ov {G^a}} \,  = (G^a -\ov G^a) = - {K}^{{G^a} \ov {B}} \, {K}_{\ov B} \\
& & {K}_A\, {K}^{{A} \ov {T_\alpha}} \, = (T_\alpha -\ov T_\alpha) = - {K}^{{T_\alpha} \ov {B}} \,  {K}_{\ov B}\nonumber
\end{eqnarray}
Note that the last piece in $V_{2}^{(c)}$ arises from the quadratic dependence of the superpotential on the odd moduli $G^a$ which is absent in the analysis of \cite{Shukla:2015hpa} where no $P$-fluxes are present. However, let us mention that this is not the only new contribution in $V_2$ as the components of $(e_\Lambda, m^\Lambda)$ (wherever they appear) also have implicit dependence on the $P$-flux as can be reminded from new flux orbits in eqns. (\ref{eq:orbits11A})-(\ref{eq:orbits11B}). Now summing everything up, we get
\bea
\label{eq:V2xxx}
& & \hskip-1.1cm V_2 = (e_\Lambda - \ov e_\Lambda) Re({\cal X}^\Lambda \ov {\cal X}^\Delta) (e_\Delta - \ov e_\Delta) + (e_\Lambda - \ov e_\Lambda) Re({\cal X}^\Lambda \ov {\cal F}_\Delta) (m^\Delta - \ov m^\Delta) \nonumber\\
& & \hskip0.3cm + (m^\Lambda - \ov m^\Lambda) Re({\cal F}_\Lambda \ov {\cal X}^\Delta) (e_\Delta - \ov e_\Delta) + (m^\Lambda - \ov m^\Lambda) Re({\cal F}_\Lambda \ov {\cal F}_\Delta) (m^\Delta - \ov m^\Delta) \\
& & - 4 \,s \biggl\{W \, \left({\hat{P}}_\Delta \, \ov {\cal X}^\Delta + {\hat{P}}^\Delta \,\ov {\cal F}_\Delta \right) + \ov W\, \left({\hat{P}}_\Lambda \, {\cal X}^\Lambda + {\hat{P}}^\Lambda \, {\cal F}_\Lambda \right) \biggr\} \nonumber
\eea 
Now, the eqn. (\ref{eq:V2xxx}) can be further re-expressed using generalized flux orbits as, 
\begin{eqnarray}
\label{eq:V2final}
& & \hskip-0.65cm V_2 = \biggl[-4\, \,Re({\cal X}^\Lambda \ov {\cal X}^\Delta) \biggl\{s^2 \, \left({\mathbb H}_\Lambda \, {\mathbb H}_\Delta + \hat{\mathbb P}_\Lambda \, \hat{\mathbb P}_\Delta \right) + \left(\hat{\mathbb Q}_\Lambda \, \hat{\mathbb Q}_\Delta +\, s^2 \, \hat{\mathbb P}_\Lambda \, \hat{\mathbb P}_\Delta \right)\nonumber\\
& & \hskip2cm  -  \, s \, {\mathbb H}_\Lambda \, \hat{\mathbb Q}_\Delta-  \, s \, {\mathbb H}_\Delta \, \hat{\mathbb Q}_\Lambda +  \, s \, {\mathbb F}_\Lambda \, \hat{\mathbb P}_\Delta +  \, s \, {\mathbb F}_\Delta \, \hat{\mathbb P}_\Lambda \biggr\} \\
& & + 4\, \,Im({\cal X}^\Lambda \ov {\cal X}^\Delta) \biggl\{ s^2 \, \left({\mathbb H}_\Lambda \, \hat{\mathbb P}_\Delta - \hat{\mathbb P}_\Lambda \, {\mathbb H}_\Delta \right) +  \, s \, \hat{\mathbb P}_\Lambda \, \hat{\mathbb Q}_\Delta-  \, s \, \hat{\mathbb P}_\Delta \, \hat{\mathbb Q}_\Lambda\biggr\} \biggr]\nonumber\\
& & + \biggl[... \biggr]  + \biggl[ ...\biggr] + \biggl[ ...\biggr] \, . \nonumber
%&& +4\, \,Re({\cal X}^\Lambda \ov {\cal F}_\Delta) \biggl[-s^2 \, {\mathbb H}_\Lambda \, {\mathbb H}^\Delta +  \, s \, {\mathbb H}_\Lambda \, \hat{\mathbb Q}^\Delta +  \, s \, {\mathbb H}^\Delta \, \hat{\mathbb Q}_\Lambda - \hat{\mathbb Q}_\Lambda \, \hat{\mathbb Q}^\Delta \biggr] \\
%& & +4\, \,Re({\cal F}_\Lambda \ov {\cal X}^\Delta) \biggl[-s^2 \, {\mathbb H}^\Lambda \, {\mathbb H}_\Delta +  \, s \, {\mathbb H}^\Lambda \, \hat{\mathbb Q}_\Delta +  \, s \, {\mathbb H}_\Delta \, \hat{\mathbb Q}^\Lambda - \hat{\mathbb Q}^\Lambda \, \hat{\mathbb Q}_\Delta \biggr] \nonumber\\
%& & + 4\, \,Re({\cal F}_\Lambda \ov {\cal F}_\Delta) \biggl[-s^2 \, {\mathbb H}^\Lambda \, {\mathbb H}^\Delta +  \, s \, {\mathbb H}^\Lambda \, \hat{\mathbb Q}^\Delta +  \, s \, {\mathbb H}^\Delta \, \hat{\mathbb Q}^\Lambda - \hat{\mathbb Q}^\Lambda \, \hat{\mathbb Q}^\Delta \biggr] \nonumber
\end{eqnarray}
where the three brackets $[....]$ have the analogous terms involving the symplectic ingredients $\left(Re({\cal X}^\Lambda \ov {\cal F}_\Delta), Im({\cal X}^\Lambda \ov {\cal F}_\Delta) \right)$, $\left(Re({\cal F}_\Lambda \ov {\cal X}^\Delta), Im({\cal F}_\Lambda \ov {\cal X}^\Delta) \right)$ and $\left(Re({\cal F}_\Lambda \ov {\cal F}_\Delta), \,Im({\cal F}_\Lambda \ov {\cal F}_\Delta) \right)$ respectively. Here we observe that in the absence of non-geometric $P$-flux, the terms with factors $Im({\cal X}^\Lambda \ov {\cal X}^\Delta), \, Im({\cal X}^\Lambda \ov {\cal F}_\Delta), Im({\cal F}_\Lambda \ov {\cal X}^\Delta)$ and $Im({\cal F}_\Lambda \ov {\cal F}_\Delta)$ are absent, and therefore have not been observed in the analysis of \cite{Shukla:2015hpa}. In addition, we note that there are two pieces of ${\mathbb P}{\mathbb P}$-type, each of them are clubbed with terms of ${\mathbb H}{\mathbb H}$- and ${\mathbb Q}{\mathbb Q}$-types as seen in the first line of eqn. (\ref{eq:V2final}). The reason for the same will be clear after considering the simplifications in the third piece, $V_3$ of the potential which we discuss now.

\subsubsection*{Simplifying $V_3$}
Now considering the inverse K\"ahler metric in eqn. (\ref{eq:InvK}) along with derivatives of the superpotential in (\ref{eq:DerW}), and using the new generalized flux orbits in eqns. (\ref{eq:orbitsB1})-(\ref{eq:orbits11B}), one gets the following rearrangement of $V_3$ after a very tedious clubbing of various pieces,
\begin{eqnarray}
\label{eq:V3final}
& & \hskip-0.6cm V_3 := K^{{A} \ov {B}} \, W_A \, \ov{W}_{\ov B} \\
& & = \biggl[4 \,Re({\cal X}^\Lambda \ov {\cal X}^\Delta) \biggl\{s^2 \left({\mathbb H}_\Lambda {\mathbb H}_\Delta + \hat{\mathbb P}_\Lambda \hat{\mathbb P}_\Delta \right) + \frac{s}{4}  {\mho}_{\Lambda a} {\cal G}^{ab} {\mho}_{b\,\Delta} + \frac{k_0^2}{9}  \tilde{\cal G}_{\alpha \beta}\, \left(\hat{\mathbb Q}_\Lambda^\alpha \hat{\mathbb Q}^\beta{}_\Delta + s^2 \hat{\mathbb P}_\Lambda^\alpha  \,\hat{\mathbb P}^\beta{}_\Delta \right) \biggr\} \nonumber\\
& & \hskip0.5cm - 4\, \,Im({\cal X}^\Lambda \ov {\cal X}^\Delta) \biggl\{ s^2 \, \left({\mathbb H}_\Lambda \, \hat{\mathbb P}_\Delta - \hat{\mathbb P}_\Lambda \, {\mathbb H}_\Delta \right) +  \, s \,  \frac{k_0^2}{9}\,\,  \tilde{\cal G}_{\alpha \beta}\, \left(\hat{\mathbb P}^\alpha{}_\Lambda \, \hat{\mathbb Q}^\beta{}_\Delta-  \, \hat{\mathbb P}^\beta{}_\Delta \, \hat{\mathbb Q}^\alpha{}_\Lambda \right)\biggr\} \biggr]\nonumber\\
& & + \biggl[... \biggr]  + \biggl[ ...\biggr] + \biggl[ ...\biggr] \, . \nonumber
%&& +4\, \,Re({\cal X}^\Lambda \ov {\cal F}_\Delta) \biggl[s^2 \,{\mathbb H}_\Lambda \, {\mathbb H}^\Delta + \frac{s}{4}  \, {\mho}_{\Lambda a}\, {\cal G}^{ab}\,{\mho}_b{}^{\Delta} + \frac{k_0^2}{9}\,\hat{\mathbb Q}_\Lambda^\alpha  \, \tilde{\cal G}_{\alpha \beta}\, \hat{\mathbb Q}^{\beta\Delta} \biggr] \\
%& & +4\, \,Re({\cal F}_\Lambda \ov {\cal X}^\Delta) \biggl[s^2 \,{\mathbb H}^\Lambda \, {\mathbb H}_\Delta + \frac{s}{4}\, {\mho}^{\Lambda}{}_{a} \, {\cal G}^{ab}\, {\mho}_{b\, \Delta} + \frac{k_0^2}{9}\,\hat{\mathbb Q}^{\Lambda\alpha}  \, \tilde{\cal G}_{\alpha \beta} \, \hat{\mathbb Q}^\beta{}_\Delta \biggr] \nonumber\\
%& & + 4\, \,Re({\cal F}_\Lambda \ov {\cal F}_\Delta) \biggl[s^2 \,{\mathbb H}^\Lambda \, {\mathbb H}^\Delta + \frac{s}{4} \, {\mho}^{\Lambda}{}_{a}\, {\cal G}^{ab} \, {\mho}_b{}^{\Delta} + \frac{k_0^2}{9}\, \hat{\mathbb Q}^{\Lambda\alpha}  \, \tilde{\cal G}_{\alpha \beta} \, \hat{\mathbb Q}^{\beta\Delta} \biggr] \nonumber
\end{eqnarray}
where the three brackets $[....]$ have analogous terms as to those of the first bracket similar to the case of $V_2$ in eqn. (\ref{eq:V2final}). Now one can observe that $V_2$ and $V_3$ have some interesting cancellations, and they sum up into the following terms,
\bea
\label{eq:V2plusV3final}
& & \hskip-0.8cm V_2 + V_3 = \biggl[4 \,Re({\cal X}^\Lambda \ov {\cal X}^\Delta) \biggl\{\frac{s}{4}  {\mho}_{\Lambda a} {\cal G}^{ab}{\mho}_{b\,\Delta} + \frac{1}{4}\left(\frac{4\, k_0^2}{9} \tilde{\cal G}_{\alpha \beta}- 4 \sigma_\alpha\, \sigma_\beta \right)\, \left(\hat{\mathbb Q}_\Lambda^\alpha  \,\hat{\mathbb Q}^\beta{}_\Delta + s^2\, \hat{\mathbb P}_\Lambda^\alpha  \,\hat{\mathbb P}^\beta{}_\Delta \right) \nonumber\\
& & \hskip2cm +  \, s \, {\mathbb H}_\Lambda \, \hat{\mathbb Q}_\Delta +  s\, {\mathbb H}_\Delta \, \hat{\mathbb Q}_\Lambda -  \, s \, {\mathbb F}_\Lambda \, \hat{\mathbb P}_\Delta -  \, s \, {\mathbb F}_\Delta \, \hat{\mathbb P}_\Lambda \biggr\} \\
& & \hskip1cm - 4\, \,Im({\cal X}^\Lambda \ov {\cal X}^\Delta) \biggl\{\frac{1}{4}\left(\frac{4\, k_0^2}{9}\, \tilde{\cal G}_{\alpha \beta}- 4 \sigma_\alpha\, \sigma_\beta \right)\, \left(\hat{\mathbb P}^\alpha{}_\Lambda \, \hat{\mathbb Q}^\beta{}_\Delta-  \, \hat{\mathbb P}^\beta{}_\Delta \, \hat{\mathbb Q}^\alpha{}_\Lambda \right)\biggr\} \biggr]\nonumber\\
& & \hskip0.75cm + \biggl[... \biggr]  + \biggl[ ...\biggr] + \biggl[ ...\biggr] \, . \nonumber
\eea
One should observe that in the absence of (non-)geometric fluxes when only standard three-form fluxes $H_3$ and $F_3$ are present, the whole contribution of $|H|^2$ type is already embedded into $V_{1}$ \cite{Taylor:1999ii, Blumenhagen:2003vr}, and hence a cancellation of pure ${\mathbb H}$-flux pieces of $V_2$ and $V_3$ has been well anticipated, and the same can be now explicitly seen from our analysis. 

It is worth to mention again at this point that the development in rearrangement of terms using new generalized flux orbits have been performed in an iterative work in a series of papers \cite{Blumenhagen:2013hva, Gao:2015nra, Shukla:2015rua, Shukla:2015bca, Shukla:2015hpa}, which have set some guiding rules for next step intuitive generalization, otherwise the rearrangement even at the intermediate steps is very peculiar and it could be harder to arrive directly into a final form without earlier motivations. 

\subsection{The main proposal of the `symplectic formulation'} 
To sum up the developments made so far, we arrive at the following collection of pieces in total $F$-term contributions to the scalar potential,

\bea
\label{eq:Vsss12345}
& & \hskip-0.7cm V_F=-\frac{1}{2} \, e^{K-K_{cs}} \,\biggl[ \biggl\{{\cal M}^{\Lambda \Delta} \left( {\mathbb F}_\Lambda \,  {\mathbb F}_\Delta  + s^2\, {\mathbb H}_\Lambda \, {\mathbb H}_\Delta + \hat{\mathbb Q}_\Lambda \,  \hat{\mathbb Q}_\Delta + s^2 \, \hat{\mathbb P}_\Lambda \,  \hat{\mathbb P}_\Delta -2\, s \, {\mathbb H}_\Lambda \,  \hat{\mathbb Q}_\Delta +2\,s\, {\mathbb F}_\Lambda \,  \hat{\mathbb P}_\Delta  \right)\nonumber\\
& & \hskip2.1cm - {\cal M}^{\Lambda}{}_{\Delta} \left( {\mathbb F}_\Lambda \,  {\mathbb F}^\Delta  + s^2\, {\mathbb H}_\Lambda \, {\mathbb H}^\Delta + \hat{\mathbb Q}_\Lambda \,  \hat{\mathbb Q}^\Delta + s^2 \, \hat{\mathbb P}_\Lambda \,  \hat{\mathbb P}^\Delta -2\, s  \,{\mathbb H}_\Lambda \,  \hat{\mathbb Q}^\Delta +2\,s\, {\mathbb F}_\Lambda \,  \hat{\mathbb P}^\Delta  \right)\nonumber\\
& & \hskip2.1cm  + {\cal M}_{\Lambda}{}^{\Delta} \left( {\mathbb F}^\Lambda \,  {\mathbb F}_\Delta  + s^2\, {\mathbb H}^\Lambda \, {\mathbb H}_\Delta + \hat{\mathbb Q}^\Lambda \,  \hat{\mathbb Q}_\Delta + s^2 \, \hat{\mathbb P}^\Lambda \,  \hat{\mathbb P}_\Delta -2\, s \, {\mathbb H}^\Lambda \,  \hat{\mathbb Q}_\Delta +2\,s\, {\mathbb F}^\Lambda \,  \hat{\mathbb P}_\Delta  \right)\nonumber\\
& & \hskip2.1cm - {\cal M}_{\Lambda \Delta} \left( {\mathbb F}^\Lambda \,  {\mathbb F}^\Delta  + s^2\, {\mathbb H}^\Lambda \, {\mathbb H}^\Delta + \hat{\mathbb Q}^\Lambda \,  \hat{\mathbb Q}^\Delta + s^2 \, \hat{\mathbb P}^\Lambda \,  \hat{\mathbb P}^\Delta -2\, s  \,{\mathbb H}^\Lambda \,  \hat{\mathbb Q}^\Delta +2\,s\, {\mathbb F}^\Lambda \,  \hat{\mathbb P}^\Delta  \right)\biggr\}\nonumber\\
& & \hskip-0.5cm + 2\biggl\{s\left(\hat{\mathbb Q}_\Lambda \, \hat{\mathbb P}^\Lambda - \hat{\mathbb Q}^\Lambda \, \hat{\mathbb P}_\Lambda \right) + s\left({\mathbb F}_\Lambda \, {\mathbb H}^\Lambda - {\mathbb F}^\Lambda \, {\mathbb H}_\Lambda \right) - \left({\mathbb F}_\Lambda \, \hat{\mathbb Q}^{\Lambda} - {\mathbb F}^\Lambda \,\hat{\mathbb Q}_\Lambda \right)  - s^2 \left({\mathbb H}_\Lambda \, \hat{\mathbb P}^{\Lambda} - {\mathbb H}^\Lambda \,\hat{\mathbb P}_\Lambda \right)\biggr\} \biggr] \nonumber\\
& & \hskip-0.5cm +e^K\, \biggl[Re({\cal X}^\Lambda \ov {\cal X}^\Delta) \biggl\{\left(\frac{4\, k_0^2}{9} \tilde{\cal G}_{\alpha \beta}- 4 \sigma_\alpha\, \sigma_\beta \right)\, \left(\hat{\mathbb Q}_\Lambda^\alpha  \,\hat{\mathbb Q}^\beta{}_\Delta + s^2\, \hat{\mathbb P}_\Lambda^\alpha  \,\hat{\mathbb P}^\beta{}_\Delta \right) +  8\, s \, {\mathbb H}_\Lambda \, \hat{\mathbb Q}_\Delta - 8 \, s \, {\mathbb F}_\Lambda \, \hat{\mathbb P}_\Delta \nonumber\\
& & \hskip1cm + s\, {\mho}_{\Lambda a} {\cal G}^{ab}{\mho}_{b\,\Delta} \biggr\}  +Re({\cal X}^\Lambda \ov {\cal F}_\Delta) \biggl\{..... \biggr\} + Re({\cal F}_\Lambda \ov {\cal X}^\Delta) \biggl\{..... \biggr\} +Re({\cal F}_\Lambda \ov {\cal F}_\Delta) \biggl\{ ..... \biggr\} \biggr] \nonumber\\
& & \hskip-0.5cm - e^K\, \biggl[Im({\cal X}^\Lambda \ov {\cal X}^\Delta) \biggl\{\left(\frac{4\, k_0^2}{9}\, \tilde{\cal G}_{\alpha \beta}- 4 \sigma_\alpha\, \sigma_\beta \right)\, \left(\hat{\mathbb P}^\alpha{}_\Lambda \, \hat{\mathbb Q}^\beta{}_\Delta-  \, \hat{\mathbb P}^\beta{}_\Delta \, \hat{\mathbb Q}^\alpha{}_\Lambda \right)\biggr\} + Im({\cal X}^\Lambda \ov {\cal F}_\Delta) \biggl\{ .....\biggr\}   \nonumber\\
& & \hskip1cm + Im({\cal F}_\Lambda \ov {\cal X}^\Delta) \biggl\{ .....\biggr\} + Im({\cal F}_\Lambda \ov {\cal F}_\Delta) \biggl\{ ..... \biggr\} \biggr] \,%-\frac{1}{2} \, e^{K-K_{cs}} \, \biggl[2\, s\, \,\left({\mathbb Q}_\Lambda \, {\mathbb P}^\Lambda - {\mathbb Q}^\Lambda \, {\mathbb P}_\Lambda \right) \nonumber\\
%& & \hskip-0.5cm +2\, s\, \,\left({\mathbb F}_\Lambda \, {\mathbb H}^\Lambda - {\mathbb F}^\Lambda \, {\mathbb H}_\Lambda \right) -2\, \left({\mathbb F}_\Lambda \, \hat{\mathbb Q}^{\Lambda} - {\mathbb F}^\Lambda \,\hat{\mathbb Q}_\Lambda \right)  - 2\,s^2\, \left({\mathbb H}_\Lambda \, \hat{\mathbb P}^{\Lambda} - {\mathbb F}^\Lambda \,\hat{\mathbb Q}_\Lambda \right)\biggr]\, .
\eea
Now for arriving at a symplectic formulation, we further need to replace the quantities $Re({\cal X}^\Lambda \ov {\cal X}^\Delta), Im({\cal X}^\Lambda \ov {\cal X}^\Delta)$ etc. into the ingredients which could be written in some symplectic manner. This has been done by introducing a new class of matrices, namely ${\cal S}$ matrices constructed by using ${\cal M}$ and ${\cal L}$ matrices (as defined in eqn. (\ref{coff5})). The subsequent final result for scalar potential pieces being quite lengthy has been presented in eqn. (\ref{eq:main1}) of the appendix \ref{sec_ExpandedVersion1}. However, assuming further that the various fluxes are constant fluctuations around the background, we can re-express the symplectic matrices into their integral forms which lead to the following very compact summary of  the full scalar potential pieces collected in eqn. (\ref{eq:Vsss12345}) or equivalently in eqn. (\ref{eq:main1}),
\begin{eqnarray}
\label{eq:main2}
& & \hskip-0.65cm V_{\rm F} = -\frac{1}{4\, s\, {\cal V}_E^2} \int_{CY_3}\,\biggl[{\mathbb F} \wedge \ast {\mathbb F} + s^2 \, {\mathbb H} \wedge \ast {\mathbb H} + \hat{\mathbb Q} \wedge \ast \hat{\mathbb Q} +s^2 \,  \hat{\mathbb P} \wedge \ast \hat{\mathbb P} + \, \frac{s}{4} \,\, {\cal G}^{ab} \, \, \tilde{\mathbb \mho}_a \wedge \ast \tilde{\mathbb \mho}_b \\
& & + \, \frac{1}{4} \, \left(\frac{4\, k_0^2}{9} \, \tilde{\cal G}_{\alpha \beta}\,   - 4\, \sigma_\alpha \, \sigma_\beta \right) \left(\tilde{\cal Q}^\alpha \wedge \ast \tilde{\cal Q}^\beta + s^2\, \tilde{\cal P}^\alpha \wedge \ast \tilde{\cal P}^\beta \right) - 2 \, s \, \left({\mathbb H} \wedge \ast \hat{\mathbb Q} -{\mathbb F} \wedge \ast \hat{\mathbb P}\right)  \nonumber\\
& & - 4 \, s \, \left({\mathbb H} \wedge \ast \tilde{\cal Q} -{\mathbb F} \wedge \ast \tilde{\cal P}\right)  + \frac{1}{4} \, \times (-2\, s)\left(\frac{4\, k_0^2}{9} \, \tilde{\cal G}_{\alpha \beta}\,  - 4\, \sigma_\alpha \, \sigma_\beta \right) \left(\hat{\mathbb P}^\alpha \wedge \tilde{\cal Q}^\beta - \hat{\mathbb Q}^\alpha \wedge \tilde{\cal P}^\beta\right)  \nonumber\\
& &  - 2 \, s \, \hat{\mathbb P} \wedge \hat{\mathbb Q} + \left( 2 \, s \, {\mathbb F} \wedge {\mathbb H} - 2 \, \, {\mathbb F} \wedge \hat{\mathbb Q} - 2 \, s^2 \, \, \, {\mathbb H} \wedge \hat{\mathbb P} \right) \biggr] \, \, \, . \nonumber
\end{eqnarray}
{\it It appears to be quite remarkable that the total four dimensional scalar potential of arbitrary number of complex structure moduli, K\"ahler moduli and odd-axions has been written out so compactly in terms of symplectic ingredients along with the moduli space metrics, and also without the need of knowing the Calabi Yau metric.} Let us note here that by using ${\cal S}$ matrices (as defined in eqn. (\ref{coff5})), we have defined a new flux combination $\tilde{\cal \mho}_a$, $\tilde{\cal Q}^\alpha$ and $\tilde{\cal P}^\alpha$ as,
\begin{eqnarray}
\label{eq:tildeQP}
 & & \tilde{\mathbb \mho}_a = -\left({{\cal S}}^{\Sigma\Delta} {\mathbb \mho}_{a\Delta} + {{\cal S}}^\Sigma{}_\Delta {\mathbb \mho}_a{}^\Delta\right) \alpha_\Sigma + \left({{\cal S}}_{\Sigma}{}^{\Delta} {\mathbb \mho}_{a \Delta} + {{\cal S}}_{\Sigma\Delta} {\mathbb \mho}_a{}^\Delta\right) \beta^\Sigma \\
& & \tilde{\cal Q}^\alpha = -\left({{\cal S}}^{\Sigma\Delta} \hat{\mathbb Q}^\alpha{}_\Delta + {{\cal S}}^\Sigma{}_\Delta \hat{\mathbb Q}^{\alpha\Delta}\right) \alpha_\Sigma + \left({{\cal S}}_{\Sigma}{}^{\Delta} \hat{\mathbb Q}^\alpha{}_\Delta + {{\cal S}}_{\Sigma\Delta} \hat{\mathbb Q}^{\alpha\Delta}\right) \beta^\Sigma \nonumber\\
& & \tilde{\cal P}^\alpha = -\left({{\cal S}}^{\Sigma\Delta} \hat{\mathbb P}^\alpha{}_\Delta + {{\cal S}}^\Sigma{}_\Delta \hat{\mathbb P}^{\alpha\Delta}\right) \alpha_\Sigma + \left({{\cal S}}_{\Sigma}{}^{\Delta} \hat{\mathbb P}^\alpha{}_\Delta + {{\cal S}}_{\Sigma\Delta} \hat{\mathbb P}^{\alpha\Delta}\right) \beta^\Sigma\,. \nonumber
\end{eqnarray}
Further, the moduli space metrices appearing in this compact version (\ref{eq:main2}) of the scalar potential are given as, 
\bea
\label{eq:metricrelationsNew}
& & \hskip2.5cm {\cal G}^{ab} = -\frac{2}{3} \, k_0\, \hat{\kappa}^{ab} = - 4 \, {\cal V}_E \, \hat{k}^{ab}, \quad \\
& & \hskip-0.8cm \left(\frac{4\, k_0^2}{9} \, \tilde{\cal G}_{\alpha \beta}\,   - 4\, \sigma_\alpha \, \sigma_\beta \right) = -\frac{2}{3} \, k_0 \, ({\hat{d}}^{-1})_\alpha{}^{\alpha'} \, {k}_{\alpha'\beta'} \, ({\hat{d}}^{-1})_\beta{}^{\beta'}= - 4 \, {\cal V}_E \, ({\hat{d}}^{-1})_\alpha{}^{\alpha'} \, {k}_{\alpha'\beta'} \, ({\hat{d}}^{-1})_\beta{}^{\beta'} \, ,\nonumber
\eea
where the original definitions in eqn. (\ref{eq:genMetrices}) have been utilized along with the various intersections defined in eqn. (\ref{eq:intersection}).

%\subsection{Some comments on the modular completion of the scalar potential obtained from the DFT reduction on CY orientifolds}
%Without going into much details, let us try to observe if this proposal can help in getting a modular completed version of the scalar potential obtained from reducing Double field theory on Calabi Yau orientifolds \cite{}. For that purpose, 
\subsection{Another representation with a new flux combination}
We will close this section by providing a more compact version of the symplectic formulation presented in eqn. (\ref{eq:main2}). Let us note that many terms of the collection given in eqn. (\ref{eq:main2}) can be further clubbed together more compactly if we define the following quantities, 
\bea
\label{eq:mathfrakFH}
& & {\mathfrak F} = {\mathbb F} + s\, \, \hat{\mathbb P} \,, \quad \quad \quad  s\, {\mathfrak H} = s\, \, {\mathbb H} - \, \, \hat{\mathbb Q}
\eea 
with appropriate indices, e.g. ${\mathfrak F}_\Lambda = {\mathbb F}_\Lambda + s\, \, \left(\hat{\mathbb P}^\alpha{}_\Lambda \, \sigma_\alpha \right)$ and similarly the other ones. Subsequently, we find that 
\begin{itemize}
\item {some of the pieces of (${\cal O}_1 \wedge \ast{\cal O}_2$) type in the rearrangement are clubbed as under,
\bea
& & \hskip-1cm -\frac{1}{4\, s\, {\cal V}_E^2} \int_{CY_3}\,\biggl[{\mathbb F} \wedge \ast {\mathbb F}  +s^2 \,  \hat{\mathbb P} \wedge \ast \hat{\mathbb P} - 2 \, s \, \left({\mathbb H} \wedge \ast \hat{\mathbb Q} -{\mathbb F} \wedge \ast \hat{\mathbb P}\right) + s^2 \, {\mathbb H} \wedge \ast {\mathbb H} + \hat{\mathbb Q} \wedge \ast \hat{\mathbb Q}  \biggr]\nonumber\\
& & \hskip3cm =  -\frac{1}{4\, s\, {\cal V}_E^2} \int_{CY_3}\,\biggl[{\mathfrak F} \wedge \ast {\mathfrak F} + s^2\, {\mathfrak H} \wedge \ast {\mathfrak H} \biggr] \, .
\eea
Moreover, some among the pieces of (${\cal O}_1 \wedge {\cal O}_2$) type are combined as under,
\bea
& & \hskip-2.0cm -\frac{1}{4\, s\, {\cal V}_E^2} \int_{CY_3}\,\biggl[- 2 \, s \, \hat{\mathbb P} \wedge \hat{\mathbb Q} + 2 \, s \, {\mathbb F} \wedge {\mathbb H} - 2 \, \, {\mathbb F} \wedge \hat{\mathbb Q} - 2 \, s^2 \, \, \, {\mathbb H} \wedge \hat{\mathbb P} \, \biggr] \\
& & = -\frac{1}{2\, s\, {\cal V}_E^2} \int_{CY_3}\, s\, \, \,{\mathfrak F} \wedge {\mathfrak H} \, . \nonumber
\eea
From these two simplifications, we find that the ten pieces of eqn. (\ref{eq:main2}) can be written into just three pieces !}
\item{ Further, considering $(\mathfrak F, \mathfrak H)$ instead of using $(\mathbb F, \mathbb H)$, we have
\bea
& & \hskip-2cm  -\frac{1}{4\, s\, {\cal V}_E^2} \int_{CY_3}\,\biggl[ - 4 \, s \, \left({\mathbb H} \wedge \ast \tilde{\cal Q} -{\mathbb F} \wedge \ast \tilde{\cal P}\right) \\
& & + \frac{1}{4} \, \left(\frac{4\, k_0^2}{9} \, \tilde{\cal G}_{\alpha \beta}\,   {\bf - 4\, \sigma_\alpha \, \sigma_\beta} \right) \left(\tilde{\cal Q}^\alpha \wedge \ast \tilde{\cal Q}^\beta + s^2 \, \, \tilde{\cal P}^\alpha \wedge \ast \tilde{\cal P}^\beta \right)  \biggr] \nonumber\\
& & \hskip-2cm = -\frac{1}{4\, s\, {\cal V}_E^2} \int_{CY_3}\,\biggl[ - 4 \, s \, \left({\mathfrak H} \wedge \ast \tilde{\cal Q} -{\mathfrak F} \wedge \ast \tilde{\cal P}\right)\nonumber\\
& & + \frac{1}{4} \, \left(\frac{4\, k_0^2}{9} \, \tilde{\cal G}_{\alpha \beta}\,   {\bf + 4\, \sigma_\alpha \, \sigma_\beta} \right) \left(\tilde{\cal Q}^\alpha \wedge \ast \tilde{\cal Q}^\beta + s^2 \, \, \tilde{\cal P}^\alpha \wedge \ast \tilde{\cal P}^\beta \right)  \biggr] .\nonumber
\eea}
\end{itemize}
This way we have an equivalent representation of the total scalar potential which is slightly more compact than the one presented in eqn. (\ref{eq:main2}),
\begin{eqnarray}
\label{eq:main3}
& & \hskip-0.50cm V_{\rm F} = -\frac{1}{4\, s\, {\cal V}_E^2} \int_{CY_3}\,\biggl[{\mathfrak F} \wedge \ast {\mathfrak F} + s^2 \, {\mathfrak H} \wedge \ast {\mathfrak H} - 4 \, s \, \left({\mathfrak H} \wedge \ast \tilde{\cal Q} -{\mathfrak F} \wedge \ast \tilde{\cal P}\right)  \\
& & + \, \frac{s}{4}\, \, {\cal G}^{ab} \, \, \tilde{\mathbb \mho}_a \wedge \ast \tilde{\mathbb \mho}_b + \frac{1}{4} \, \left(\frac{4\, k_0^2}{9} \, \tilde{\cal G}_{\alpha \beta}\,  + 4\, \sigma_\alpha \, \sigma_\beta \right) \left(\tilde{\cal Q}^\alpha \wedge \ast \tilde{\cal Q}^\beta + s^2\, \tilde{\cal P}^\alpha \wedge \ast \tilde{\cal P}^\beta \right) \nonumber\\
& &   + \frac{1}{4} \, \times (-2\, s)\left(\frac{4\, k_0^2}{9} \, \tilde{\cal G}_{\alpha \beta}\,  - 4\, \sigma_\alpha \, \sigma_\beta \right) \left(\hat{\mathbb P}^\alpha \wedge \tilde{\cal Q}^\beta - \hat{\mathbb Q}^\alpha \wedge \tilde{\cal P}^\beta\right) + 2 \, s \, {\mathfrak F} \wedge {\mathfrak H} \biggr] \, \, \, . \nonumber
\end{eqnarray}
%Notice that in this compact version all the terms analogous to the 3/7-brane tadpoles are nicely fit in a single piece of type ${\mathfrak F} \wedge {\mathfrak H}$, and moreover in a completely modular invariant fashion. As argued earlier, this statement holds subject to satisfying a subset of Bianchi identities as given in eqn. (\ref{eq:BIs1}). 
Now we demonstrate the validity of our symplectic formulation in two concrete examples.

\section{Examples to demonstrate the applicability of the `symplectic formulation'}
\label{sec_2examples}
Here we will consider two explicit examples to illustrate the insights of our symplectic formulation of the four dimensional scalar potential. These examples are developed in the framework of type IIB flux compactifications on ${\mathbb T}^6/({\mathbb Z}_2\times{\mathbb Z}_2$) and ${\mathbb T}^6/{\mathbb Z}_4$ orientifolds. 
 
\subsection{Example A: Type IIB $\hookrightarrow$ ${\mathbb T}^6/({\mathbb Z}_2\times{\mathbb Z}_2$)-orientifold}
Let us fix our conventions for this setup of type IIB superstring theory compactified on  an orientifold of a sixfold $X_6 ={\mathbb T}^6 / \left(\mathbb Z_2\times \mathbb Z_2\right)$. The complex coordinates $z_i$'s on each of the tori in ${\mathbb T}^6={\mathbb T}^2\times {\mathbb T}^2\times {\mathbb T}^2$ are defined as
\bea
z^1=x^1+ U_1  \, x^2, ~ z^2=x^3+ U_2 \, x^4,~ z^3=x^5+ U_3 \, x^6 ,
\eea
where the three complex structure moduli $U_i$'s can be written as
$U_i= v_i + i\, u_i,\,\,i=1,2,3$. Further, the two $\mathbb Z_2$ orbifold actions are defined as
\bea
\label{thetaactions}
& & \theta:(z^1,z^2,z^3)\to (-z^1,-z^2,z^3), \quad \quad \quad   \ov\theta:(z^1,z^2,z^3)\to (z^1,-z^2,-z^3)\, . 
\eea
Moreover, the full orientifold action is: ${\cal O} \equiv  (\Omega_p\,(-1)^{F_L}\,  \sigma)$ in which the holomorphic involution $\sigma$ is being defined as
\bea
\label{eq:orientifold}
& & \sigma : (z^1,z^2,z^3)\rightarrow (-z^1,-z^2,-z^3)\,,
\eea
resulting in a setup with the presence of $O3/O7$-plane. The complex structure moduli dependent prepotential is given as,
\bea
\label{eq:prepotentialA}
& & {\cal F} = \frac{{\cal X}^1 \, {\cal X}^2 \, {\cal X}^3}{{\cal X}^0} = U_1 \, U_2 \, U_3
\eea 
which results in the following period-vectors,
\bea
& & {\cal X}^0=1\,, \quad   {\cal X}^1=U_1\, , \quad {\cal X}^2=U_2\, , \quad  {\cal X}^3=U_3\, , \\
& & {\cal F}_0=\, -\, \, U_1 \, U_2 \, U_3\, , \quad  {\cal F}_1= U_2 \, U_3\, , \quad  {\cal F}_2=U_3 \, U_1\, , \quad  {\cal F}_3=U_1 \, U_2\,  \nonumber
\eea
Now, the holomorphic three-form $\Omega_3=dz^1\wedge dz^2\wedge dz^3$ can be expanded as,
\bea
& & \hskip-1.5cm  \Omega_3\,  = \alpha_0 +  \, U_1 \, \alpha_1 + U_2 \, \alpha_2 + U_3 \alpha_3 +\, U_1 \, U_2 \,  U_3 \, \beta^0 -U_2 \, U_3 \, \beta^1- U_1 \, U_3 \, \beta^2 - U_1\, U_2 \, \beta^3 \, ,
\eea
where we have chosen the following basis of closed three-forms
\bea
\label{formbasis}
& & \hskip-1cm \alpha_0=1\wedge 3\wedge 5\,, \, \alpha_1=2\wedge 3\wedge 5\, , \,  \alpha_2=1\wedge 4\wedge 5\, , \, \alpha_3=1\wedge 3\wedge 6\, , \nonumber\\
& & \hskip-1cm \beta^0= 2\wedge 4\wedge 6,\, \beta^1= -1\wedge 4\wedge 6,\,  \beta^2=-2\wedge 3\wedge 6,\, \beta^3=-\, 2\wedge 4\wedge 5\, ,\nonumber
\eea
where $1\wedge 3\wedge 5 = dx^1\wedge dx^3\wedge dx^5$ etc. Subsequently we find that 
\bea
& & K_{cs} = -\ln\biggl[i \, \left( \ov{\cal X}^\Lambda {\cal F}_\Lambda - {\cal X}^\Lambda \ov{\cal F}_\Lambda \right)\biggr] = -\sum_{j = 1}^{3} \ln\left(i (U_j - \ov U_j)\right).
\eea
This also demands that $Im(U_i) <0$ which is rooted from the condition of being inside the physical domain \cite{Ceresole:1995ca,Blumenhagen:2003vr}, and this is equally important as to demand $Im(\tau) > 0$ and $Im(T_\alpha) < 0$ which are related to string coupling and volume moduli to take positive values. %, or in other words positive definiteness of moduli space metrices.
Now, the basis of orientifold even two-forms and four-forms are as under,
\bea
& & \hskip3cm \mu_1 = dx^1 \wedge dx^2,  \quad \mu_2 = dx^3 \wedge dx^4, \quad  \mu_3 = dx^5 \wedge dx^6 \\
& & \hskip-0.7cm \tilde{\mu}^1 = dx^3 \wedge dx^4 \wedge dx^5 \wedge dx^6,  \quad \tilde{\mu}^2 = dx^1 \wedge dx^2 \wedge dx^5 \wedge dx^6, \quad  \tilde{\mu}^3 = dx^1 \wedge dx^2 \wedge dx^3 \wedge dx^4 \nonumber
\eea
implying that $\hat{d}_\alpha{}^\beta = \delta_\alpha{}^\beta$. The only non-trivial triple intersection number ($k_{\alpha\beta\gamma}$) is given as $\kappa_{123}=k_{123} = 1$ which implies the volume form to be ${\cal V}_E = t_1 \, t_2 t_3$. Let us mention that for this example there are no two-forms
anti-invariant under the orientifold projection, i.e. $h^{1,1}_-(X_6) = 0$, and  therefore no ${B}_2$ and $C_2$ moduli as well as no geometric flux components such as $\omega_{a \Lambda}, \omega_{a}{}^\Lambda$ are present. 

Now, the expressions for K\"ahler potential and the generalized flux-induced superpotential take the following simplified forms,
\bea
\label{eq:Kexample1}
& & \hskip-1.6cm K = -\ln\left(-i(\tau-\ov\tau)\right) -\sum_{j=1}^{3} \ln\left(i(U_j - \ov U_j)\right) - \sum_{\alpha=1}^{3} \ln\left(\frac{i\,(T_\alpha - \ov T_\alpha)}{2}\right)  \\
\label{eq:Wexample1}
& & \hskip-1.6cm W = \biggl[\biggl(F_\Lambda + \tau\, H_\Lambda \biggr)  + \left(\hat{Q}^{\alpha}{}_{\Lambda} + \tau \, \hat{P}^{\alpha}{}_{\Lambda}\right) \, T_\alpha\biggr]\, {\cal X}^\Lambda \\
& &  \hskip2cm + \biggl[\biggl(F^\Lambda + \tau\, H^\Lambda\biggr)  + \biggl(\hat{Q}^{\alpha \Lambda} + \tau \,\hat{P}^{\alpha \Lambda} \biggr) T_\alpha \biggr]\, {\cal F}_\Lambda\, , \nonumber
\eea
where $\Lambda = 0, 1, 2, 3$ and $\alpha = 1,2,3$ implying the presence of 8 components for each of three-form fluxes $H_3$ and $F_3$, %given as,
%\bea
%\label{eq:fluxconversionA1}
%& & H_0 , \, \quad H_1, \, \quad H_2, \, \quad H_3, \, \, \quad H^0 , \, \quad H^1 , \, \quad H^2 , \, \quad H^3 \nonumber\\
%& & F_0 , \,  \quad F_1 , \, \quad F_2 , \, \quad F_3 , \, \, \quad F^0 , \, \quad F^1 , \, \quad F^2 , \, \quad F^3 \nonumber
%\eea
and similarly 24 flux components for each of the $Q/P$ fluxes. % can be written with appropriate cohomology indices as: $\hat{Q}_\Lambda^\alpha, \hat{P}_\Lambda^\alpha, \hat{Q}^{\alpha\Lambda}$ and $\hat{P}^{\alpha\Lambda}$. 
Now to analyze and express the scalar potential in our symplectic formulation, first we utilize the K\"ahler potential (\ref{eq:Kexample1}) and superpotential (\ref{eq:Wexample1}) which results in 9661 terms in the total $F$-term contributions. Subsequently, using our symplectic rearrangement in eqn. (\ref{eq:main2}) results in a precise recovery of the total 9661 terms of the scalar potential. To reflect the involvement of terms within each kind of pieces, we express the counting of terms distributed as under,
\bea
\label{eq:sympVtotExample1}
& & \hskip-0.4cm V_{\mathbb F \mathbb F} = {\bf  -}\frac{1}{4\, s\, {\cal V}_E^2} \, \,\int_{X_6} \,{\mathbb F} \wedge \ast {\mathbb F}, \hskip7.5cm \#(V_{\mathbb F \mathbb F})= 4108 \nonumber\\
& & \hskip-0.4cm V_{\mathbb H \mathbb H} ={\bf  -}\frac{s}{4\, {\cal V}_E^2} \, \,\int_{X_6} \, {\mathbb H} \wedge \ast {\mathbb H}, \hskip7.5cm \#(V_{\mathbb H \mathbb H})= 1054  \nonumber\\
& & \hskip-0.4cm V_{\mathbb H \hat{\mathbb Q}} ={\bf  -}\frac{1}{4\,\, {\cal V}_E^2} \, \,\int_{X_6}\left(-2 \, {\mathbb H} \wedge \ast \hat{\mathbb Q}  - 4 \, {\mathbb H} \wedge \ast \tilde{\cal Q} \right) , \hskip4.2cm  \#(V_{\mathbb H \mathbb Q})=  1350 \nonumber\\
& & \hskip-0.4cm V_{\mathbb F \hat{\mathbb P}} ={\bf  -}\frac{1}{4\,\, {\cal V}_E^2} \, \,\int_{X_6}\left(2 \, {\mathbb F} \wedge \ast \hat{\mathbb P}  + 4 \, {\mathbb F} \wedge \ast \tilde{\cal P} \right) , \hskip4.9cm  \#(V_{\mathbb F \mathbb P})=  1350 \nonumber\\
& & \hskip-0.4cm V_{\hat{\mathbb Q} \hat{\mathbb Q}} ={\bf  -}\frac{1}{4\, s\, {\cal V}_E^2} \, \,\int_{X_6} \left( \hat{\mathbb Q} \wedge \ast \hat{\mathbb Q} -({\cal V}_E\, \hat{\kappa}_{\alpha\beta})\, \, \tilde{\cal Q}^\alpha \wedge \ast \tilde{\cal Q}^\beta\right) , \hskip2.80cm \#(V_{\mathbb Q \mathbb Q})= 1551 \nonumber\\
& & \hskip-0.4cm V_{\hat{\mathbb P} \hat{\mathbb P}} ={\bf  -}\frac{s}{4\, {\cal V}_E^2} \, \,\int_{X_6} \left( \hat{\mathbb P} \wedge \ast \hat{\mathbb P} -({\cal V}_E\, \hat{\kappa}_{\alpha\beta})\, \, \tilde{\cal P}^\alpha \wedge \ast \tilde{\cal P}^\beta\right) ,\hskip3.50cm \#(V_{\mathbb P \mathbb P})= 408 \nonumber\\
& & \hskip-0.4cm V_{\hat{\mathbb P} \hat{\mathbb Q}} ={\bf  -}\frac{1}{4\, {\cal V}_E^2} \, \,\int_{X_6} (-2) \biggl[ \hat{\mathbb P} \wedge \hat{\mathbb Q} -({\cal V}_E\, \hat{\kappa}_{\alpha\beta})\, \, \left(\hat{\mathbb P}^\alpha \wedge \tilde{\cal Q}^\beta - \hat{\mathbb Q}^\alpha \wedge \tilde{\cal P}^\beta\right) \biggr] ,\hskip0.60cm \#(V_{\hat{\mathbb P} \hat{\mathbb P}})= 972 \nonumber\\
& & \hskip-0.4cm V_{\mathbb H \mathbb F} ={\bf  -}\frac{1}{4\, {\cal V}_E^2} \, \,\int_{X_6} (+2) \times \left(\, \mathbb F \wedge \mathbb H \right), \hskip2.3cm \#(V_{\mathbb H \mathbb F})=  128 \nonumber\\
& & \hskip-0.4cm V_{\mathbb F \hat{\mathbb Q}} ={\bf  -}\frac{1}{4\, s\, {\cal V}_E^2} \, \,\int_{X_6} (-2) \times \left(\, \mathbb F \wedge \hat{\mathbb Q} \right) , \hskip2.0cm \#(V_{\mathbb F \hat{\mathbb Q}})=  288 \\
& & \hskip-0.4cm V_{\mathbb H \hat{\mathbb P}} ={\bf  -}\frac{s}{4\, {\cal V}_E^2} \, \,\int_{X_6} (-2) \times \left(\, \mathbb H \wedge \hat{\mathbb P} \right) , \hskip2.3cm \#(V_{\mathbb H \hat{\mathbb P}})=  72 \, . \nonumber
\eea
While collecting the various pieces, here we  have used the following simplified relations for the current example, 
\bea
& & \hskip-1.5cm \left(\frac{4}{9}\, k_0^2 \tilde{\cal G}_{\alpha \beta}\right) \equiv 4 \, \sigma_\alpha \sigma_\beta - 4 {\cal V}_E\, k_{\alpha\beta}= \left(
 \begin{array}{ccc}
4\, \sigma_1^2 & 0 & 0\\
0 & 4\, \sigma_2^2  & 0 \\
0 &0 & 4\, \sigma_3^2 \\
\end{array}
\right)
\eea
and 
\bea
& & \hskip-1.5cm  {\cal V}_E \, k_{\alpha\beta} = \left(
 \begin{array}{ccc}
0&  \sigma_1 \, \sigma_2 &  \sigma_1 \, \sigma_3\\
 \sigma_1 \, \sigma_2 &0 &  \sigma_2 \, \sigma_3 \\
 \sigma_1 \, \sigma_3 &  \sigma_2 \, \sigma_3 & 0 \\
\end{array}
\right)
\eea
Note that in the aforementioned collection, $V_{\mathbb H \hat{\mathbb Q}}$ and $V_{\mathbb F \hat{\mathbb P}}$ pieces have some terms which are not part of $F$-term contributions, i.e. do not follow within the 9661 terms obtained from $K$ and $W$ in eqns. (\ref{eq:Kexample1})-(\ref{eq:Wexample1}). However such terms are absent in $(V_{\mathbb H \hat{\mathbb Q}} + V_{\mathbb F \hat{\mathbb P}})$ due to internal cancellations. The same results in $\#(V_{\mathbb H \hat{\mathbb Q}} + V_{\mathbb F \hat{\mathbb P}})=  1080$ instead of their individual sum being as $2700$. Taking the same into account, one finds that the total pieces sum up into 9661 terms as is the case for the total $F$-term contributions. It is worth to mention the following points,
\begin{itemize}
\item{We observe that we are able to rewrite the total F-term scalar potential in terms of symplectic ingredients and {\it without using the internal background metric}. Therefore we can compare the results with previous studies of \cite{Gao:2015nra} in which the rearrangement was performed via using the internal background metric, and not the symplectic ingredients.}
\item{In this context, one also observes that all the pieces except $V_{\mathbb H \mathbb F}, V_{\mathbb F \hat{\mathbb Q}}, V_{\mathbb H \hat{\mathbb P}}$ and $V_{\hat{\mathbb P} \hat{\mathbb Q}}$ are written as $ {\cal O}_1 \wedge \ast {\cal O}_2$ indicating that the analogous pieces involving internal metric can be written with all the real six-dimensional indices properly contracted through the internal background metric as in \cite{Gao:2015nra}.}
\item{The last four pieces $V_{\mathbb H \mathbb F}, V_{\mathbb F \hat{\mathbb Q}}, V_{\mathbb H \hat{\mathbb P}}$ and $V_{\hat{\mathbb P} \hat{\mathbb Q}}$ written as $ {\cal O}_1 \wedge {\cal O}_2$ are naively looking like tadpole terms. However, note that they are not, especially the $V_{\hat{\mathbb P} \hat{\mathbb Q}}$ piece which is peculiar in many sense ! For example, this term has some information about the internal background via period/metric inputs. This can also be seen from their analogous parts in \cite{Gao:2015nra} which explicitly involves the internal metric implying that the $V_{\hat{\mathbb P} \hat{\mathbb Q}}$ piece is not topological. For the current symplectic representation, this information gets incorporated via introducing ${\cal S}$ matrices within $\tilde{\cal P}$ and $\tilde{\cal Q}$ redefinitions as in eqn. (\ref{eq:tildeQP}).}
%\item{$V_{\mathbb F \mathbb F}, V_{\mathbb H \mathbb H}$ along with the tadpole pieces $V_{\mathbb H \mathbb F}, V_{\mathbb F \mathbb Q}, V_{\mathbb H \mathbb P}$ have the same number of terms. }
\end{itemize}

\subsubsection*{Comments on generalized tadpole terms}
Let us make some comments about the tadpole terms. In this regard, note that although the explicit forms of Bianchi identities in the presence of ``all'' fluxes along with $P$-flux is not known, however for the cases which correspond to the type IIB orientifold examples with $h^{1,1}_-(CY/\sigma) = 0 = h^{2,1}_+(CY/\sigma)$, the same is known \cite{Aldazabal:2006up,Aldazabal:2008zza, Guarino:2008ik}.  Nevertheless, based on modular completion arguments, we expect that a subset of the NS-NS Bianchi identities, should be the following, 
\bea
\label{eq:BIs1}
& & \hskip-1.0cm H^\Lambda \, \hat{Q}_\Lambda{}^\alpha - H_\Lambda \hat{Q}^{\alpha \Lambda} = 0, \quad \quad  F^\Lambda \, \hat{P}_\Lambda{}^\alpha - F_\Lambda \hat{P}^{\alpha \Lambda} =0, \quad \quad  H_\Lambda \, \omega_{a}{}^{\Lambda} - H^\Lambda \, \omega_{a k} = 0, %\quad \quad F_\Lambda \, \omega_{a}{}^{\Lambda} - F^\Lambda \, \omega_{a k} = 0 
\\
& & \hskip-1cm \hat{Q}^{\alpha \Lambda} \hat{Q}^\beta{}_{\Lambda} - \hat{Q}^{\beta \Lambda} \hat{Q}^\alpha{}_{\Lambda} = 0, \quad \quad \omega_{a}{}^{\Lambda} \omega_{b \Lambda} - \omega_{b}{}^{\Lambda} \omega_{a k} =0, \quad \quad \hat{P}^{\alpha \Lambda} \hat{P}^\beta{}_{\Lambda} - \hat{P}^{\beta \Lambda} \hat{P}^\alpha{}_{\Lambda} = 0 \nonumber\\
& & \hskip-1cm  \omega_{a \Lambda} \hat{Q}^{\alpha \Lambda} - \omega_{a}{}^{\Lambda} \hat{Q}^\alpha{}_{\Lambda} = 0, \quad \quad \omega_{a \Lambda} \hat{P}^{\alpha \Lambda} - \omega_{a}{}^{\Lambda} \hat{P}^\alpha{}_{\Lambda} = 0, \quad \quad \hat{P}^{\alpha \Lambda} \hat{Q}^{\beta}{}_{\Lambda} - \hat{Q}^{\beta \Lambda} \hat{P}^{\alpha}{}_{\Lambda} = 0\nonumber
%& & \hskip1.0cm R^K \, \hat{\omega}_{\alpha K} - R_K \hat{\omega}_{\alpha}{}^{K} = 0, \quad \quad \quad R_K \, Q^{a K} - R^K \, Q^{a}{}_{K} = 0 \nonumber\\
%& & \hskip-1cm \hat{\omega}_{\alpha}{}^{K} \hat{\omega}_{\beta K} - \hat{\omega}_{\beta}{}^{K} \hat{\omega}_{\alpha K} = 0, \quad  Q^{a K} Q^{b}{}_{K} -Q^{b K} Q^{a}{}_{K} =0 , \quad Q^{a K} \hat{\omega}_{\alpha K} - Q^{a}{}_{K} \hat{\omega}_{\alpha}^{K} = 0\nonumber
\eea
On the lines of studies made for this toroidal example in a non-symplectic approach of \cite{Gao:2015nra}, we expect that the last three pieces of the scalar potential rearrangement in eqn. (\ref{eq:main2}) should generically correspond to the generalized 3/7-brane tadpoles to be nullified via a combination of NS-NS and RR Bianchi identities\footnote{Moreover, the first piece of ($\hat{\mathbb P} \wedge \hat{\mathbb Q}$)-type in the last line of eqn. (\ref{eq:main2}) should also be generically canceled by a subset of the Bianchi identities in eqn. (\ref{eq:BIs1}). However the same is not generically true for all the other two contributions of ($\hat{\mathbb P} \wedge \tilde{\cal Q}$)- and ($\hat{\mathbb Q} \wedge \tilde{\cal P}$)-types as could be indirectly seen from the analysis of \cite{Gao:2015nra}.}, i.e. by adding contributions from the local sources such as branes/orientifold planes. % which can be given as under,
%\bea
%\label{eq:VD}
%& & \hskip-1cm {V_{local}} = - V_{{\mathbb F}{\mathbb H}} - V_{{\mathbb F}\hat{\mathbb Q}} - V_{{\mathbb H}\hat{\mathbb P}}\,.
%\eea
%Here we Note that, the guess in eqn. (\ref{eq:BIs1}) are purely motivated by modular completion arguments, and may be stronger than the actual Bianchi identities; for example in the absence of geometric flux, the modular completion of ``$HQ = 0$'' type identity is known to be of ``$HQ-FP = 0$" type in \cite{Aldazabal:2006up} though it was later reformulated into a different form after studying the flux algebra in a setup with $h^{11}_-(X) =0$ \cite{Aldazabal:2008zza, Guarino:2008ik}. 
%\subsubsection*{Comments on modular invariant tadpole terms}
%As we have discussed earlier as well as  the last three pieces $V_{\mathbb H \mathbb F}$, $V_{\mathbb F \hat{\mathbb Q}}$ and $V_{\mathbb H \hat{\mathbb P}}$ should correspond to {\it generalized tadpole contributions} and these have to be canceled by local sources plus satisfying a subset of NS-NS Bianchi identities given in (\ref{eq:BIs1}). 
To elaborate more on it, we find that the total of 488 terms of the last three pieces of (\ref{eq:sympVtotExample1}) corresponds to 128 terms for  RR tadpoles, while remaining ones can be eliminated via generalized modular completed version of NS-NS Bianchi identities (\ref{eq:BIs1}) as seen in the analysis of \cite{Gao:2015nra, Shukla:2016xdy}. For example, the actual 3/7-brane tadpoles contributions can be written in terms of the older generalized flux orbits as under,
\bea
& & V_3 = \frac{1}{2\, {\cal V}_E^2} \, (F_\Lambda H^\Lambda - H_\Lambda F^\Lambda) \, , \quad \quad \quad  V_7 = V_7^{(1)} + V_7^{(2)} + V_7^{(3)}
\eea
where
\bea
& & \hskip-0.7cm V_7^{(1)} = - \frac{1}{2\, s\, {\cal V}_E^2} \, (F_\Lambda \hat{Q}^{\alpha\Lambda} - \hat{Q}^\alpha{}_\Lambda F^\Lambda)\, \sigma_\alpha \, , \quad  V_7^{(2)} = - \frac{\left(C_0^2+s^2\right)}{2\, s\, {\cal V}_E^2} \, (H_\Lambda \hat{P}^{\alpha\Lambda} - \hat{P}^\alpha{}_\Lambda H^\Lambda)\, \sigma_\alpha \nonumber\\
& &  \hskip1.0cm V_7^{(3)} = - \frac{1}{2\, {\cal V}_E^2} \, \left(\frac{C_0}{s}\right) \biggl[(H_\Lambda \hat{Q}^{\alpha\Lambda} + F_\Lambda \hat{P}^{\alpha\Lambda} )-(\hat{Q}^\alpha{}_\Lambda H^\Lambda + \hat{P}^\alpha{}_\Lambda F^\Lambda) \biggr] \, \sigma_\alpha \, 
\eea
The three pieces of 7-brane tadpoles correspond to the triplet of RR eight-form potential $(C_8, \tilde{C_8}, C_8^\prime)$ which transform under S-duality as,
\bea
& & C_8 \to \tilde{C_8}, \quad \quad \tilde{C_8} \to C_8, \quad \quad C_8^\prime \to - \, C_8^\prime
\eea
The triplet of eight-form potentials couples to $D7$, $NS_{7_i}$ and $I_{7_i}$ branes resulting in $FQ$, $HP$ and $(HQ+FP)$ type of tadpole terms \cite{Aldazabal:2006up,Aldazabal:2008zza, Guarino:2008ik}. Also, observe that under S-duality $V_3 \leftrightarrow V_3$, $V_7^{(1)} \leftrightarrow V_7^{(2)}$ while $V_7^{(3)}$ remains invariant even though the combinations $(HQ+FP)$ is anti-invariant. It is because of an additional anti-invariant factor $C_0/s \to - \, C_0/s$ which makes the third piece $V_7^{(3)}$ survive. Further, out of the total of 488 number of terms, the ``actual" tadpole terms are 128 enumerated as: $\#(V_3) = 8, \#(V_7^{(1)}) = 24, \#(V_7^{(1)}) = 48$ and $\#(V_7^{(1)}) = 48$. All the remaining pieces are nullified by the flux constraints given in eqn. (\ref{eq:BIs1}). However, the use of new generalized flux orbits has helped in nicely combining the three $7$-brane tadpole pieces into two pieces $-(V_{\mathbb F \hat{\mathbb Q}}+V_{\mathbb H \hat{\mathbb P}})$ by taking appropriate care of $C_0$ factors as well. Moreover, had it not been inspired by the collection through new generalized flux orbits, it would have been rather difficult to invoke the explicit form of  $I_{7_i}$-brane tadpoles which has a peculiar nature to appear with a factor $C_0/s$ being anti-invariant under S-duality.

\subsection{Example B: Type IIB $\hookrightarrow$ ${\mathbb T}^6/({\mathbb Z}_4$)-orientifold}
\label{sec_detailedV2}
%This example can be considered as a complimentary example to the previous one. Both of them have their own advantages and limitations. Where on the one hand, Example A has given a good demonstration in the symplectic sector via non-trivial complex structure moduli dependence, for example B, one has $h^{2,1}_-(CY) = 0$ leading to a trivial symplectic sector as all the matrices will be simply constant. On the other hand, the example A is too simple to illustrate all the features of proposed symplectic formalism in eqns. (\ref{eq:main1}), (\ref{eq:main2}) and/or (\ref{eq:main3}) because it does not have odd axions due to $h^{1,1}_-(X_6)=0$ and so use of generalized flux orbits corrected via odd-axions ${B}_2/C_2$ have not been demonstrated. This was the first motivation to consider this ``Example B'' in \cite{Shukla:2015bca}. However, we will not reconsider it in the full detail here as it has been already studied in the context of utility of new generalized flux orbits. Though having $h^{2,1}_-(CY) = 0$ makes this example less interesting from the point of view of the complex structure moduli dependent symplectic ingredients, nevertheless, let us point out that the the coefficients of $PQ$ piece were not invoked in their most generic form in \cite{Shukla:2015bca}.
As this toroidal orientifolds example has been considered previously with the presence of $P$-flux in \cite{Shukla:2015bca}, we will not reconsider the same in full detail rather we aim to provide the aspects which were not understood in \cite{Shukla:2015bca} from a generic point of view. %Moreover we also note that in \cite{Shukla:2015bca}, this example has been centered more towards illustrating the utility of new generalized flux orbits as it does not have interesting enough symplectic sector to show up due to no complex structure moduli being present. For these reasons we shift the discussions on the second example in the appendix \ref{sec_detailedV}.}.

Leaving all the orientifold construction related details to be directly referred from \cite{Shukla:2015bca,Robbins:2007yv}, let us start with the following explicit expressions of the K\"ahler- and super-potentials for analyzing F-term scalar potential,  
\bea
& & \hskip-1cm K = -\ln2\, -\ln\left(-i(\tau-\ov \tau)\right) - 2 \, \ln{\cal V}_E(T_\alpha, \tau, G^a; \ov T_\alpha, \ov \tau, \ov G^a) \, \\
& & \hskip-1cm W = \biggl[\left(F_0 + \tau\, H_0 + \omega_{a0} \, G^a + \hat{Q}^{\alpha}{}_{0} \, T_\alpha + \tau\, \hat{P}^{\alpha}{}_{0} \, T_\alpha -\,\hat{P}^{\alpha}{}_{0} \left(\frac{1}{2} \hat{\kappa}_{\alpha a b} G^a G^b\right)\right) \nonumber\\
& & -i\,  \left(F^0 + \tau\, H^0 + \omega_a{}^0 \, G^a + \hat{Q}^{\alpha 0} \, T_\alpha +\tau\, \hat{P}^{\alpha 0} \, T_\alpha  -\,\hat{P}^{\alpha 0} \left(\frac{1}{2} \hat{\kappa}_{\alpha a b} G^a G^b\right)\right)\biggr]\,, \nonumber
\eea
where $a = \{1,2\}$, $\alpha=\{1,2,3\}$ showing the presence of 2 complexified odd axions ($G^a$) and 3 complexified K\"ahler moduli ($T_\alpha$). Moreover, the Einstein frame sixfold volume written in terms of chiral variables is given as,
\bea
\label{eq:Volume}
& & {\cal V}_E\equiv {\cal V}_E(T_\alpha, S, G^a) = \frac{1}{4}\, \, \left(\frac{i(T_3 -\ov T_3)}{2} - \frac{i}{4 (\tau -\ov \tau)} \, \hat{\kappa}_{3 a b} \, (G^a -\ov G^a) (G^b -\ov G^b)\right)^{1/2} \nonumber\\
& & \hskip2cm \times  \biggl[\biggl(\frac{i(T_1 -\ov T_1)}{2}\biggr)^2 -2\, \biggl(\frac{i(T_2 -\ov T_2)}{2}\biggr)^2\biggr]^{1/2} \, \, 
\eea
Now, we will use the following new generalized flux orbits,
\begin{subequations}
\bea
\label{eq:orbitsB1Ex2}
& \hskip-1.0cm {\mathbb H}_0 =  {\bf h}_0~, \, \, \, \, \, \hat{\mathbb Q}^{\alpha}{}_{0} ={ \bf{\hat{q}^{\alpha}{}_{0} }} + C_0 \, {\bf \hat{p}^{\alpha}{}_{0}} ~,  \quad \quad \quad {\mathbb F}_0=   {\bf f}_0 + C_0 \, ~{\bf h}_0~, \, \, \, \, \, \hat{\mathbb P}^{\alpha}{}_{0} = {\bf \hat{p}^{\alpha}{}_{0}}~; 
%& & \\
%&& {\cal H}^0 =  {\bf h}^0~, \, \, \, \, \, {\hat{\cal Q}}^{\alpha 0} = {\bf \hat{q}^{\alpha 0}} + C_0 \, {\bf \hat{p}^{\alpha 0}} ~, \nonumber\\
%&& {\cal F}^0=   {\bf f}^0 + C_0 \, ~{\bf h}^0~, \, \, \, \, \, {\hat{\cal P}}^{\alpha 0} = {\bf {\hat{p}}^{\alpha 0}}~, \nonumber
\eea
where
\bea
\label{eq:orbitsB2Ex2}
& & \hskip-0.5cm {\bf h}_0 = H_0 + (\omega_{01} \, {b}^1+\omega_{02} \, {b}^2) + \frac{1}{2} \hat{Q}^3{}_0 \left(\hat{\kappa}_{3 11} (b^1)^2 + \hat{\kappa}_{3 22} (b^2)^2\right)  +  \left(\hat{P}^1{}_0 \,{\rho}_1 + \hat{P}^2{}_0 \,{\rho}_2 + \hat{P}^3{}_0 \,{\rho}_3\right) \nonumber\\
& & \hskip-0.5cm  {\bf f}_0 = F_0 + (\omega_{01} \, {c}^1 + \omega_{02} \, {c}^2) - \hat{P}^3{}_0 \, \left(\frac{1}{2}\, \hat{\kappa}_{3 11} (c^1)^2+\frac{1}{2}\, \hat{\kappa}_{3 22} (c^2)^2 \right) \, \\
& &  \hskip2cm + \hat{Q}^1{}_0 \,{\rho}_1 + \hat{Q}^2{}_0 \,{\rho}_2+ \hat{Q}^3{}_0\, \left({\rho}_3 + \hat{\kappa}_{3 11} c^1 b^1 + \hat{\kappa}_{3 22} c^2 b^2\right) \,\nonumber\\
%&& \\
& & \hskip-0.5cm {\cal \mho}_{01} = \omega_{01} + \hat{Q}^3{}_0 \, \left(\hat{\kappa}_{311}\, b^1\right) -\, \hat{P}^3{}_0 \, \left(\hat{\kappa}_{311}\, c^1\right), \,  {\cal \mho}_{02} = \omega_{02} + \hat{Q}^3{}_0 \, \left(\hat{\kappa}_{322}\, b^2\right) -\, \hat{P}^3{}_0 \, \left(\hat{\kappa}_{322}\, c^2\right)\nonumber\\
%& &  {\bf h}^0 = H^0 + (\omega_{a}{}^{0} \, {b}^a) + \hat{Q}^{\alpha 0} \, \left(\frac{1}{2}\, \hat{\kappa}_{\alpha a b} b^a b^b\right)  +\hat{P}^{\alpha 0} \, \left({\rho}_\alpha \right)\nonumber\\
%& &  {\bf f}^0 = F^0 + (\omega_a{}^0 \, {c}^a) -  \hat{P}^{\alpha 0} \, \left(\frac{1}{2}\, \hat{\kappa}_{\alpha a b} c^a c^b\right) + \hat{Q}^{\alpha 0} \, \left({\rho}_\alpha + \hat{\kappa}_{\alpha a b} c^a b^b\right)\, \, \nonumber \\
& &  \hskip-0.5cm {\bf \hat{q}^{1}{}_{0}} = \hat{Q}^{1}{}_{0} , \quad {\bf \hat{q}^{2}{}_{0}} = \hat{Q}^{2}{}_{0}, \quad {\bf \hat{q}^{3}{}_{0}} = \hat{Q}^{3}{}_{0}, \quad \quad  {\bf \hat{p}^{1}{}_{0}} = \hat{P}^{1}{}_{0}, \quad {\bf \hat{p}^{2}{}_{0}} = \hat{P}^{2}{}_{0}, \quad {\bf \hat{p}^{3}{}_{0}} = \hat{P}^{3}{}_{0}% \quad {\bf {\hat{p}}^{\alpha 0}} = {\hat{P}}^{\alpha 0} 
\nonumber
\eea
\end{subequations}
and similarly all the other components with upper index being `$\Lambda=0$' can be written. Now, the huge $F$-term scalar potential, which results in a total of 960 terms, can be very compactly rewritten as under \cite{Shukla:2015bca}. 
\bea
\label{eq:5termsEx2}
& & \hskip-0.75cm V_{{\mathbb F}{\mathbb F}} =  \frac{{\mathbb F}_0^2+{({\mathbb F}^0)}^2}{4 \, s\,{\cal V}_E^2}\,, \quad V_{{\mathbb H}{\mathbb H}} =  \frac{s \left({\mathbb H}_0^2 +\, {({\mathbb H}^0)}^2 \right)}{4 \, {\cal V}_E^2}\, , \quad V_{{\mathbb H} \hat{\mathbb Q}} =  \frac{3\left({\mathbb H}^0 \, \hat{\mathbb Q}^{0} + \hat{\mathbb Q}_0 \, {\mathbb H}_0 \right)}{2  \, {\cal V}_E^2}, \\
& & \hskip-0.75cm V_{{\cal \mho}{\cal \mho}} = \frac{\sigma_3 \left(\mho_{01} \mho_{01} + \mho_{1}{}^{0} \mho_{1}{}^{0}\right) + 2\sigma_3 \left(\mho_{02} \mho_{02} + \mho_{2}{}^{0} \mho_{2}{}^{0}\right)}{4 {\cal V}_E^2}, \quad V_{{\mathbb F} \hat{\mathbb P}} =  -\frac{3 \left({\mathbb F}^0 \, \hat{\mathbb P}^{0} + \hat{\mathbb P}_0 {\mathbb F}_0 \right)}{2 {\cal V}_E^2}  ,\nonumber\\
& & \hskip-0.75cm V_{ {\mathbb F} {\mathbb H}} =  \frac{{\mathbb H}_0 {\mathbb F}^0-{\mathbb F}_0 {\mathbb H}^0}{2 \, {\cal V}_E^2}, \quad V_{{\mathbb F} \hat{\mathbb Q}} =  \frac{{\mathbb F}_0 \hat{\mathbb Q}^{0} - \hat{\mathbb Q}_0 \, {\mathbb F}^0}{2 \, s \, {\cal V}_E^2}, \quad  V_{{\mathbb H} \hat{\mathbb  P}} =  \frac{s\, \left({\mathbb H}_0 \, \hat{\mathbb P}^{0} - \hat{\mathbb P}_0\, {\mathbb H}^0 \right)}{2 \, {\cal V}_E^2} \, . \nonumber
\eea
\bea
\label{eq:5termsEx22}
& & \hskip-1.2cm V_{\hat{\mathbb Q} \hat{\mathbb Q}} =\frac{1}{4 \,s\, {\cal V}_E^2} \biggl[\left(4 \sigma_2^2 - \sigma_1^2\right) \left(\hat{\mathbb Q}^1{}_0 \hat{\mathbb Q}^1{}_0 + \hat{\mathbb Q}^{10} \hat{\mathbb Q}^{10} \right) + \left(\sigma_1^2 - \sigma_2^2\right) \left(\hat{\mathbb Q}_0{}^2 \hat{\mathbb Q}_0{}^2 + \hat{\mathbb Q}^{20} \hat{\mathbb Q}^{20} \right) \nonumber\\
& &  + \sigma_3^2 \left(\hat{\mathbb Q}_0{}^3 \hat{\mathbb Q}_0{}^3 + \hat{\mathbb Q}^{30} \hat{\mathbb Q}^{30} \right) + 2 \, \sigma_1 \sigma_2 \left(\hat{\mathbb Q}_0{}^1 \hat{\mathbb Q}_0{}^2 + \hat{\mathbb Q}^{10} \hat{\mathbb Q}^{20} \right)- 6 \, \sigma_2 \sigma_3 \left(\hat{\mathbb Q}_0{}^2 \hat{\mathbb Q}_0{}^3 +  \hat{\mathbb Q}^{20} \hat{\mathbb Q}^{30} \right) \nonumber\\
& & - 6 \, \sigma_1 \sigma_3 \left(\hat{\mathbb Q}_0{}^1 \hat{\mathbb Q}_0{}^3 + \hat{\mathbb Q}^{10} \hat{\mathbb Q}^{30} \right) \biggr] \nonumber\\
& & \hskip-1.2cm V_{\hat{\mathbb P} \hat{\mathbb P}} =\frac{s}{4\, {\cal V}_E^2} \biggl[\left(4 \sigma_2^2 - \sigma_1^2\right) \left(\hat{\mathbb P}_0{}^1 \hat{\mathbb P}_0{}^1 + \hat{\mathbb P}^{10} \hat{\mathbb P}^{10} \right) + \left(\sigma_1^2 - \sigma_2^2\right) \left(\hat{\mathbb P}_0{}^2 \hat{\mathbb P}_0{}^2 + \hat{\mathbb P}^{20}  {\mathbb P}^{20} \right) \nonumber\\
& &  + \sigma_3^2 \left(\hat{\mathbb P}_0{}^3 \hat{\mathbb P}_0{}^3 + \hat{\mathbb P}^{30} \hat{\mathbb P}^{30} \right) + 2 \, \sigma_1 \sigma_2 \left(\hat{\mathbb P}_0{}^1 \hat{\mathbb P}_0{}^2 + \hat{\mathbb P}^{10} \hat{\mathbb P}^{20} \right)- 6 \, \sigma_2 \sigma_3 \left(\hat{\mathbb P}_0{}^2 \hat{\mathbb P}_0{}^3 +  \hat{\mathbb P}^{20} \hat{\mathbb P}^{30} \right) \nonumber\\
& & - 6 \, \sigma_1 \sigma_3 \left(\hat{\mathbb P}_0{}^1 \hat{\mathbb P}_0{}^3 + \hat{\mathbb P}^{10} \hat{\mathbb P}^{30} \right) \biggr] \, \\
& &  \hskip-1.2cm V_{\hat{\mathbb P} \hat{\mathbb Q}} =\frac{1}{2 \, {\cal V}_E^2} \biggl[ (3\sigma_1^2-4\sigma_2^2)\, (\hat{\mathbb P}_0{}^1 \hat{\mathbb Q}^{10}-\hat{\mathbb  Q}_0{}^1 \hat{\mathbb P}^{10}) -(\sigma_1^2-3 \sigma_2^2)\, (\hat{\mathbb P}_0{}^2 \hat{\mathbb Q}^{20}- \hat{\mathbb Q}_0{}^2 \hat{\mathbb P}^{20}) \nonumber\\
& &  \hskip0.5cm + \, \sigma_3^2\, (\hat{\mathbb P}_0{}^3  \hat{\mathbb Q}^{30}- \hat{\mathbb Q}_0{}^3 \hat{\mathbb P}^{30}) + \sigma_1\, \sigma_2\, (\hat{\mathbb P}_0{}^2 \hat{\mathbb Q}^{10} + \hat{\mathbb P}_0{}^1 \hat{\mathbb Q}^{20} - \hat{\mathbb Q}_0{}^2  \hat{\mathbb P}^{10} - \hat{\mathbb Q}_0{}^1 \hat{\mathbb P}^{20}) \nonumber\\
& &  \hskip0.5cm + \, 5 \sigma_2\, \sigma_3\, (\hat{\mathbb P}_0{}^2 \hat{\mathbb Q}^{30} + \hat{\mathbb P}_0{}^3 \hat{\mathbb Q}^{20} - \hat{\mathbb Q}_0{}^3 \hat{\mathbb P}^{20} -\hat{\mathbb Q}_0{}^2 \hat{\mathbb P}^{30}) \nonumber\\
& &  \hskip0.5cm +\, 5 \sigma_1\, \sigma_3\, (\hat{\mathbb P}_0{}^1 \hat{\mathbb Q}^{30} + \hat{\mathbb P}_0{}^3 \hat{\mathbb Q}^{10} - \hat{\mathbb Q}_0{}^3 \hat{\mathbb P}^{10}-\hat{\mathbb Q}_0{}^1 \hat{\mathbb P}^{30}) \biggr] \,. \nonumber
\eea
Now, using the following symplectic ingredients, 
\bea
\label{eq:periodMexampleB}
& & {\cal M}^{00} = -1, \quad {\cal M}^{0}{}_{0} = 0, \quad  {\cal M}_0{}^0 = 0, \quad {\cal M}_{00} = 1 \nonumber\\ 
& & {\cal L}^{00} = -1,  \quad {\cal L}^{0}{}_{0} = 0, \quad  {\cal L}_0{}^0 = 0, \quad {\cal L}_{00} = 1  \, \\ 
& & {{\cal S}}^{00} = 0, \quad {{\cal S}}^{0}{}_{0} = 2, \quad  {{\cal S}}_0{}^0 =-2, \quad {{\cal S}}_{00} = 0 \nonumber
\eea
it is obvious to read off the pieces in eqn. (\ref{eq:5termsEx2})  from our symplectic formulation given in eqn. (\ref{eq:main2}). For example, the  most complicated looking piece of eqn. (\ref{eq:5termsEx2}), which is $V_{{\mathbb \mho}{\mathbb \mho}}$, can be known simply by considering ${\cal G}^{ab} = - 4 \, {\cal V}_E \, \hat{k}^{ab}$. Now $\hat{k}_{ab} = \hat{k}_{\alpha a b} t^\alpha$, so one has the only non-zero components given as ${\cal G}^{11} = \sigma_3$ and ${\cal G}^{22} = 2\, \sigma_3$. Here we recall that $\sigma_\alpha = \frac{1}{2}\, \kappa_{\alpha\beta\gamma} t^\alpha \, t^\beta \, t^\gamma$ and results in $\sigma_1 = t_1\, t_3, \sigma_2 = t_2 \, t_3$ and $\sigma_3=(t_1^2-2\, t_2^2)$ following from the intersection numbers $k_{311} = 1/2, k_{322} = -1, \hat{k}_{311} = -1, \hat{k}_{322} = -1/2$ along with $\hat{d}_\alpha{}^\beta = diag\{1/2, -1, 1/4\}$ and $d^a{}_b=diag\{-1,-1/2\}$ \cite{Robbins:2007yv}.

To see the same for the complicated pieces $V_{\hat{\mathbb Q}\hat{\mathbb Q}}$, $V_{\hat{\mathbb P}\hat{\mathbb P}} $ and $V_{\hat{\mathbb P}\hat{\mathbb Q}}$ is not that straight. However,  a careful observation tells us how the volume moduli dependent factors of $V_{\hat{\mathbb Q}\hat{\mathbb Q}}$, $V_{\hat{\mathbb P}\hat{\mathbb P}} $ as well as $V_{\hat{\mathbb P}\hat{\mathbb Q}}$ can be read off from the following coefficient matrix depending on the K\"ahler moduli-space metric,
\bea
\label{eq:exBqqpp}
& & \hskip-1.5cm {\rm QQ/PP:} \quad \left(\frac{4}{9}\, k_0^2 \tilde{\cal G}_{\alpha \beta} - 4 \sigma_\alpha \, \sigma_\beta\right) + \sigma_\alpha \, \sigma_\beta = \left(
 \begin{array}{ccc}
4 \sigma_2^2 -  \sigma_1^2& \sigma_1 \, \sigma_2 & -3 \sigma_1 \, \sigma_3\\
\sigma_1 \, \sigma_2 & \sigma_1^2 -  \sigma_2^2 & -3 \sigma_2 \, \sigma_3 \\
-3 \sigma_1 \, \sigma_3 & -3 \sigma_2 \, \sigma_3 & \sigma_3^2 \\
\end{array}
\right)
\eea
and
\bea
\label{eq:exBpq}
& & \hskip-1.5cm {\rm PQ:}\,\, (-2)\biggl[\left(\frac{4}{9}\, k_0^2 \tilde{\cal G}_{\alpha \beta} - 4 \sigma_\alpha \, \sigma_\beta\right) - \sigma_\alpha \, \sigma_\beta \biggr] = (-2)\left(
 \begin{array}{ccc}
4 \sigma_2^2 -  3\, \sigma_1^2& -\sigma_1 \, \sigma_2 & -5 \sigma_1 \, \sigma_3\\
-\sigma_1 \, \sigma_2 & \sigma_1^2 -  3 \sigma_2^2 & -5 \sigma_2 \, \sigma_3 \\
-5 \sigma_1 \, \sigma_3 & -5 \sigma_2 \, \sigma_3 & - \sigma_3^2 \\
\end{array}
\right)
\eea
{\it This origin of correct volume moduli dependent pieces was not clear from the analysis of \cite{Shukla:2015bca} in which the use of modular completed orbits was main focus rather than looking at the other structures.} Note that this analysis demonstrates the fact that by merely knowing the intersection numbers and eqn. (\ref{eq:periodMexampleB}), we can directly {\it read off} all the pieces of eqns. (\ref{eq:5termsEx2}) and (\ref{eq:5termsEx22}) from our symplectic formulation given in eqn. (\ref{eq:main2}).

\section{Expressing the symplectic ingredients in terms of the c.s. moduli}
\label{sec_Intersections1}
By the end of the previous section, in our analysis made for the second toroidal model ``Example B'', we realized that all the scalar potential pieces can be directly read off if the various intersection numbers (along with the even/odd hodge numbers) are known, and there is no need to perform the detailed computation for the scalar potential by starting form the K\"ahler and superpotential expressions. Let us mention that this observation with ``Example B'' can be extended for arbitrary compactifications with rigid CY threefolds as one can always read off the (modular invariant) scalar potential directly from our proposal in eqn. (\ref{eq:main2}) just by knowing the various new flux orbits from eqns. (\ref{eq:orbitsB1})-(\ref{eq:orbits11B}) (via looking at the non-vanishing intersection numbers) and the simplified version of the symplectic quantities given in eqn. (\ref{eq:periodMexampleB}).

Though, the aforementioned class of examples has been quite simple due to the absence of complex structure moduli, nevertheless it motivates us to look for a new form of the scalar potential in which one could directly read off the complex structure moduli dependent pieces as well. On these lines of motivation, now we provide a more explicit version of the generic non-geometric scalar potential\footnote{Here by being ``generic", we mean to be considering arbitrary Calabi Yau compactification with arbitrary number of complex structure and K\"ahler moduli, along with odd-axions. It is worth to mention that we are still assuming the large volume and large complex structure limits to avoid the non-perturbative effects.} by explicitly expanding out the symplectic ingredients in such a way that one could directly write down the scalar potential via merely knowing the topological data of the compactifying Calabi Yau and its mirror. As we have already argued the applicability of our proposal for the rigid CY cases separately, the computation now onwards is focused on the cases where $h^{2,1}_-(X)$ is non zero, i.e. at least one complex structure modulus is present.

\subsection{Symplectic quantities through the axionic/saxionic components of c.s. moduli}
\label{sec_symplectic2}
Let us consider the following form of the prepotential,
\bea
\label{eq:prepotentialNew}
& & {\cal F} = \frac{l_{ijk} \, {\cal X}^i\, {\cal X}^j \, {\cal X}^k}{{\cal X}^0} +  \frac{1}{2} \,{a_{ij} \, {\cal X}^i\, {\cal X}^j} +  \,{b_{i} \,{\cal X}^0\,  {\cal X}^i} +  \frac{1}{2} \,{i\, \gamma}\, ({\cal X}^0)^2
\eea
which still remain quite generic though we have neglected the non-perturbative contributions assuming the large complex structure limit. Also, as compared to the prepotential in eqn. (\ref{eq:prepotential}), a factor of $3!$ has been absorbed in the first term via a redefinition of the mirror triple-intersection numbers as $l_{ijk} = \frac{1}{3!}\, \hat{l}_{ijk}$. This has been done just to avoid the repetitive appearance of the rational pre-factors at the intermediate stage of computations, e.g. in the various derivatives of the prepotential ${\cal F}_0, {\cal F}_i, {\cal F}_{00}$ etc. With this normalization, the first derivatives of the prepotential ${\cal F}$ are given as under,
\bea
\label{eq:Prepder1}
& & {\cal F}_0 = -\, l_{ijk}\, U^i \, U^j\, U^k + b_i \, U^i + i\, \gamma, \\
& & \hskip0cm {\cal F}_i = 3\, l_{ijk} \, U^j\, U^k + a_{ij}\, U^j + b_i \nonumber
\eea
Here ${\cal X}^0 = 1$ and ${\cal X}^i = U^i$ have been utilized. Now the complex structure moduli dependent part of the K\"ahler potential is simplified as,

\bea
& & \hskip-1cm K_{cs} \equiv -\ln\left(i\int_{X}\Omega_3\wedge{\bar\Omega_3}\right) = -\ln\biggl[i\, \left(\ov {\cal X}^\Lambda \, {\cal F}_\Lambda - {\cal X}^\Lambda \, \ov {\cal F}_\Lambda \right)\biggr]\\
& & = -\ln \biggl[i\, \, l_{ijk} \, \left(-U^i U^j U^k + 3\, \ov U^i U^j U^k -3 \,U^i \ov U^j \ov U^k + \ov U^i \ov U^j \ov U^k\right) - 2 \, \gamma \biggr]\nonumber
\eea
Further considering the saxionic/axionic parts of complex structure moduli to be given as $U^i= v^i + i\, u^i$, one finds that 
\bea 
\label{eq:KcsSimp}
K_{cs} = -\ln \left(-8 \, l_{ijk}\, u^i u^j u^k - 2 \gamma \right) = -\ln\left(-2(4\, l + \gamma)\right).
\eea
where we have used the short hand notations $l= l_{ijk}\,u^i\, u^j\, u^k, \, l_{i}= l_{ijk}\, u^j\, u^k$ and $l_{ij}= l_{ijk}\, u^k$. Let us note that the real parameters $a_{ij}$ and $b_i$'s do not appear in the K\"ahler potential, however $\gamma$ parameter, which has a perturbative origin on the mirror side, does appear in the simplified K\"ahler potential. %Also, recall that in our conventions, the physical domain of validity demands $(l_{ijk}\, u^i u^j u^k) <0$. %\footnote{Also note that the normalization factors are important to check the symplectic identities as we will see later, and the presence of factor 8 indicates that we are using the following normalization of holomorphic three form $\Omega_3$ on a generic Caleb Yau,
%\bea 
%\frac{i}{8}\, \left(\Omega_3 \wedge \ov \Omega_3\right) = \frac{1}{3 !} \, J \wedge J \wedge J\,.
%\eea}.
Now, the double derivatives of prepotential (\ref{eq:prepotentialNew}) are given as, 
\bea
\label{eq:Prepder2}
& & \hskip0cm {\cal F}_{00} = 2\, l_{ijk}\, U^i \, U^j\, U^k + i\, \gamma, \quad {\cal F}_{0i} = -3\, l_{ijk} \, U^j\, U^k + b_i, \quad {\cal F}_{ij} = 6\, l_{ijk} \, U^k\,+ a_{ij} . 
\eea
Subsequently, the real and imaginary parts are separated out as under,
\bea
& \hskip-1cm Re({\cal F}_{00}) = 2\, l_{ijk}\, v^i \, v^j\, v^k - 6\, l_i\, v^i, & \qquad Im({\cal F}_{00}) = 6\, l_{ij}\, v^i \, v^j - 2\, l + \gamma \nonumber\\
& \hskip-0.75cm Re({\cal F}_{0i}) = 3\, l_i - 3\, l_{ijk}\, v^j\, v^k + b_i, & \qquad    Im({\cal F}_{0i}) = -6\, l_{ij}\, v^j \nonumber\\
&  \hskip-2.20cm Re({\cal F}_{ij}) = 6\, l_{ijk}\, v^k+ a_{ij}, & \qquad Im({\cal F}_{ij}) = 6\, l_{ij}\,.
\eea
Now the various components of the inverse matrix $Im({\cal F}^{\Lambda\Delta})$ are given as under,
\bea
& & \hskip-2cm  Im({\cal F}^{00}) = -\frac{1}{(2\, l-\gamma)}, \qquad Im({\cal F}^{0i}) = -\frac{v^i}{(2\, l-\gamma)},\quad Im({\cal F}^{ij}) =\frac{1}{6}\, l^{ij} -\frac{v^i\, v^j}{(2\, l-\gamma)}\,.
\eea
Using the derivatives of the prepotential in period matrix eqn. (\ref{eq:periodN}), one gets the following form of the period matrix components,
\bea
& \hskip-0.5cm Re({\cal N}_{00}) = 2\, l_{ijk}\, v^i \, v^j\, v^k - 2\,\beta \, l_i\, v^i, & \qquad Im({\cal N}_{00}) = x_0-x_{ij} \, v^i \, v^j \nonumber\\
& \hskip-0.5cm Re({\cal N}_{0i}) = - 3\, l_{ijk}\, v^j\, v^k + b_i + \beta\, l_i, & \qquad    Im({\cal N}_{0i}) = x_{ij} \, v^j \\
&  \hskip-2.2cm Re({\cal N}_{ij}) = 6\, l_{ijk}\, v^k +a_{ij} , & \qquad Im({\cal N}_{ij}) = -x_{ij} \,, \nonumber
\eea
where we have used the following redefinitions,
\bea 
\label{eq:simpperiodF1}
& & \beta= \left(\frac{9 \gamma}{8\, l - \gamma}\right), \quad x_0 = \frac{(4\, l + \gamma)(2\, l - \gamma)}{(8\, l - \gamma)}, \quad x_{ij} = 6\biggl[l_{ij} -\frac{12\, l_i \, l_j}{(8\, l - \gamma)} \biggr]\,.
\eea
Now the various components of the inverse period matrix $Im({\cal N}^{\Lambda\Delta})$ are given as under,
\bea
& & \hskip-2cm Im({\cal N}^{00}) = \frac{1}{\, x_0}, \qquad Im({\cal N}^{0i}) = \frac{v^i}{\, x_0}, \qquad Im({\cal N}^{ij}) = \frac{v^i\, v^j}{\, x_0} - x^{ij}
\eea
where the inverse of $x_{ij}$ turns out to be having the following form,
\bea
\label{eq:simpperiodF2}
& x^{ij}= \frac{1}{6}\, l^{ij} - \frac{2}{4\, l + \gamma} \, u^i\, u^j
\eea
Here, the inverse quantity $l^{ij}$ is defined from the relation $l^{ij}\, l_j = u^i$. Utilizing the simplifications derived in this section, now one can simplify the complicated expressions of the various matrices ${\cal L}, {\cal M}, ({\cal L}+{\cal M})$ and ${\cal S}$ which, being too lengthy, are presented in the appendix \ref{sec_proof}. Moreover, we have also shown in the appendix \ref{sec_proof} that the two main symplectic identities in eqns. (\ref{eq:symp10})-(\ref{eq:symp11}) generically hold. Here we just provide the compact versions of the relevant matrices. For that, we have defined the following quantities, which can be directly read-off from the topological data of the mirror CY threefolds,
\bea 
\label{eq:parameters4}
& & \hskip-0.8cm r_0 = \left(b_i \, v^i -\beta \, l_i\, v^i - l_{ijk} v^i v^j v^k  \right), \quad \qquad \quad \epsilon_i = \left(b_i + \beta\, l_i\,- 3\, l_{ijk} v^j v^k\right), \\
& & \hskip-0.8cm r_i = \left(b_i + a_{ij} v^j +\beta \, l_i + 3 \, l_{ijk}\, v^j v^k\right), \qquad \quad \eta_{ij} = \left(a_{ij} + 6 \,l_{ijk}\, v^k \right), \nonumber\\
& & \hskip-0.8cm \phi_0 = \left(b_i \, v^i + 3\, l_i\, v^i - l_{ijk} v^i v^j v^k  \right), \quad \qquad \quad \psi_0 = \left(l + \gamma + b_i\, u^i - 3\, l_{ij} v^i v^j\right) ,\nonumber\\
& & \hskip-0.8cm \phi_i = \left(b_i + a_{ij} v^j - 3 \, l_i + 3 \, l_{ijk}\, v^j v^k\right), \qquad \quad \psi_i = \left(a_{ij}\, u^j + 6 \,l_{ij}\, v^j \right), \nonumber\\
& & \nonumber\\
& & \hskip-0.8cm \beta=\frac{9 \gamma}{8\, l - \gamma}, \qquad x_0 = \frac{(4\, l + \gamma)(2\, l - \gamma)}{(8\, l - \gamma)}, \qquad y_0 = -\frac{2}{(4\, l + \gamma)}\, ,\nonumber\\
& & \hskip-0.8cm x_{ij} = 6 l_{ij} -\frac{72\, l_i \, l_j}{(8\, l - \gamma)}, \qquad x^{ij}= \frac{1}{6}\, l^{ij} - \frac{2}{4\, l + \gamma} \, u^i\, u^j\,. \nonumber
\eea
\subsubsection*{Components of $\cal M$-matrix :}
\bea
\label{eq:M-matrixCompact}
& & \hskip-1.5cm {\bf (i).} \quad  {\cal M}^{00} = \frac{1}{x_0}, \qquad {\cal M}^{0i} = \frac{v^i}{x_0} = {\cal M}^{i0}, \qquad {\cal M}^{ij} = \frac{v^i\, v^j}{x_0} -x^{ij}\, ,\\
& & \hskip-1.5cm {\bf (ii).} \quad {\cal M}^0{}_0 = -\frac{r_0}{x_0}, \quad {\cal M}^i{}_0 = -\frac{r_0\, v^i}{x_0} + x^{ij}\, \epsilon_j , \quad {\cal M}_0{}^i = -\frac{r_i}{x_0}, \quad {\cal M}^i{}_j = -\frac{r_j\, v^i}{x_0} + x^{ik}\, \eta_{kj}\, ,\nonumber\\
& & \hskip-1.5cm {\bf (iii).} \quad {\cal M}_0{}^0 = \frac{r_0}{x_0}, \quad {\cal M}_i{}^0 = \frac{r_i}{x_0}, \quad {\cal M}_0{}^i = \frac{r_0\, v^i}{x_0} - x^{ij}\, \epsilon_j, \quad {\cal M}_i{}^j = \frac{r_i\, v^j}{x_0} - x^{jk}\, \eta_{ki}\, ,\nonumber\\
& & \hskip-1.5cm {\bf (iv).} \quad {\cal M}_{00} = -\frac{r_0^2 + x_0^2}{x_0} + x_{ij} \,v^i\, v^j + \epsilon_i x^{ij} \epsilon_j, \quad {\cal M}_{0i} = -\frac{r_0\, r_i}{x_0} - x_{ij}\, v^j + \eta_{ij} x^{jk} \epsilon_k = {\cal M}_{i0} \nonumber\\
& & \hskip1cm {\cal M}_{ij} = -\frac{r_i \, r_j}{x_0} + x_{ij}  + \eta_{im} x^{mn} \eta_{nj}\, . \nonumber
\eea
\subsubsection*{Components of ${\cal M}_1 = (\cal L+ \cal M)$-matrix :}
Similarly, we can compactly rewrite a new kind of ${\cal M}_1$-matrices defined as a sum of ${\cal M}$ and ${\cal L}$ matrices as under, 
\bea
\label{eq:M1-matrixCompact}
& & \hskip-1.5cm {\bf (i).} \quad ({\cal M}_1)^{00} = -y_0, \quad ({\cal M}_1)^{0i} = -y_0 \, v^i = ({\cal M}_1)^{i0}, \quad ({\cal M}_1)^{ij} = -y_0 \left(u^i \, u^j + v^i \, v^j\right)\\
& & \hskip-1.5cm {\bf (ii).} \quad ({\cal M}_1)^0{}_0 = y_0 \, \phi_0, \quad  ({\cal M}_1)^0{}_i = y_0 \, \phi_i , \quad ({\cal M}_1)^i{}_0 = y_0\left(\phi_0 \, v^i +\psi_0 \, u^i\right) , \nonumber\\
& & \hskip2cm ({\cal M}_1)^i{}_j =y_0\left(u^i \, \psi_j +v^i \, \phi_j \right) \nonumber\\
& & \hskip-1.5cm {\bf (iii).} \quad ({\cal M}_1)_0{}^0 = -y_0\, \phi_0, \quad  ({\cal M}_1)_0{}^i = -y_0 \left(\psi_0 \, u^i+\phi_0 \, v^i\right), \quad ({\cal M}_1)_i{}^0 = -y_0\,\phi_i, \nonumber\\
& & \hskip2cm ({\cal M}_1)_i{}^j = -y_0 \left(v^j \, \phi_i+ u^j \, \psi_i \right) \nonumber\\
& & \hskip-1.5cm {\bf (iv).} \quad ({\cal M}_1)_{00} = y_0\left(\phi_0^2+\psi_0^2\right), \quad ({\cal M}_1)_{0i} = y_0\left(\phi_0 \, \phi_i+\psi_0 \, \psi_i\right)=({\cal M}_1)_{i0}, \nonumber\\
& & \hskip2cm ({\cal M}_1)_{ij} = y_0\left(\phi_i \, \phi_j+\psi_i \, \psi_j \right)\,. \nonumber
\eea
\subsubsection*{Components of $\cal S$-matrix :}
\bea 
\label{eq:S-matrixCompact}
& & \hskip-1.0cm {\bf (i).} \quad {\cal S}^{00} = 0, \quad {\cal S}^{0i} = -y_0 \, u^i = -{\cal S}^{i0}, \quad {\cal S}^{ij} = y_0 \left(u^i \, v^j - v^i \, u^j\right)\\
& & \hskip-1.0cm {\bf (ii).} \quad {\cal S}^0{}_0 = -y_0 \, \psi_0, \quad  {\cal S}^0{}_i = -y_0 \, \psi_i , \quad {\cal S}^i{}_0 = y_0\left(\phi_0 \, u^i -\psi_0 \, v^i\right) , \quad {\cal S}^i{}_j =y_0\left(u^i \, \phi_j -v^i \, \psi_j \right) \nonumber\\
& & \hskip-1.0cm {\bf (iii).} \quad {\cal S}_0{}^0 = y_0\, \psi_0, \quad  {\cal S}_0{}^i = y_0 \left(\psi_0 \, v^i-\phi_0 \, u^i\right), \quad {\cal S}_i{}^0 = y_0\,\psi_i, \quad {\cal S}_i{}^j = y_0 \left(v^j \, \psi_i- u^j \, \phi_i \right) \nonumber\\
& & \hskip-1.0cm {\bf (iv).} \quad {\cal S}_{00} = 0, \quad {\cal S}_{0i} = y_0\left(\psi_0 \, \phi_i-\phi_0 \, \psi_i\right) = -{\cal S}_{i0}, \quad {\cal S}_{ij} = y_0\left(\psi_i \, \phi_j-\phi_i \, \psi_j \right)\, . \nonumber
\eea
Note that the tilde flux-matrices  $\tilde{\cal Q}^\alpha, \tilde{\cal P}_\alpha$ and $\tilde{\cal \mho}_a$ as defined in eqn.(\ref{eq:tildeQP}) can be further simplified using these compactly written $\cal S$-matrix components in eqn. (\ref{eq:S-matrixCompact}). For example, the electric and magnetic components of $\tilde{\cal Q}^{\alpha}$ which are given as, 
\bea
 & & \tilde{\cal Q}^{\alpha\Sigma} = -\left({{\cal S}}^{\Sigma\Delta} {\mathbb Q}^\alpha{}_{\Delta} + {{\cal S}}^\Sigma{}_\Delta {\mathbb Q}^\alpha{}^\Delta\right), \qquad  \tilde{\cal Q}^\alpha{}_\Sigma = + \left({{\cal S}}_{\Sigma}{}^{\Delta} {\mathbb Q}^\alpha{}_{\Delta} + {{\cal S}}_{\Sigma\Delta} {\mathbb Q}^{\alpha\Delta} \right)\,, \nonumber
\eea
are simplified as under,
\bea 
\label{eq:tildeflux-matrixCompact}
& & \hskip-1cm \tilde{\cal Q}^{\alpha0} = y_0 \, \left(u^i\, \hat{\mathbb Q}_i{}^\alpha + \psi_0 \, \hat{\mathbb Q}^{\alpha 0} + \psi_i \, \hat{\mathbb Q}^{\alpha i} \right) \\
& & \hskip-1cm \tilde{\cal Q}^{\alpha j} = y_0 \biggl[ v^j \left(u^i\, \hat{\mathbb Q}_i{}^\alpha + \psi_0 \, \hat{\mathbb Q}^{\alpha 0} + \psi_i \, \hat{\mathbb Q}^{\alpha i} \right)- u^j \left(\hat{\mathbb Q}_0{}^\alpha + v^i\, \hat{\mathbb Q}_i{}^\alpha + \phi_0 \, \hat{\mathbb Q}^{\alpha 0} + \phi_i \, \hat{\mathbb Q}^{\alpha i} \right)\biggr]\nonumber\\
& & \hskip-1cm \tilde{\cal Q}^{\alpha}{}_{0} = -y_0 \biggl[ \phi_0 \left(u^i\, \hat{\mathbb Q}_i{}^\alpha + \psi_0 \, \hat{\mathbb Q}^{\alpha 0} + \psi_i \, \hat{\mathbb Q}^{\alpha i} \right)- \psi_0 \left(\hat{\mathbb Q}_0{}^\alpha + v^i\, \hat{\mathbb Q}_i{}^\alpha + \phi_0 \, \hat{\mathbb Q}^{\alpha 0} + \phi_i \, \hat{\mathbb Q}^{\alpha i} \right)\biggr]\nonumber\\
& & \hskip-1cm \tilde{\cal Q}^{\alpha}{}_{j} =-y_0 \biggl[ \phi_j \left(u^i\, \hat{\mathbb Q}_i{}^\alpha + \psi_0 \, \hat{\mathbb Q}^{\alpha 0} + \psi_i \, \hat{\mathbb Q}^{\alpha i} \right)- \psi_j \left(\hat{\mathbb Q}_0{}^\alpha + v^i\, \hat{\mathbb Q}_i{}^\alpha + \phi_0 \, \hat{\mathbb Q}^{\alpha 0} + \phi_i \, \hat{\mathbb Q}^{\alpha i} \right)\biggr]. \nonumber
\eea
Similar components for other tilde fluxes $\tilde{\cal P}_\alpha$ and $\tilde{\cal \mho}_a$ can also be analogously written. 

Using these explicit ingredients in eqns. (\ref{eq:M-matrixCompact}), (\ref{eq:M1-matrixCompact}), (\ref{eq:S-matrixCompact}) and (\ref{eq:tildeflux-matrixCompact}), {\it the collection of scalar potential pieces given in eqn. (\ref{eq:main2}) can be explicitly written out in terms of real moduli and axions}, however that results in a quite lengthy collection and therefore has been places as eqn. (\ref{eq:main4}) in the appendix \ref{sec_csVExpanded}. Moreover, this lengthy collection in eqn. (\ref{eq:main4}) suggests for using the following flux combinations which mix not only the volume moduli and RR axions but also the complex structure moduli and their axionic partners,
\bea 
\label{eq:parameters3}
& & \hskip-2cm \Theta^0[{\mathbb Y}] = {\mathbb Y}^0, \hskip5.3cm \Theta_0[{\mathbb Y}] =  {\mathbb Y}_0 + v^i \,  {\mathbb Y}_i + r_0  {\mathbb Y}^0 + r_i  {\mathbb Y}^i\, , \nonumber\\
& & \hskip-2cm \Theta^i[{\mathbb Y}] = {\mathbb Y}^i - v^i \, {\mathbb Y}^0, \hskip4.0cm \Theta_i[{\mathbb Y}] =  {\mathbb Y}_i + \epsilon_i\,  {\mathbb Y}^0 + \eta_{ij} \,  {\mathbb Y}^j\, ,\\
& & \hskip-1.9cm \Phi[{\mathbb Y}] = {\mathbb Y}_0 +v^i \, {\mathbb Y}_i +\phi_0\,  {\mathbb Y}^0+\phi_i \, {\mathbb Y}^i, \quad \qquad \Psi[{\mathbb Y}] =u^i \, {\mathbb Y}_i +\psi_0\,  {\mathbb Y}^0+\psi_i \, {\mathbb Y}^i\,, \nonumber
\eea
where the flux parameters ${\mathbb Y} \in \{ {\mathbb F}, {\mathbb H}, {\mathbb \mho}_a, \hat{\mathbb Q}^\alpha, \hat{\mathbb P}^\alpha \}$ have appropriate upper and lower $h^{21}_-(CY)$-indices. Moreover, here we have denoted $\Theta[\mathbb Y]$ etc. in such a way so that one could distinguish among various flux orbits; for example, $\Theta^0[{\mathbb H}] = {\mathbb H}^0$ and $\Theta_0[{\mathbb F}] =  {\mathbb F}_0 + v^i \,  {\mathbb F}_i + r_0  {\mathbb F}^0 + r_i  {\mathbb F}^i$ etc.

\subsection{Scalar potential with explicit dependence on saxions and axions of c.s. moduli}
Utilizing the additional simplifications through the collection (\ref{eq:main4}) and the flux-redefinitions given in eqn. (\ref{eq:parameters3}), we arrive at the following compact version of the previous scalar potential rearrangement given in eqn. (\ref{eq:main2}),
\bea
\label{eq:main5}
& & \hskip-1cm V_{\mathbb F \mathbb F} = -\frac{1}{4\,s\,{\cal V}_E^2}\biggl[\frac{(\Theta_0[{\mathbb F}])^2}{x_0}  - \Theta^i[{\mathbb F}]\, x_{ij} \, \Theta^j[{\mathbb F}] - \Theta_i[{\mathbb F}] \, x^{ij} \Theta_j[{\mathbb F}]+ x_0\, (\Theta^0[{\mathbb F}])^2\biggr] \, \\
& & \hskip-1cm V_{\mathbb H \mathbb H} = -\frac{s}{4\,{\cal V}_E^2}\biggl[\frac{(\Theta_0[{\mathbb H}])^2}{x_0}  - \Theta^i[{\mathbb H}]\, x_{ij} \, \Theta^j[{\mathbb H}] - \Theta_i[{\mathbb H}] \, x^{ij} \Theta_j[{\mathbb H}]+ x_0\, (\Theta^0[{\mathbb H}])^2\biggr] \nonumber\\
& & \hskip-1cm V_{\hat{\mathbb Q} \hat{\mathbb Q}} = -\frac{1}{4\,s\,{\cal V}_E^2}\biggl[\biggl\{\frac{(\Theta_0[\hat{\mathbb Q}])^2}{x_0}  - \Theta^i[\hat{\mathbb Q}]\, x_{ij} \, \Theta^j[\hat{\mathbb Q}] - \Theta_i[\hat{\mathbb Q}] \, x^{ij} \Theta_j[\hat{\mathbb Q}]+ x_0\, (\Theta^0[\hat{\mathbb Q}])^2 \biggr\} \nonumber\\
& & \hskip0cm  + \left(\frac{1}{4\,l+\gamma}\right) \left(\frac{4\, k_0^2}{9} \tilde{\cal G}_{\alpha \beta}- 4 \sigma_\alpha\, \sigma_\beta \right)\,\biggl\{\Psi[\hat{\mathbb Q}^\alpha] \,  \Psi[\hat{\mathbb Q}^\beta]+\Phi[\hat{\mathbb Q}^\alpha] \,  \Phi[\hat{\mathbb Q}^\beta] \biggr\}\biggr]  \nonumber\\
& & \hskip-1cm V_{\hat{\mathbb P} \hat{\mathbb P}} = -\frac{s}{4\,{\cal V}_E^2}\biggl[\biggl\{\frac{(\Theta_0[\hat{\mathbb P}])^2}{x_0}  - \Theta^i[\hat{\mathbb P}]\, x_{ij} \, \Theta^j[\hat{\mathbb P}] - \Theta_i[\hat{\mathbb P}] \, x^{ij} \Theta_j[\hat{\mathbb P}]+ x_0\, (\Theta^0[\hat{\mathbb P}])^2 \biggr\} \nonumber\\
& & \hskip0cm  + \left(\frac{1}{4\,l+\gamma}\right) \left(\frac{4\, k_0^2}{9} \tilde{\cal G}_{\alpha \beta}- 4 \sigma_\alpha\, \sigma_\beta \right)\,\biggl\{\Psi[\hat{\mathbb P}^\alpha] \,  \Psi[\hat{\mathbb P}^\beta]+\Phi[\hat{\mathbb P}^\alpha] \,  \Phi[\hat{\mathbb P}^\beta] \biggr\}\biggr]  \nonumber\\ 
&  & \hskip-1cm V_{\mathbb \mho \mathbb \mho} = -\frac{1}{4\,{\cal V}_E^2}\biggl[\left(\frac{1}{4\,l+\gamma}\right)\, {\cal G}^{ab}\,\biggl\{\Psi[{\mathbb \mho}_a] \,  \Psi[{\mathbb \mho}_b]+\Phi[{\mathbb \mho}_a] \,  \Phi[{\mathbb \mho}_b] \biggr\}\biggr]  \nonumber\\
& & \hskip-1cm V_{\mathbb H \hat{\mathbb Q}} = +\frac{1}{2\,{\cal V}_E^2}\biggl[\biggl\{\frac{\Theta_0[{\mathbb H}]\, \Theta_0[\hat{\mathbb Q}]}{x_0}  - \Theta^i[{\mathbb H}]\, x_{ij} \, \Theta^j[\hat{\mathbb Q}] - \Theta_i[{\mathbb H}] \, x^{ij} \Theta_j[\hat{\mathbb Q}]+ x_0\, \Theta^0[{\mathbb H}]\,\Theta^0[\hat{\mathbb Q}]\biggr\} \nonumber\\
& & \hskip0cm - \left(\frac{4}{4\,l+\gamma}\right)\,\biggl\{\Psi[{\mathbb H}] \,  \Psi[\hat{\mathbb Q}]+\Phi[{\mathbb H}] \,  \Phi[\hat{\mathbb Q}] \biggr\}\biggr]\nonumber\\
& & \hskip-1cm V_{\mathbb F \hat{\mathbb P}} = -\frac{1}{2\,{\cal V}_E^2}\biggl[\biggl\{\frac{\Theta_0[{\mathbb F}]\, \Theta_0[\hat{\mathbb P}]}{x_0}  - \Theta^i[{\mathbb F}]\, x_{ij} \, \Theta^j[\hat{\mathbb P}] - \Theta_i[{\mathbb F}] \, x^{ij} \Theta_j[\hat{\mathbb P}]+ x_0\, \Theta^0[{\mathbb F}]\,\Theta^0[\hat{\mathbb P}]  \biggr\} \nonumber\\
& & \hskip0cm - \left(\frac{4}{4\,l+\gamma}\right)\,\biggl\{\Psi[{\mathbb F}] \,  \Psi[\hat{\mathbb P}]+\Phi[{\mathbb F}] \,  \Phi[\hat{\mathbb P}] \biggr\}\biggr]\nonumber\\
& & \hskip-1cm V_{\hat{\mathbb P} \hat{\mathbb Q}} =\frac{1}{2 \, {\cal V}_E^2} \biggl[\frac{1}{4} \, \left(\frac{4\, k_0^2}{9} \tilde{\cal G}_{\alpha \beta}\,   - 4\, \sigma_\alpha \sigma_\beta \right) \left(\frac{2}{4\,l+\gamma}\right) \biggl\{\Psi[\hat{\mathbb Q}^\alpha] \,  \Phi[\hat{\mathbb P}^\beta]-\Psi[\hat{\mathbb P}^\alpha] \,  \Phi[\hat{\mathbb Q}^\beta] \nonumber\\
& & +\Phi[\hat{\mathbb P}^\alpha] \,  \Psi[\hat{\mathbb Q}^\beta]-\Phi[\hat{\mathbb Q}^\alpha] \,  \Psi[\hat{\mathbb P}^\beta] \biggr\} \biggr]\,, \nonumber
\eea
where the new flux combinations defined in eqn. (\ref{eq:parameters3}) involve the quantities which entirely depend on the topological data of the mirror CY as defined in eqn. (\ref{eq:parameters4}). 

\subsubsection*{Summary of the second proposal for the scalar potential}
Analogous to the compact symplectic version in eqn. (\ref{eq:main2}) and the alternate version in eqn. (\ref{eq:main3}), now the scalar potential in eqn. (\ref{eq:main5}) can be further, very compactly, rewritten as
\bea
\label{eq:main7}
& & \hskip-0.8cm V = -\frac{1}{4\,s\,{\cal V}_E^2} \Biggl[\sum_{{\mathbb Y} \in \bigl\{ \mathfrak F, \, (s\, \mathfrak H) \bigr\} }\left(\frac{(\Theta_0[{\mathbb Y}])^2}{x_0}  - \Theta^i[{\mathbb Y}]\, x_{ij} \, \Theta^j[{\mathbb Y}] - \Theta_i[{\mathbb Y}] \, x^{ij} \Theta_j[{\mathbb Y}]+ x_0\, (\Theta^0[{\mathbb Y}])^2 \right) \, \\
& & \hskip-0.7cm + \left(\frac{1}{4\,l+\gamma}\right) \left(\frac{4\, k_0^2}{9} \tilde{\cal G}_{\alpha \beta}- 4 \sigma_\alpha\, \sigma_\beta \right)\,\biggl\{\left(\Phi[\hat{\mathbb Q}^\alpha] - s\, \Psi[\hat{\mathbb P}^\alpha]\right) \,\left(\Phi[\hat{\mathbb Q}^\beta] - s\, \Psi[\hat{\mathbb P}^\beta]\right)\nonumber\\
& & \hskip-0.7cm+  \left(\Psi[\hat{\mathbb Q}^\alpha] + s\, \Phi[\hat{\mathbb P}^\alpha]\right) \,\left(\Psi[\hat{\mathbb Q}^\beta] + s\, \Phi[\hat{\mathbb P}^\beta]\right) \biggr\} + \left(\frac{1}{4\,l+\gamma}\right) \bigg\{ s\, {\cal G}^{ab}\,\biggl(\Psi[{\mathbb \mho}_a] \,  \Psi[{\mathbb \mho}_b]+\Phi[{\mathbb \mho}_a] \,  \Phi[{\mathbb \mho}_b] \biggr) \nonumber\\
& & \hskip-0.7cm  + 8\, s \biggl( \Psi[{\mathbb H}] \,  \Psi[\hat{\mathbb Q}]+\Phi[{\mathbb H}] \,  \Phi[\hat{\mathbb Q}]\biggr) - 8\,s \biggl( \Psi[{\mathbb F}] \,  \Psi[\hat{\mathbb P}]+\Phi[{\mathbb F}] \,  \Phi[\hat{\mathbb P}] \, \biggr)\biggr\} \Biggr] + V_{tad} \, . \nonumber
\eea
Here, the tadpole terms denoted as $V_{tad}$ are to be nullified via adding additional sources, and are given in eqn. (\ref{eq:tadpoleT}). Further, motivated by the alternate representation of the symplectic formulation given in eqn. (\ref{eq:main3}), the flux combinations $\mathbb Y$ appearing in the summation of the first line represent ${\mathfrak F} = {\mathbb F} +  (s\, \, \hat{\mathbb P})$ and $(s\, {\mathfrak H}) = (s\, \, {\mathbb H}) - \, \, \hat{\mathbb Q}$ with appropriate indices. Moreover, all the flux orbits appearing via ${\mathbb Y}$ are defined through eqns. (\ref{eq:orbitsB1})-(\ref{eq:orbits11B}) while those with are ${\Theta}, \Phi$ and $\Psi$ are defined through the eqns. (\ref{eq:parameters3}) and (\ref{eq:parameters4}).

Note that, for an arbitrary non-geometric flux compactification setup, in order to write down the total $F$-term scalar potential from our proposal in eqn. (\ref{eq:main7}) we just need the topological data of the compactifying (CY) threefold and its mirror. The data on the compactifying threefold side help in writing the flux orbits in eqns. (\ref{eq:orbitsB1})-(\ref{eq:orbits11B}) while data on the mirror threefold side help in expressing the parameters given in eqn. (\ref{eq:parameters4}) to be used for writing ${\Theta}, \Phi$ and $\Psi$ which are defined through the eqns. (\ref{eq:parameters3}), and these are all what one needs to express the total $F$-term scalar potential.

\section{Reading off scalar potential via the topological data of CY and its mirror}
\label{sec_Intersections}
%\subsection{Demonstration of the applicability of the final proposal}
To demonstrate how one could directly use the generic results in eqn. (\ref{eq:main7}) to know the scalar potential just via knowing some topological data on the (CY) threefold and its mirror, let us reconsider our first toroidal model, i.e. Example A\footnote{Recall that, we have already demonstrated the reading off scalar potential pieces for simpler topological data for Example B with no complex structure moduli, which has motivated us for a more general analysis.}.
\subsubsection*{Step 0: Given data}
In this example, the relevant data for the untwisted sector of the toroidal orientifold ($X$) is,
\bea
& & \hskip -1.8cm h^{11}_+(X) = 3, \, \, h^{11}_-(X) = 0, \quad h^{21}_-(X) = 3, \, \, h^{21}_+(X) = 0, \quad k_{123} = 1, \, \hat{d}^\alpha{}_\beta = \delta^\alpha{}_\beta\,,\nonumber\\
& & \hskip 2.0cm \hat{l}_{123} = 1, \quad b_i = 0, \quad  a_{ij} = 0, \quad \gamma = 0\,,
\eea
where the data in the second line corresponds to ${\cal F} = U_1 \,U_2 \,U_3$ which we have used earlier. Also recall that we have set a normalization of $3!$ for $l_{ijk}$, and so $l_{123}= \hat{l}_{123}/6 = 1/6$.
\subsubsection*{Step 1: Checking the counting of fluxes, moduli and axions} The very first implication this data provides is the fact that odd axions $G^a$,  geometric flux $\omega$ and the non-geometric $R$-flux are absent, while there are eight components of $F_3/H_3$ flux each, along with 24 components for $Q/P$-fluxes each. These can be denoted as 
\bea
\label{eq:modela11}
F^0, F^i, F_i, F_0, \quad H^0, H^i, H_i, H_0, \quad {\hat Q}^{\alpha0}, {\hat Q}^{\alpha i}, {\hat Q}^{\alpha}{}_{0}, {\hat Q}^{\alpha}{}_{i}, \quad {\hat P}^{\alpha0}, {\hat P}^{\alpha i}, {\hat P}^{\alpha}{}_{0}, {\hat P}^{\alpha}{}_{i}, \,\, .
\eea
where $ \{i, \alpha\} \in \{1,2,3\}$ corresponds to the three c.s. moduli and K\"ahler moduli each. 
\subsubsection*{Step 2: Writing out the new generalized flux orbits}
Now considering the new flux orbits from eqns. (\ref{eq:orbitsB1})-(\ref{eq:orbits11B}), we have 
\bea
\label{eq:modela33}
& & \hskip-1.5cm {\mathbb H}^0 = H^0 + \rho_\alpha\, {\hat P}^{\alpha 0}, \quad {\mathbb H}^i = H^i + \rho_\alpha\, {\hat P}^{\alpha i}, \quad {\mathbb H}_i = H_i + \rho_\alpha\, {\hat P}^\alpha{}_i, \quad {\mathbb H}_0 = H_0 + \rho_\alpha\, {\hat P}^\alpha{}_0 \nonumber\\
& & \hskip-0.00cm {\mathbb F}^0 = \left(F^0 + \rho_\alpha\, {\hat Q}^{\alpha 0}\right) + C_0\,{\mathbb H}^0, \quad {\mathbb F}^i  = \left(F^i + \rho_\alpha\, {\hat Q}^{\alpha i}\right) + C_0\, {\mathbb H}^i, \quad \, \\
& & {\mathbb F}_i = \left(F_i + \rho_\alpha\, {\hat Q}^\alpha{}_i \right)+ C_0\, {\mathbb H}_i, \quad {\mathbb F}_0 = \left(F_0 + \rho_\alpha\, {\hat Q}^\alpha{}_0 \right) + C_0\, {\mathbb H}_0\nonumber\\
& & \hskip-1.5cm \hat{\mathbb Q} = \hat{Q} + C_0\, \hat{P}, \qquad \hat{\mathbb P} = \hat{P}\, \qquad {\rm with} \, \, {\rm similar}\, \, {\rm appropriate} \, \, {\rm indices}. \nonumber
\eea
\subsubsection*{Step 3: Writing out the ingredients with complex structure moduli}
With a form of the prepotential to be ${\cal F} = U_1 \,U_2 \,U_3$, we find that the complex structure moduli dependent intermediate quantities defined in eqn. (\ref{eq:parameters4}) are reduced into,
\bea
\label{eq:modela22}
& &  x_0 = l\, = u_1\, u_2\, u_3, \qquad y_0 = -1/(2\, u_1\, u_2\, u_3),\qquad \beta = 0 = \gamma, \qquad \qquad r_0 = -\, v_1\, v_2\, v_3,  \nonumber\\
& & \phi_0= u_1\, u_2\, v_3 + u_1\, v_2\, u_3+ v_1\, u_2\, u_3 -v_1\, v_2\, v_3, \quad \psi_0=u_1\, u_2\, u_3 -v_1\, v_2\, u_3 - v_1\, u_2\, v_3- u_1\, v_2\, v_3, \nonumber\\
& & \phi_i =  \left( \begin{array}{c}
v_2\, v_3 -u_2\, u_3  \\
v_3\, v_1 - u_3\, u_1  \\
v_1\, v_2- u_1\, u_2  \end{array} \right), \quad \psi_i =  \left( \begin{array}{c}
u_2\, v_3 +v_2\, u_3  \\
u_3\, v_1 + v_3\, u_1  \\
u_1\, v_2 + v_1\, u_2  \end{array} \right), \quad r_i =  \left( \begin{array}{c}
v_2\, v_3  \\
v_3\, v_1   \\
v_1\, v_2  \end{array} \right) = -\epsilon_i \, \quad, \\
& &  \eta_{ij} =  \left( \begin{array}{ccc}
0 & v_3 & v_2 \\
v_3 & 0 & v_1 \\
v_2 & v_1 & 0 \end{array} \right), \quad  x_{ij} =  \left( \begin{array}{ccc}
-\frac{u_2\, u_3}{u_1} & 0 & 0 \\
0 & -\frac{u_1\, u_3}{u_2} & 0 \\
0 & 0 & -\frac{u_1\, u_2}{u_3} \end{array} \right)\, . \nonumber
\eea
Using these simplified relations in eqns. (\ref{eq:modela33})-(\ref{eq:modela22}) along with the flux combinations in eqn. (\ref{eq:parameters3}), we have subsequently verified that the proposed form of the scalar potential given in eqn. (\ref{eq:main7}) indeed reproduces all the 9661 terms generated from the K\"ahler- and super- potentials as given by eqns. (\ref{eq:Kexample1})-(\ref{eq:Wexample1}). To conclude, it may be attractive to mention that, for this particular example, we have obtained the same scalar potential having a total of 9661 terms via three different routes \footnote{One may count a fourth route of arriving at the same scalar potential (up to satisfying a subset of NS-NS Bianchi identities) via utilizing the metric of the compactifying toroidal sixfold as shown in \cite{Gao:2015nra}. However, the lack of knowledge about CY metric has motivated us to find alternatives via utilizing the period matrices, and the moduli space matrices involving the complex structure moduli and the K\"ahler moduli.},
\begin{itemize}
\item{first one being obtained from the K\"ahler and super-potentials in eqns. (\ref{eq:Kexample1})-(\ref{eq:Wexample1}).}
\item{second one being obtained from `symplectic proposal' in eqn. (\ref{eq:main2}). This has been elaborated in eqn. (\ref{eq:sympVtotExample1}).}
\item{third one being directly read off from the final proposal given in eqn. (\ref{eq:main7}).}
\end{itemize}
This appears to be quite remarkable !  %On the lines of these studies and illustrations, we argue that the generic form of scalar potential proposed in eqn. (\ref{eq:main5}) helps in directly reading off the various pieces via merely knowing the topological data of the compactifying (CY) threefold and its mirror.

\section{Summary and conclusions}
\label{sec_conclusion}
In this article, we have proposed a modular invariant symplectic formulation for the non-geometric scalar potential obtained from a modular completed flux superpotential within the framework of type IIB orientifold compactification. We provide two {\it additional} ways of representing the total $F$-term scalar potential arising from a generic K\"ahler potential ($K$) and a modular completed version of the flux superpotential ($W$) involving (non-)geometric fluxes on top of standard NS-NS flux $H_3$ and RR flux $F_3$.

In the first proposal, we have presented the scalar potential in a symplectic formulation via utilizing a new set of generalized flux orbits along with some crucial symplectic identities. Motivated by the recent studies in \cite{Blumenhagen:2013hva, Gao:2015nra, Shukla:2015rua, Shukla:2015bca}, this representation has generalized the results of \cite{Shukla:2015hpa} by including the non-geometric $P$-flux leading to modular completion of the scalar potential. We have been able to reduce the scalar potential into a form from which their ten-dimensional origin could be invoked. This argument is based on the recent dimensional oxidation proposal on the lines of  \cite{ Shukla:2015hpa, Blumenhagen:2013hva, Gao:2015nra, Shukla:2015rua, Shukla:2015bca} where one considers all the fluxes as constant parameters. This generalization with $P$-flux inclusion results in the following form, 
\begin{eqnarray}
\label{eq:main6}
& & \hskip-1.0cm V = -\frac{1}{4\, s\, {\cal V}_E^2} \int_{X_6}\,\biggl[{\mathbb F} \wedge \ast {\mathbb F} + s^2 \, {\mathbb H} \wedge \ast {\mathbb H} + \hat{\mathbb Q} \wedge \ast \hat{\mathbb Q} +s^2 \,  \hat{\mathbb P} \wedge \ast \hat{\mathbb P} + \, \frac{s}{4} \,\, {\cal G}^{ab} \, \, \tilde{\mathbb \mho}_a \wedge \ast \tilde{\mathbb \mho}_b \\
& & + \, \frac{1}{4} \, \left(\frac{4\, k_0^2}{9} \, \tilde{\cal G}_{\alpha \beta}\,   - 4\, \sigma_\alpha \, \sigma_\beta \right) \left(\tilde{\cal Q}^\alpha \wedge \ast \tilde{\cal Q}^\beta + s^2\, \tilde{\cal P}^\alpha \wedge \ast \tilde{\cal P}^\beta \right) - 2 \, s \, \left({\mathbb H} \wedge \ast \hat{\mathbb Q} -{\mathbb F} \wedge \ast \hat{\mathbb P}\right)  \nonumber\\
& & - 4 \, s \, \left({\mathbb H} \wedge \ast \tilde{\cal Q} -{\mathbb F} \wedge \ast \tilde{\cal P}\right)  + \frac{1}{4} \, \times (-2\, s)\left(\frac{4\, k_0^2}{9} \, \tilde{\cal G}_{\alpha \beta}\,  - 4\, \sigma_\alpha \, \sigma_\beta \right) \left(\hat{\mathbb P}^\alpha \wedge \tilde{\cal Q}^\beta - \hat{\mathbb Q}^\alpha \wedge \tilde{\cal P}^\beta\right)  \nonumber\\
& &  - 2 \, s \, \hat{\mathbb P} \wedge \hat{\mathbb Q} + \left( 2 \, s \, {\mathbb F} \wedge {\mathbb H} - 2 \, \, {\mathbb F} \wedge \hat{\mathbb Q} - 2 \, s^2 \, \, \, {\mathbb H} \wedge \hat{\mathbb P} \right) \biggr] \, , \nonumber
\end{eqnarray}
where the various flux-combinations $\{ {\mathbb F}, {\mathbb H}, \hat{\mathbb Q}, \hat{\mathbb P} \}$ are defined in eqns. (\ref{eq:orbitsB1})-(\ref{eq:orbits11B}) while the other ones $\{\tilde{\cal \mho}, \tilde{{\cal Q}}, \tilde{{\cal P}} \}$ are defined in eqn. (\ref{eq:tildeQP}). 

Though our current analysis has not been intended to find a concrete ten-dimensional theory which, after dimensional reduction, could result in the generic modular completed scalar potential in eqn. (\ref{eq:main6}), however on the lines of recent analysis based on the reduction of Double Field Theory (DFT) on CYs without $P$-flux, it would be interesting to look for a modular completion of the proposal made in \cite{Blumenhagen:2015lta}, and then compare the same with those of ours. For looking at the connections between the approach of \cite{Blumenhagen:2015lta} and that of ours, see \cite{Shukla:2015hpa}. 

In the first proposal, all the information about the complex structure moduli have been encoded into the symplectic quantities, and so the full scalar potential was still not explicitly written in terms of real scalars coming from the complex structure moduli. For that purpose, in the second step we have explicitly investigated the saxionic/axionic dependence of complex structure moduli by expanding out the sympletic sector of the first proposal. A summary of the final version of the scalar potential can be presented as under, 
\bea
\label{eq:main8}
& & \hskip-0.8cm V = -\frac{1}{4\,s\,{\cal V}_E^2} \Biggl[\sum_{{\mathbb Y} \in \bigl\{ \mathfrak F, \, (s\, \mathfrak H) \bigr\} }\left(\frac{(\Theta_0[{\mathbb Y}])^2}{x_0}  - \Theta^i[{\mathbb Y}]\, x_{ij} \, \Theta^j[{\mathbb Y}] - \Theta_i[{\mathbb Y}] \, x^{ij} \Theta_j[{\mathbb Y}]+ x_0\, (\Theta^0[{\mathbb Y}])^2 \right) \, \\
& & \hskip-0.7cm + \left(\frac{1}{4\,l+\gamma}\right) \left(\frac{4\, k_0^2}{9} \tilde{\cal G}_{\alpha \beta}- 4 \sigma_\alpha\, \sigma_\beta \right)\,\biggl\{\left(\Phi[\hat{\mathbb Q}^\alpha] - s\, \Psi[\hat{\mathbb P}^\alpha]\right) \,\left(\Phi[\hat{\mathbb Q}^\beta] - s\, \Psi[\hat{\mathbb P}^\beta]\right)\nonumber\\
& & \hskip-0.7cm+  \left(\Psi[\hat{\mathbb Q}^\alpha] + s\, \Phi[\hat{\mathbb P}^\alpha]\right) \,\left(\Psi[\hat{\mathbb Q}^\beta] + s\, \Phi[\hat{\mathbb P}^\beta]\right) \biggr\} + \left(\frac{1}{4\,l+\gamma}\right) \bigg\{ s\, {\cal G}^{ab}\,\biggl(\Psi[{\mathbb \mho}_a] \,  \Psi[{\mathbb \mho}_b]+\Phi[{\mathbb \mho}_a] \,  \Phi[{\mathbb \mho}_b] \biggr) \nonumber\\
& & \hskip-0.7cm  + 8\, s \biggl( \Psi[{\mathbb H}] \,  \Psi[\hat{\mathbb Q}]+\Phi[{\mathbb H}] \,  \Phi[\hat{\mathbb Q}]\biggr) - 8\,s \biggl( \Psi[{\mathbb F}] \,  \Psi[\hat{\mathbb P}]+\Phi[{\mathbb F}] \,  \Phi[\hat{\mathbb P}] \, \biggr)\biggr\} \Biggr] + V_{tad} \, . \nonumber
\eea
Here flux combinations $\mathbb Y$ appearing in the summation of the first line represent ${\mathfrak F} = {\mathbb F} +  (s\, \, \hat{\mathbb P})$ and $(s\, {\mathfrak H}) = (s\, \, {\mathbb H}) - \, \, \hat{\mathbb Q}$ with appropriate indices. Moreover, all the flux orbits represented via ${\mathbb Y}$ are defined through eqns. (\ref{eq:orbitsB1})-(\ref{eq:orbits11B}) while those with ${\Theta}, \Phi$ and $\Psi$ are defined through the eqns. (\ref{eq:parameters3}) and (\ref{eq:parameters4}). As we have demonstrated, all these combinations can indeed be directly read-off from some topological data of the compactifying (CY) threefolds and their mirrors. It will be interesting to find some fundamental reasons behind the new flux combinations in eqn. (\ref{eq:parameters3}) similar to the previous flux orbits given in eqns. (\ref{eq:orbitsB1})-(\ref{eq:orbits11B}) which have been motivated by T-duality and S-duality arguments. 

In a nutshell, we have expressed the generic tree level $F$-term non-geometric scalar potential into a very compact form {\it written explicitly in terms of the real moduli and axions}. Subsequently we have illustrated how the compact form of the generic scalar potential proposed in eqn. (\ref{eq:main8}) can be useful for writing down the full scalar potential via merely knowing some topological data of the orientifold. The proposal being valid for a generic Calabi Yau compactification with arbitrary number of complex structure moduli as well as K\"ahler moduli may help in utilizing the same towards some model independent applications, e.g. for making attempts in moduli stabilization and realization of de-Sitter vacua. We hope to get back to these issues through a detailed investigation in a future work.

\section*{Acknowledgments}
I am grateful to Michele Cicoli and Fernando Quevedo for their kind support and encouragements. I am very thankful to Michele Cicoli, David Ciupke, Gustavo A. Duran,  Francesco Muia and Damian M. Pena for useful discussions. I would like to thank Ralph Blumenhagen for many useful discussions and enlightening learning of the subject during earlier collaborations. I also thank the High Energy Physics group at University of Bologna and INFN-Bologna, Bologna (Italy), for their kind hospitality during short visits to the centers where some parts of this work have been done.

%\newpage

\appendix

\section{A summary of various symbols and redefinitions}
\label{sec_symbolsList}
As we have utilized various distinct symbols for fluxes and moduli at different intermediate stages, let us list them here at one place for the convenience to the readers\footnote{We thank the referee for her/his important suggestions to list all the distinct symbols, and for an overall improvement of the presentation of the article.},
\subsection*{Moduli and axions}
\begin{itemize}
\item{The three types of chiral variables $\tau, G^a$ and $T_\alpha$ are defined in eqn. (\ref{eq:N=1_coords}). Here the indices $a$ and $\alpha$ are counted via odd/even Hodge numbers $h^{11}_-(CY)$ and $h^{11}_+(CY)$ respectively.}
\item{The S-duality invariant quantities $\sigma_\alpha$ and $\tilde{\rho}_\alpha$ are defined below eqn. (\ref{eq:N=1_coords2}) which are, respectively the four-cycle volume moduli and an axionic combination involving $C_2, C_4$ and $B_2$ axions. These quantities are utilized for constructing the flux orbits as given in eqn. (\ref{eq:orbitsA0}) and eqn. (\ref{eq:orbits11A}). }
%\item{The letter $\sigma$ (without an index $\alpha$) is denoted as the holomorphic involution acting on the internal Calabi Yau manifold. However, this letter $\sigma$ does not have any explicit appearance/relevance while writing out the scalar potential.}
\item{We have also used short hand notations: $k_{0} = k_{\alpha\beta\gamma} t^\alpha t^\beta t^\gamma \equiv 6 {\cal V}_E, k_{\alpha} = k_{\alpha\beta\gamma} t^\beta t^\gamma, k_{\alpha\beta} = k_{\alpha\beta\gamma}  t^\gamma$ and $\hat{k}_{ab} = \hat{k}_{\alpha ab} t^\alpha$. Moreover, moduli space matrices (${\cal G}_{\alpha\beta}, {\cal G}_{ab}$) and their respective inverses (${\cal G}^{\alpha\beta}, {\cal G}^{ab}$)  are defined in eqns. (\ref{eq:genMetrices}). }
\item{We have used $\tilde{\cal G}_{\alpha \beta}=\left(({\hat{d}^{-1}})_{\alpha}{}^{\alpha'}\, {\cal G}_{\alpha' \beta'}\, ({\hat{d}^{-1}})_{\beta}{}^{\beta'}\right)$ and $\hat{\kappa}_{\alpha a b} = (\hat{d}^{-1})_\alpha{}^{\beta}\hat{k}_{\beta a b}$. Note that, from eqn. (\ref{eq:intersection}), in case of four-forms being dual to the respective two-forms, these redefinitions will be trivial as $d_a^b = \delta_a^b$ and $\hat{d}_\alpha^\beta = \delta_\alpha^\beta$, e.g. in Example A. However, for Example B, we have $\hat{d}_\alpha{}^\beta = diag\{1/2, -1, 1/4\}$ and $d^a{}_b=diag\{-1,-1/2\}$ (following the notation of \cite{Robbins:2007yv}), and so this distinction is relevant otherwise we will not end up having the result in eqns. (\ref{eq:exBqqpp})-(\ref{eq:exBpq}) needed for matching the scalar potential pieces. }
\end{itemize}

\subsection*{Fluxes and superpotential}
\begin{itemize}
\item{The svarious NS-NS and RR fluxes are denoted as,
\bea
& & \hskip-0.5cm  H\equiv \left(H_\Lambda, H^\Lambda\right), \, \, \omega\equiv \left({\omega}_a{}^\Lambda, {\omega}_{a \Lambda} , \hat{\omega}_\alpha{}^K, \hat{\omega}_{\alpha K}\right), \, \, Q\equiv \left({Q}^{a{}K}, \, {Q}^{a}{}_{K}, \, \hat{Q}^{\alpha{}\Lambda} , \, \hat{Q}^{\alpha}{}_{\Lambda}\right), \,  \nonumber\\
& & \hskip0.5cm R\equiv \left(R_K, R^K \right), \, \, F\equiv \left(F_\Lambda, F^\Lambda\right), \quad P\equiv \left({P}^{a{}K}, \, {P}^{a}{}_{K}, \, \hat{P}^{\alpha{}\Lambda} , \, \hat{P}^{\alpha}{}_{\Lambda}\right). 
\eea}
\item{All these fluxes are counted via the various hodge numbers $\Lambda \in \{0,1, 2, .., h^{21}_-(CY)\}$, $K \in \{1, 2, .., h^{21}_+(CY)\}$, $\alpha \in \{1, 2, .., h^{11}_+(CY)\}$ and $a \in \{1, 2, .., h^{11}_-(CY)\}$. Later on, we split the indices $\Lambda$ as: $\Lambda = \{ 0, i\}$ where $i$'s are counted via $h^{21}_-(CY)$.}
\item{Flux components counted via $h^{2,1}_+(CY)$ indices do not appear in our current ($F$-term) analysis, i.e. fluxes $\{ \hat{\omega}_\alpha{}^K, \hat{\omega}_{\alpha K}, {Q}^{a{}K}, \, {Q}^{a}{}_{K}, {P}^{a{}K}, \, {P}^{a}{}_{K} \}$ are not relevant in this article.}
%\item{Here we also note that geometric flux components with ``even" (2,1)-cohomology indices ($K$) and the non-geometric $Q/P$ flux components with ``odd" (2,1)-cohomology indices ($\Lambda$) appear with a ``hat" while for the cases otherwise there is no ``hat".}
\item{Flux combinations denoted with small and bold letters, e.g. $({\bf h^{\Lambda}, h_\Lambda; \omega_{a\Lambda}, \omega_a{}^\Lambda; \hat{q}^{\alpha\Lambda}}$ etc.) as defined in eqns. (\ref{eq:orbits11A})-(\ref{eq:orbits11B}) are motivated by the $SL(2,\mathbb Z)$ arguments as seen from their nice transformation structures given in eqns. (\ref{eq:orbits11C})-(\ref{eq:orbits11D}). However, these fluxes do not explicitly appear in our final scalar potential rearrangements.}
\item{{\it The explicit flux combinations which are directly utilized for representing the scalar potential terms} turn out to be the ones denoted by capital letters like ${\mathbb F}_\Lambda$, ${\mathbb H}_\Lambda, {\mathbb \mho}_{a \Lambda}, \hat{\mathbb Q}^\alpha{}_\Lambda, \hat{\mathbb P}^\alpha{}_\Lambda$ etc. as defined in eqns. (\ref{eq:orbitsB1}) and (\ref{eq:orbits11B}).}
\item{If a $Q$-flux component appears in the scalar potential piece as $\{\hat{\mathbb Q}^\Lambda, \hat{\mathbb Q}_\Lambda \}$, i.e. without an $(1,1)$-cohomology index $\alpha$, this means that $\{ \hat{\mathbb Q}^\Lambda = \hat{\mathbb Q}^{\Lambda \alpha} \sigma_\alpha, \, \hat{\mathbb Q}_\Lambda = \hat{\mathbb Q}^\alpha_{\Lambda} \sigma_\alpha \}$ has been considered, and similar is the case for $P$-flux components as well. This helps in denoting the non-geometric fluxes in the same way as the standard flux $F/H$ being expanded in the basis of odd-(2,1) forms, namely $\hat{\mathbb Q} = \hat{\mathbb Q}^\Lambda {\cal A}_\Lambda +  \hat{\mathbb Q}_\Lambda {\cal B}^\Lambda$.}
\item{For presenting an alternate compact way of the scalar potential pieces in eqn. (\ref{eq:main3}), we have used another flux combination denoted as $({\mathfrak F}, s\, {\mathfrak H})$ as defined in eqn. (\ref{eq:mathfrakFH}).}
\item{The superpotential involves two symplectic vectors, namely $(e_\Lambda, m^\Lambda)$ as defined in eqns. (\ref{eq:W_gen})- (\ref{eq:eANDm}) and $({\cal X}^\Lambda, {\cal F}_\Lambda)$ as defined from the (3,0)-form $\Omega_3$ in eqn. (\ref{eq:Omega3}).}
\end{itemize}

\subsection*{Pre-potential and periods}
\begin{itemize}
\item{It is worth to recall that while $({\mathbb F}_\Lambda, {\mathbb F}^\Lambda)$ denote the components of generalized new flux orbits for $F_3$-flux, we denote $({\cal F}_\Lambda, {\cal F}_{\Lambda \Sigma})$ as the derivatives of the prepotential denoted as ${\cal F}$ and defined in eqn. (\ref{eq:prepotential}) or equivalently in eqn. (\ref{eq:prepotentialNew}).}
\item{The prepotentials in eqns. (\ref{eq:prepotential}) and  (\ref{eq:prepotentialNew}) have a slight difference between the mirror triple-intersection numbers as we have $l_{ijk} = \frac{1}{3!}\, \hat{l}_{ijk}$. This is just to avoid the repetitive appearance of the rational pre-factors at many places in the intermediate stages of computations presented in section \ref{sec_Intersections1}.}
\item{Period vectors ${\cal N}_{\Lambda \Delta}$ to be determined by the prepotential derivatives ${\cal F}_{\Lambda \Delta}$ are defined in eqn. (\ref{eq:periodN}). Moreover, the reverse case of ${\cal F}_{\Lambda \Delta}$ being also determined via ${\cal N}_{\Lambda \Delta}$ corresponds to an analogous relation as given in eqn. (\ref{eq:periodF}).}
\end{itemize}

\subsection*{Other symplectic quantities}
\begin{itemize}
\item{The set of matrices $\{{\cal M}^{\Lambda\Delta}, {\cal M}^{\Lambda}{}_{\Delta}, {\cal M}_{\Lambda}{}^{\Delta}, {\cal M}_{\Lambda\Delta}\}$, which appear in the definition of Hodge start of odd three-forms $\{{\cal A}_\Lambda, {\cal B}^\Lambda \}$ as given eqn. (\ref{stardef}), are defined in eqn. (\ref{coff}). Moreover, an analogous set of matrices  denoted as $\{{\cal L}^{\Lambda\Delta}, {\cal L}^{\Lambda}{}_{\Delta}, {\cal L}_{\Lambda}{}^{\Delta}, {\cal L}_{\Lambda\Delta}\}$ are defined in eqn. (\ref{coff2}) via exchanging the role of ${\cal N}_{\Lambda\Delta}$ with ${\cal F}_{\Lambda\Delta}$ in eqn. (\ref{coff}).}
\item{These two sets of matrices ${\cal M}$ and ${\cal L}$ are utilized to define a new collection of matrices $\{{\cal S}^{\Lambda\Delta}, {\cal S}^{\Lambda}{}_{\Delta}, {\cal S}_{\Lambda}{}^{\Delta}, {\cal S}_{\Lambda\Delta}\}$ as defined in eqn. (\ref{coff5}).}
\item{Using these sets of four ${\cal S}$ matrices and $\{{\mathbb \mho}_a, \hat{\mathbb Q}^\alpha, \hat{\mathbb P}^\alpha \}$, we have defined another set of new flux components with calligraphic letters and tildes, being denoted $\{ \tilde{\cal \mho}_a, \tilde{\cal Q}^\alpha, \tilde{\cal P}^\alpha \}$ as mentioned in eqn. (\ref{eq:tildeQP}).}
\item{Finally, there are only two classes ($\{{\mathbb F}, {\mathbb H}, {\mathbb \mho}_a, \hat{\mathbb Q}^\alpha, \hat{\mathbb P}^\alpha \}$ and $\{ \tilde{\cal \mho}_a, \tilde{\cal Q}^\alpha, \tilde{\cal P}^\alpha \}$) of flux combinations which appear in the final symplectic version of the potential in eqn. (\ref{eq:main2}). }
\end{itemize}

\subsection*{Some redefinitions for invoking explicit c.s. moduli dependence}
For rewriting the scalar potential pieces using explicit dependencies on the saxionic components ($u^i$) and the axionic components ($v^i$) of complex structure moduli as given in eqn. (\ref{eq:main7}), we have introduced the following more redefinitions,
\bea 
& & \hskip-2cm \Theta^0[{\mathbb Y}] = {\mathbb Y}^0, \hskip5.3cm \Theta_0[{\mathbb Y}] =  {\mathbb Y}_0 + v^i \,  {\mathbb Y}_i + r_0  {\mathbb Y}^0 + r_i  {\mathbb Y}^i\, , \nonumber\\
& & \hskip-2cm \Theta^i[{\mathbb Y}] = {\mathbb Y}^i - v^i \, {\mathbb Y}^0, \hskip4.0cm \Theta_i[{\mathbb Y}] =  {\mathbb Y}_i + \epsilon_i\,  {\mathbb Y}^0 + \eta_{ij} \,  {\mathbb Y}^j\, ,\\
& & \hskip-1.9cm \Phi[{\mathbb Y}] = {\mathbb Y}_0 +v^i \, {\mathbb Y}_i +\phi_0\,  {\mathbb Y}^0+\phi_i \, {\mathbb Y}^i, \quad \qquad \Psi[{\mathbb Y}] =u^i \, {\mathbb Y}_i +\psi_0\,  {\mathbb Y}^0+\psi_i \, {\mathbb Y}^i\,, \nonumber
\eea
where the flux parameters ${\mathbb Y} \in \{ {\mathbb F}, {\mathbb H}, {\mathbb \mho}_a, \hat{\mathbb Q}^\alpha, \hat{\mathbb P}^\alpha \}$ have appropriate upper and lower $h^{21}_-(CY)$-indices. Moreover, here we have denoted $\Theta[\mathbb Y]$ etc. in such a way so that one could distinguish among various flux orbits; for example, $\Theta^0[{\mathbb H}] = {\mathbb H}^0$ and $\Theta_0[{\mathbb F}] =  {\mathbb F}_0 + v^i \,  {\mathbb F}_i + r_0  {\mathbb F}^0 + r_i  {\mathbb F}^i$ etc.  In addition, we have used the followings,
\bea 
& & \hskip-0.8cm r_0 = \left(b_i \, v^i -\beta \, l_i\, v^i - l_{ijk} v^i v^j v^k  \right), \quad \qquad \quad \epsilon_i = \left(b_i + \beta\, l_i\,- 3\, l_{ijk} v^j v^k\right) \\
& & \hskip-0.8cm r_i = \left(b_i + a_{ij} v^j +\beta \, l_i + 3 \, l_{ijk}\, v^j v^k\right), \qquad \quad \eta_{ij} = \left(a_{ij} + 6 \,l_{ijk}\, v^k \right). \nonumber\\
& & \hskip-0.8cm \phi_0 = \left(b_i \, v^i + 3\, l_i\, v^i - l_{ijk} v^i v^j v^k  \right), \quad \qquad \quad \psi_0 = \left(l + \gamma + b_i\, u^i - 3\, l_{ij} v^i v^j\right) \nonumber\\
& & \hskip-0.8cm \phi_i = \left(b_i + a_{ij} v^j - 3 \, l_i + 3 \, l_{ijk}\, v^j v^k\right), \qquad \quad \psi_i = \left(a_{ij}\, u^j + 6 \,l_{ij}\, v^j \right). \nonumber\\
& & \nonumber\\
& & \hskip-1.3cm \beta=\frac{9 \gamma}{8\, l - \gamma}, \, x_0 = \frac{(4\, l + \gamma)(2\, l - \gamma)}{(8\, l - \gamma)}, \qquad x_{ij} = 6 l_{ij} -\frac{72\, l_i \, l_j}{(8\, l - \gamma)}, \, \, x^{ij}= \frac{1}{6}\, l^{ij} - \frac{2}{4\, l + \gamma} \, u^i\, u^j\, , \nonumber
\eea
where we have utilized short hand notations: $l=l_{ijk}u^i u^j u^k, l_i = l_{ijk} u^j u^k$ and $l_{ij} = l_{ijk} u^k$.

\section{Expanded version of the symplectic formulation of the scalar potential}
\label{sec_ExpandedVersion1}
For arriving at a symplectic formulation for the scalar potential collection presented in eqn. (\ref{eq:Vsss12345}), all we need to do is to replace the quantities $Re({\cal X}^\Lambda \ov {\cal X}^\Delta), Im({\cal X}^\Lambda \ov {\cal X}^\Delta)$ etc. into the ingredients which could be written in terms of the symplectic matrices ${\cal M}$ and ${\cal L}$ defined in (\ref{coff}) and (\ref{coff2}) respectively. Apart from the relations (\ref{eq:symp10})-(\ref{eq:symp11}), one can further show that the following non-trivial relations holds which will be more directly useful (as in \cite{Shukla:2015hpa}),
\bea
\label{eq:symp121}
& & \hskip-1.1cm 8 \, e^{K_{cs}} Re({\cal X}^\Gamma \, \ov{\cal X}^\Delta) = {{\cal S}}^{\Gamma}{}_{\Lambda} \left({\cal M}^{\Lambda \Sigma} {{\cal S}}_{\Sigma}^{\, \, \, \, \, \Delta} + {\cal M}^\Lambda_{\, \, \, \, \Sigma} {{\cal S}}^{\Sigma \Delta}\right) - {{\cal S}}^{\Gamma \Lambda} \left({\cal M}_{\Lambda}^{\,\,\, \Sigma} {{\cal S}}_{\Sigma}^{\, \, \, \, \, \Delta} + {\cal M}_{\Lambda \, \Sigma} {{\cal S}}^{\Sigma \Delta}\right) \, \nonumber\\
& & \hskip-1.1cm 8 \, e^{K_{cs}} Re({\cal X}^\Gamma \, \ov{\cal F}_\Delta) = {{{\cal S}}}^{\Gamma}{}_{\Lambda} \left({\cal M}^{\Lambda \Sigma} {{\cal S}}_{\Sigma \Delta} + {\cal M}^\Lambda_{\, \, \, \, \Sigma} {{{\cal S}}}^{\Sigma}_{\,\,\,\, \,\,\Delta}\right) - {{{\cal S}}}^{\Gamma \Lambda} \left({\cal M}_{\Lambda}^{\,\,\, \Sigma} {{\cal S}}_{\Sigma \, \Delta} + {\cal M}_{\Lambda \, \Sigma} {{{\cal S}}}^{\Sigma}_{\,\,\,\,\, \Delta}\right) \nonumber\\
& & \hskip-1.1cm 8 \, e^{K_{cs}} Re({\cal F}_\Gamma \, \ov{\cal X}^\Delta) = {{\cal S}}_{\Gamma \Lambda} \left({\cal M}^{\Lambda \Sigma} {{{\cal S}}}_{\Sigma}^{\, \, \, \, \, \Delta} + {\cal M}^\Lambda_{\, \, \, \, \Sigma} {{\cal S}}^{\Sigma \Delta}\right) -{{{\cal S}}}_{\Gamma}{}^{\Lambda} \left({\cal M}_{\Lambda}^{\,\,\, \Sigma} {{{\cal S}}}_{\Sigma}^{\, \, \, \, \, \Delta} + {\cal M}_{\Lambda \, \Sigma} {{\cal S}}^{\Sigma \Delta}\right) \, \\
& & \hskip-1.1cm 8 \, e^{K_{cs}} Re({\cal F}_\Gamma \, \ov{\cal F}_\Delta) = {{\cal S}}_{\Gamma \Lambda} \left({\cal M}^{\Lambda \Sigma} {{\cal S}}_{\Sigma \Delta} + {\cal M}^\Lambda_{\, \, \, \, \Sigma} \, {{{\cal S}}}^{\Sigma}_{\,\,\,\, \,\,\Delta}\right) - {{{\cal S}}}_{\Gamma}{}^{\Lambda} \left({\cal M}_{\Lambda}^{\,\,\, \Sigma} {{\cal S}}_{\Sigma \, \Delta} + {\cal M}_{\Lambda \, \Sigma} \, {{{\cal S}}}^{\Sigma}_{\,\,\,\,\, \Delta}\right) \, ,\nonumber
\eea
where
\begin{eqnarray}
\label{coff5}
&& \hskip-1.5cm {{\cal S}}^{\Lambda \Delta} = \left({\cal M}^{\Lambda}{}_{ \Sigma} \, {\cal L}^{\Sigma \Delta} + {\cal M}^{\Lambda \Sigma} \, {\cal L}_{\Sigma}{}^{ \Delta} \right), \quad \quad \quad \quad \quad {{\cal S}}_\Lambda^{\, \, \, \, \Delta} =\left({\cal M}_{\Lambda}{}_{ \Sigma} \, {\cal L}^{\Sigma \Delta} + {\cal M}_{\Lambda}{}^{ \Sigma} \, {\cal L}_{\Sigma}{}^{ \Delta} \right)  - \delta_\Lambda{}^\Delta\, \, \\
&& \hskip-1.5cm {{\cal S}}^{\Lambda}{}_{ \Delta}  = - \left({\cal M}^{\Lambda}{}_{ \Sigma} \, {\cal L}^{\Sigma}{}_\Delta + {\cal M}^{\Lambda \Sigma} \, {\cal L}_{\Sigma \Delta}\right) + \delta^\Lambda{}_\Delta, \quad \quad \quad {{\cal S}}_{\Lambda \Delta}\, =  - \left({\cal M}_{\Lambda \Sigma} \, {\cal L}^{\Sigma}{}_{ \Delta} + {\cal M}_{\Lambda}{}^{\Sigma} \, {\cal L}_{\Sigma \Delta} \right) \nonumber
\end{eqnarray}
Now utilizing the relations given in eqns. (\ref{eq:symp10})-(\ref{eq:symp11}) and (\ref{eq:symp121})-(\ref{coff5}), we can summarize the various pieces of the scalar potential in eqn. (\ref{eq:Vsss12345}) to be given as,
\bea
\label{eq:main1}
& & \hskip-0.5cm V_{{\mathbb F} {\mathbb F}} = -\frac{1}{4\, s \, {\cal V}_E^2} \, \,\biggl[{\mathbb F}_\Lambda \, {\cal M}^{\Lambda \Delta} \,  {\mathbb F}_\Delta - {\mathbb F}_\Lambda \, {\cal M}^{\Lambda}_{\, \, \, \Delta} \, {\mathbb F}^\Delta+  {\mathbb F}^\Lambda \, {\cal M}_{\Lambda}^{\, \, \, \Delta} \, {\mathbb F}_\Delta- {\mathbb F}^\Lambda \, {\cal M}_{\Lambda \Delta} \, {\mathbb F}^\Delta \biggr] \\
& & \hskip-0.5cm V_{{\mathbb H} {\mathbb H}} =  -\frac{s}{4\, \, {\cal V}_E^2}  \, \biggl[{\mathbb H}_\Lambda \, {\cal M}^{\Lambda \Delta} \,  {\mathbb H}_\Delta - {\mathbb H}_\Lambda \, {\cal M}^{\Lambda}_{\, \, \, \Delta} \, {\mathbb H}^\Delta+  {\mathbb H}^\Lambda \, {\cal M}_{\Lambda}^{\, \, \, \Delta} \, {\mathbb H}_\Delta- {\mathbb H}^\Lambda \, {\cal M}_{\Lambda \Delta} \, {\mathbb H}^\Delta \biggr]\nonumber\\
& & \hskip-0.50cm V_{\hat{\mathbb Q} \hat{\mathbb Q}}=  -\frac{1}{4\, s \, {\cal V}_E^2}  \, \biggl[\left(\hat{\mathbb Q}_\Lambda \, {\cal M}^{\Lambda \Delta} \,  \hat{\mathbb Q}_\Delta - \hat{\mathbb Q}_\Lambda \, {\cal M}^{\Lambda}_{\, \, \, \Delta} \, \hat{\mathbb Q}^\Delta+  \hat{\mathbb Q}^\Lambda \, {\cal M}_{\Lambda}^{\, \, \, \Delta} \, \hat{\mathbb Q}_\Delta- \hat{\mathbb Q}^\Lambda \, {\cal M}_{\Lambda \Delta} \, \hat{\mathbb Q}^\Delta\right) \nonumber\\
& & \hskip0.75cm +\frac{1}{4} \, \left(\frac{4\, k_0^2}{9} \, \tilde{\cal G}_{\alpha \beta}\,   - 4\, \sigma_\alpha \, \sigma_\beta \right) \biggl(\tilde{\cal Q}^\alpha{}_\Lambda \, {\cal M}^{\Lambda \Delta} \,  \tilde{\cal Q}^\beta{}_\Delta - \tilde{\cal Q}^\alpha{}_\Lambda \, {\cal M}^{\Lambda}_{\, \, \, \Delta} \, \tilde{\cal Q}^{\beta\Delta}\nonumber\\
& & \hskip7cm+  \tilde{\cal Q}^{\alpha\Lambda} \, {\cal M}_{\Lambda}^{\, \, \, \Delta} \, \tilde{\cal Q}^\beta{}_\Delta- \tilde{\cal Q}^{\alpha\Lambda} \, {\cal M}_{\Lambda \Delta} \, \tilde{\cal Q}^{\beta\Delta}\biggr) \biggr] \,\nonumber\\
& & \hskip-0.50cm V_{\hat{\mathbb P} \hat{\mathbb P}}=  -\frac{s}{4 \, {\cal V}_E^2}  \, \biggl[\left(\hat{\mathbb P}_\Lambda \, {\cal M}^{\Lambda \Delta} \,  \hat{\mathbb P}_\Delta - \hat{\mathbb P}_\Lambda \, {\cal M}^{\Lambda}_{\, \, \, \Delta} \, \hat{\mathbb P}^\Delta+  \hat{\mathbb P}^\Lambda \, {\cal M}_{\Lambda}^{\, \, \, \Delta} \, \hat{\mathbb P}_\Delta- \hat{\mathbb P}^\Lambda \, {\cal M}_{\Lambda \Delta} \, \hat{\mathbb P}^\Delta\right)\nonumber\\
& & \hskip0.75cm +\frac{1}{4} \, \left(\frac{4\, k_0^2}{9} \, \tilde{\cal G}_{\alpha \beta}\,   - 4\, \sigma_\alpha \, \sigma_\beta \right) \biggl(\tilde{\cal P}^\alpha{}_\Lambda \, {\cal M}^{\Lambda \Delta} \,  \tilde{\cal P}^\beta{}_\Delta - \tilde{\cal P}^\alpha{}_\Lambda \, {\cal M}^{\Lambda}_{\, \, \, \Delta} \, \tilde{\cal P}^{\beta\Delta}\nonumber\\
& & \hskip7cm+  \tilde{\cal P}^{\alpha\Lambda} \, {\cal M}_{\Lambda}^{\, \, \, \Delta} \, \tilde{\cal P}^\beta{}_\Delta- \tilde{\cal P}^{\alpha\Lambda} \, {\cal M}_{\Lambda \Delta} \, \tilde{\cal P}^{\beta\Delta}\biggr) \biggr] \,\nonumber\\
& & \hskip-0.50cm V_{{\mathbb H} \hat{\mathbb Q}} =  -\frac{1}{4\, \, {\cal V}_E^2}  \,\bigg[ (-2) \, \, \biggl\{ \left({\mathbb H}_\Lambda \, {\cal M}^{\Lambda \Delta} \,  \hat{\mathbb Q}_\Delta - {\mathbb H}_\Lambda \, {\cal M}^{\Lambda}_{\, \, \, \Delta} \, \hat{\mathbb Q}^\Delta+  {\mathbb H}^\Lambda \, {\cal M}_{\Lambda}^{\, \, \, \Delta} \, \hat{\mathbb Q}_\Delta- {\mathbb H}^\Lambda \, {\cal M}_{\Lambda \Delta} \, \hat{\mathbb Q}^\Delta\right) \nonumber\\
& & \hskip2.5cm +\, 2\, \left({\mathbb H}_\Lambda \, {\cal M}^{\Lambda \Delta} \,  \tilde{{\cal Q}}_\Delta - {\mathbb H}_\Lambda \, {\cal M}^{\Lambda}_{\, \, \, \Delta} \, \tilde{{\cal Q}}^\Delta+  {\mathbb H}^\Lambda \, {\cal M}_{\Lambda}^{\, \, \, \Delta} \, \tilde{{\cal Q}}_\Delta- {\mathbb H}^\Lambda \, {\cal M}_{\Lambda \Delta} \, \tilde{{\cal Q}}^\Delta\right) \biggr\}  \biggr] \nonumber\\
& & \hskip-0.50cm V_{{\mathbb F} \hat{\mathbb P}} =  -\frac{1}{4 \, {\cal V}_E^2}  \,\bigg[ (+2) \, \, \biggl\{ \left({\mathbb F}_\Lambda \, {\cal M}^{\Lambda \Delta} \,  \hat{\mathbb P}_\Delta - {\mathbb F}_\Lambda \, {\cal M}^{\Lambda}_{\, \, \, \Delta} \, \hat{\mathbb P}^\Delta+  {\mathbb F}^\Lambda \, {\cal M}_{\Lambda}^{\, \, \, \Delta} \, \hat{\mathbb P}_\Delta- {\mathbb F}^\Lambda \, {\cal M}_{\Lambda \Delta} \, \hat{\mathbb P}^\Delta\right) \nonumber\\
& & \hskip2.5cm +\, 2\, \left({\mathbb F}_\Lambda \, {\cal M}^{\Lambda \Delta} \,  \tilde{{\cal P}}_\Delta - {\mathbb F}_\Lambda \, {\cal M}^{\Lambda}_{\, \, \, \Delta} \, \tilde{{\cal P}}^\Delta+  {\mathbb F}^\Lambda \, {\cal M}_{\Lambda}^{\, \, \, \Delta} \, \tilde{{\cal P}}_\Delta- {\mathbb F}^\Lambda \, {\cal M}_{\Lambda \Delta} \, \tilde{{\cal P}}^\Delta\right) \biggr\}  \biggr] \nonumber\\
& & \hskip-0.50cm V_{{\mathbb \mho} {\mathbb \mho}} = -\frac{1}{4\, \, {\cal V}_E^2}  \,\biggl[\frac{{\cal G}^{ab}}{4}\, \biggl(\tilde{\mathbb \mho}_{a\Lambda} \, {\cal M}^{\Lambda \Delta} \,  \tilde{\mathbb \mho}_{b \Delta} - \tilde{\mathbb \mho}_{a \Lambda} \, {\cal M}^{\Lambda}_{\, \, \, \Delta} \, \tilde{\mathbb \mho}_b{}^\Delta +  \tilde{\mathbb \mho}_a{}^\Lambda \, {\cal M}_{\Lambda}^{\, \, \, \Delta} \, \tilde{\mathbb \mho}_{b \Delta}- \tilde{\mathbb \mho}_a{}^\Lambda \, {\cal M}_{\Lambda \Delta} \, \tilde{\mathbb \mho}_b{}^\Delta\biggr) \biggr]\nonumber\\
& & \hskip-0.5cm V_{\hat{\mathbb P} \hat{\mathbb Q}} = -\frac{1}{4\, \, {\cal V}_E^2}  \, \biggl[(-2)\, \biggl\{\left(\hat{\mathbb P}_\Lambda \, \hat{\mathbb Q}^{\Lambda} - \hat{\mathbb P}^\Lambda \,\hat{\mathbb Q}_\Lambda \right) +\frac{1}{4} \, \left(\frac{4\, k_0^2}{9} \, \tilde{\cal G}_{\alpha \beta}\,   - 4\, \sigma_\alpha \, \sigma_\beta \right)\nonumber\\
& & \hskip5cm \times \left(\hat{\mathbb P}^\alpha{}_\Lambda \,  \tilde{\cal Q}^{\beta \Lambda} + \hat{\mathbb Q}^{\alpha\Lambda} \,  \, \tilde{\cal P}^{\beta}{}_{\Lambda} - \hat{\mathbb Q}^\alpha{}_\Lambda \,  \tilde{\cal P}^{\beta \Lambda} - \hat{\mathbb P}^{\alpha\Lambda} \,  \, \tilde{\cal Q}^{\beta}{}_{\Lambda}   \right) \biggr\} \biggr] \nonumber\\
& & \hskip-0.5cm V_{{\mathbb F} {\mathbb H}} =  -\frac{1}{4\, \, {\cal V}_E^2}  \, \biggl[(+2)\,\, \,\left({\mathbb F}_\Lambda \, {\mathbb H}^\Lambda - {\mathbb F}^\Lambda \, {\mathbb H}_\Lambda \right) \biggr]\nonumber\\
& & \hskip-0.5cm V_{{\mathbb F} \hat{\mathbb Q}} = -\frac{1}{4\, s \, {\cal V}_E^2}  \, \biggl[(-2)\, \left({\mathbb F}_\Lambda \, \hat{\mathbb Q}^{\Lambda} - {\mathbb F}^\Lambda \,\hat{\mathbb Q}_\Lambda \right) \biggr] \nonumber\\
& & \hskip-0.5cm V_{{\mathbb H} {\mathbb P}} = -\frac{s}{4\, \, {\cal V}_E^2}  \, \biggl[(-2)\, \left({\mathbb H}_\Lambda \, \hat{\mathbb P}^{\Lambda} - {\mathbb H}^\Lambda \,\hat{\mathbb P}_\Lambda \right) \biggr] \nonumber
\eea
where we have defined $\tilde{\cal \mho}_a$, $\tilde{\cal Q}^\alpha$ and $\tilde{\cal P}^\alpha$ as in eqn. (\ref{eq:tildeQP}). The good thing about the symplectic rearrangement in eqn. (\ref{eq:main1}), via introducing symplectic matrices $\cal M$ and ${\cal S}$, is the fact that now one can express various pieces either as ${\cal O}_1\wedge \ast {\cal O}_2$ or ${\cal O}_1\wedge{\cal O}_2$ form. This has further helped to rewrite this huge collection, given in eqn. (\ref{eq:main1}), in a couple of lines as presented in eqn. (\ref{eq:main2}).

{\it Let us emphasize here that although a naive looking at the various pieces of this collection in eqn. (\ref{eq:main1}) gives an impression that they are very similar to those of \cite{Shukla:2015hpa} upto the places where $P$-flux is explicit seen to be present, however this is not true because as mentioned before, the crucial superpotential quantities $e_\Lambda$ and $m^\Lambda$ are internally very different after the inclusion of $P$-fluxes, which have also changed each of the similar looking flux orbits with the appearance of $P$-flux. Moreover, the searching/clubbing of various pieces together gets much more tedious, but it is the beauty of our new S-duality respecting flux-orbits which guides us throughout to help in collecting things in such a way that the overall symplectic structures within analogous pieces look quite similar to those of the case without $P$-flux !}

\section{Proof of the crucial symplectic relations for generic prepotentials}
\label{sec_proof}
\subsection{Verifying the peculiar symplectic relation given in eqn. (\ref{eq:periodF})}

Using the results in eqns. (\ref{eq:Prepder1})-(\ref{eq:simpperiodF2}), let us quickly verify the following interesting symplectic relation (which was given as eqn. (\ref{eq:periodF}),
\bea
{\cal F}_{\Lambda\Delta} = \ov{\cal N}_{\Lambda\Delta} + 2 \, i \, \frac{Im({\cal N}_{\Lambda\Gamma}) \, {\cal X}^\Gamma X^\Sigma \, (Im{\cal N}_{\Sigma \Delta}) }{Im({\cal N}_{\Gamma\Sigma}) {\cal X}^\Gamma X^\Sigma} \nonumber
\eea 
Note that the same holds true in a very analogous manner as to the period matrices in eqn. (\ref{eq:periodN}) which are written in terms of derivatives of the prepotential ${\cal F}$. One can show that the denominator of the second piece turns out to be,
\bea 
& & Im({\cal N}_{\Gamma\Sigma}) {\cal X}^\Gamma X^\Sigma = \left(x_0+x_{ij}\,u^i\,u^j\right)\,, 
\eea
and also one has,
\bea 
& & Im({\cal N}_{0\Gamma}) \, {\cal X}^\Gamma X^\Sigma \, (Im{\cal N}_{\Sigma 0}) = \left(x_0 + i\, x_{ij}\,u^i\,v^j\right)^2 \nonumber\\
& & Im({\cal N}_{0\Gamma}) \, {\cal X}^\Gamma X^\Sigma \, (Im{\cal N}_{\Sigma i}) = -i\, (x_{ij}\, u^j)\,\left(x_0 + i\, x_{i'j'}\,u^{i'}\,v^{j'}\right) \\
& & Im({\cal N}_{i\Gamma}) \, {\cal X}^\Gamma X^\Sigma \, (Im{\cal N}_{\Sigma j}) = -(x_{ii'}\, u^{i'})\, (x_{jj'}\, u^{j'})\nonumber
\eea
Now using 
\bea 
& & \hskip-1cm x_0 = \frac{(4\, l + \gamma)(2\, l - \gamma)}{(8\, l - \gamma)}, \quad x_{ij} = 6\biggl[l_{ij} -\frac{12\, l_i \, l_j}{(8\, l - \gamma)} \biggr],\, \quad x^{ij}= \frac{1}{6}\, l^{ij} - \frac{2}{4\, l + \gamma} \, u^i\, u^j \, ,
\eea
 one can show that following relations hold, 
\bea
\label{eq:simpperiodF3}
& & x_{ij}\, u^j = -\frac{6\, l_i\, (4\, l + \gamma)}{(8\,l-\gamma)}, \\
& & x_{ij}\, u^i \, u^j = -\frac{6\, l\, (4\, l + \gamma)}{(8\,l-\gamma)}, \nonumber\\
& & x_{ij}\, u^i \, v^j = -\frac{6\, (l_i \, v^i)\, (4\, l + \gamma)}{(8\,l-\gamma)}, \nonumber\\
& & \hskip-0.0cm x_{ij}\, v^i \, v^j = 6\,(l_{ij}\, v^i v^j)-\frac{72\, (l_i \, v^i)^2}{(8\,l-\gamma)}\,.
\nonumber
\eea 
Using these relations, we find that the various components of the RHS of eqn. (\ref{eq:periodF}) are the same as those of respective LHS terms as we see below,
\bea 
& & \hskip-1cm \ov{\cal N}_{00} +2\, i\,\frac{Im({\cal N}_{0\Gamma}) \, {\cal X}^\Gamma X^\Sigma \, (Im{\cal N}_{\Sigma0})}{Im({\cal N}_{\Gamma\Sigma}) {\cal X}^\Gamma X^\Sigma}  = \left(2\, l_{ijk}\, v^i \, v^j\, v^k - 6\, l_i\, v^i\right) + i \left(6\, l_{ij}\, v^i \, v^j - 2\, l + \gamma\right) \equiv {\cal F}_{00}\nonumber\\
& & \hskip-1cm \ov{\cal N}_{0i} +2\, i\,\frac{Im({\cal N}_{0\Gamma}) \, {\cal X}^\Gamma X^\Sigma \, (Im{\cal N}_{\Sigma i})}{Im({\cal N}_{\Gamma\Sigma}) {\cal X}^\Gamma X^\Sigma}= \left(3\, l_i - 3\, l_{ijk}\, v^j\, v^k + b_i\right) + i \left( -6\, l_{ij}\, v^j\right) \equiv {\cal F}_{0i} \, \\
& & \hskip-1cm \ov{\cal N}_{ij} +2\, i\, \frac{Im({\cal N}_{i\Gamma}) \, {\cal X}^\Gamma X^\Sigma \, (Im{\cal N}_{\Sigma j})}{Im({\cal N}_{\Gamma\Sigma}) {\cal X}^\Gamma X^\Sigma}= \left( 6\, l_{ijk}\, v^k+ a_{ij} \right) + i \left(6 \, l_{ij} \right) \equiv {\cal F}_{ij}\, , \nonumber
\eea
where a simplification with $Im({\cal N}_{\Gamma\Sigma}) {\cal X}^\Gamma X^\Sigma = -\frac{(4\,l+\gamma)^2}{(8\, l-\gamma)}$ have been utilized. This completes the proof of the symplectic relation in eqn. (\ref{eq:periodF}).

\subsection{Verifying the other non-trivial symplectic identities given in eqns. (\ref{eq:symp10})-(\ref{eq:symp11})}
Now the main goal is to prove the useful symplectic relations given in eqns. (\ref{eq:symp10})-(\ref{eq:symp11}). Note that the first set of identities given in eqn. (\ref{eq:symp10}) has been the most crucial one for the symplectic rearrangement in \cite{Shukla:2015hpa}, and after verifying these identities for $h^{2,1}_-(CY) \in \{0,1,2,3\}$, the same were conjectured to be true for arbitrary number of complex structure moduli, which now we prove to be generically true. For that we have to first collect the intermediate quantities, like $\cal L$ and $\cal M$ matrices obtained by using the results of eqn. (\ref{eq:Prepder1})-(\ref{eq:simpperiodF2}) in their respective definitions.

\subsubsection*{Explicit expressions for ${\cal L}$-matrices using eqn. (\ref{coff2})}
\bea
\label{eq:explicitLmatrix1}
& & \hskip-1.1cm {\bf (i). \quad {\cal L}^{\Lambda\Sigma}: }\qquad \nonumber\\
& & \hskip1.00cm {\cal L}^{00} = -\, \frac{1}{(2\, l-\gamma)}, \quad {\cal L}^{0i} =-\frac{v^i}{(2\, l-\gamma)} \equiv {\cal L}^{i0} , \quad {\cal L}^{ij}  = \frac{1}{6}\, l^{ij} -\frac{v^i\, v^j}{(2\, l-\gamma)} 
\eea
\bea
\label{eq:explicitLmatrix2}
& & \hskip-0.5cm {\bf (ii). \quad {\cal L}_{\Lambda}{}^{\Sigma}: }\qquad \nonumber\\
& & \hskip1cm {\cal L}_{0}{}^{0} = -\, \, \frac{\left(b_i\, v^i - 3\, l_i\, v^i - l_{ijk}\, v^i \, v^j\, v^k \right)}{(2\, l-\gamma)}, \\
& & \hskip1cm {\cal L}_{i}{}^{0} = -\, \frac{\left(3\, l_i + 3\, l_{ijk} \, v^j\, v^k + a_{ij}\, v^j + b_i\right) }{(2\, l-\gamma)}, \nonumber\\
& & \hskip1cm {\cal L}_{0}{}^{i} = -\frac{v^i \, \left(b_j\, v^j-l_{i'j'k'}\, v^{i'} \, v^{j'}\, v^{k'} -3\, l_j \, v^j \right)}{(2\, l-\gamma)}+ \left(\frac{u^i}{2} - \frac{l^{ij}\, l_{jkp}\, v^k\, v^p}{2} + \frac{1}{6}\, b_j\, l^{ji}\right)\nonumber\\
%& & \hskip5cm -\frac{(2 l-\gamma)}{2}\, \left(u^i - l^{ij}\, l_{jkp}\, v^k\, v^p + \frac{1}{3}\, b_j\, l^{ji}\right)\biggr],  \nonumber\\
& & \hskip1cm {\cal L}_{i}{}^{j} = -\frac{\left(3\,  l_{ij'k'}\,  v^{j'}\, v^{k'} \, v^{j} \, + 3 \, l_i\, v^j +b_i\, v^j + a_{ik}\,v^k \,v^j\right)}{(2\, l-\gamma)} +\, l_{ikp}\, v^p\, l^{kj}\, + \frac{1}{6} a_{ik}\,l^{kj}, \nonumber
\eea
\bea
\label{eq:explicitLmatrix3}
& & \hskip-9.3cm {\bf (iii). \quad {\cal L}^{\Lambda}{}_{\Sigma}: }\, \, \, \, \quad {\cal L}^{\Lambda}{}_{\Sigma} = - \left({\cal L}_{\Lambda}{}^{\Sigma}\right)^{T}\, 
\eea
\bea
\label{eq:explicitLmatrix4}
& & \hskip-1.2cm {\bf (iv). \quad {\cal L}_{\Lambda\Sigma}: }\qquad \nonumber\\
& &  \hskip-1cm {\cal L}_{00} =  \frac{\left(b_i v^i -l_{ijk}\, v^i v^j v^k - 3 \, l_i v^i\right)^2}{(2\, l -\gamma)}  + \left(2\,l - \gamma - 6\, l_{ij} \, v^i v^j\right) \nonumber\\
& & \hskip1.5cm  -\frac{1}{6} \left(b_i -3\, l_{ij'k'}\, v^{j'} v^{k'} + 3 \, l_i \right) \, l^{ij}\, \left(b_j - 3\, l_{jmn}\, v^m v^n + 3\, l_j \right) \nonumber\\
& & \hskip-1cm {\cal L}_{0i} =  \frac{\left(b_i + a_{im}\,v^m + 3\, l_{imn}\, v^m v^n + 3\, l_i\right)\, \left(b_j v^j -l_{jj'k'}\, v^j v^{j'} v^{k'} - 3 \, l_j v^j\right)}{(2\, l -\gamma)} + 6\, l_{ij}\,v^j\nonumber\\
& & \hskip1cm  - \frac{1}{6} \,\left(6\, l_{imm'} v^{m'} + a_{im}\right)\, l^{mn}\, \left(b_n -3\, l_{nj'k'}\, v^{j'} v^{k'} + 3 \, l_n \right) \equiv {\cal L}_{i0} \, \\
& & \hskip-1cm {\cal L}_{ij} =  \frac{\left(b_i + a_{im}\,v^m + 3\, l_{imn}\, v^m v^n + 3\, l_i\right)\, \left(b_j + a_{jm'}\,v^{m'} + 3\, l_{jm'n'}\, v^{m'} v^{n'} + 3\, l_j\right)}{(2\, l - \gamma)} \, \nonumber\\
& & \hskip1.5cm -6\, l_{ij} - \frac{1}{6} \,\left(6\, l_{imm'} v^{m'} + a_{im}\right)\, l^{mn} \, \left(6\, l_{jnn'} v^{n'} + a_{jn}\right). \nonumber
\eea
\subsubsection*{Explicit expressions for ${\cal M}$-matrices using eqn. (\ref{coff})}
\bea
\label{eq:explicitMmatrix1}
& & \hskip-1.5cm {\bf (i). \quad {\cal M}^{\Lambda\Sigma}: }\qquad \nonumber\\
& & \hskip1.5cm {\cal M}^{00} = \frac{1}{\, x_0}, \qquad {\cal M}^{0i} =\frac{v^i}{\, x_0} \equiv {\cal M}^{i0} , \qquad {\cal M}^{ij}  = \frac{v^i\, v^j}{\, x_0} - x^{ij} \,,
\eea
\bea 
\label{eq:explicitMmatrix2}
& & \hskip-0.5cm {\bf (ii). \quad {\cal M}_{\Lambda}{}^{\Sigma}: }\qquad \nonumber\\
& & \hskip1.5cm {\cal M}_{0}{}^{0} = \frac{ \left(b_i\,v^i -l_{ijk}\, v^i \, v^j\, v^k - \beta\,\, l_i\,v^i\right)}{\, x_0}, \\
& & \hskip1.5cm {\cal M}_{i}{}^{0} = \, \frac{\left(b_i + a_{ij}\,v^j + 3\left(l_{ijk} \, v^j\, v^k \right) + \beta\,\, l_i \right)}{\, x_0}, \nonumber\\
& & \hskip1.5cm {\cal M}_{0}{}^{i} = \frac{v^i \, \left(b_j\, v^j -l_{i'j'k'}\, v^{i'} \, v^{j'}\, v^{k'} - \beta\,\, l_i\, v^i\right)}{\, x_0} \, -x^{ij} \bigg[\beta\,\, l_j - 3\, l_{jpq}\, v^p\, v^q) + b_j \biggr],  \nonumber\\
& & \hskip1.5cm {\cal M}_{i}{}^{j} = \frac{ \left(3\,  l_{ij'k'}\,  v^{j'}\, v^{k'} \,+ \beta\, l_i+ b_i + a_{ik}\, v^k \right)\, v^j}{\, x_0} \, -\biggl[\left(6 l_{ikp}\, v^p + a_{ik} \right)\,x^{jk}\biggr]\,; \nonumber
\eea
\bea
\label{eq:explicitMmatrix3}
& & \hskip-8.8cm {\bf (iii). \quad {\cal M}^{\Lambda}{}_{\Sigma}: }\, \, \, \, \quad {\cal M}^{\Lambda}{}_{\Sigma} = - \left({\cal M}_{\Lambda}{}^{\Sigma}\right)^{T}\, 
\eea
\bea
\label{eq:explicitMmatrix4}
& & \hskip-1.0cm {\bf (iv). \quad {\cal M}_{\Lambda\Sigma}: }\qquad \nonumber\\
& &  \hskip-1cm {\cal M}_{00} = -\left(x_0 - x_{ij}\, v^i v^j\right) -\frac{\left(b_i\, v^i - l_{ijk}\, v^i v^j v^k - \beta \, l_i \, v^i \right)^2}{x_0} \nonumber\\
& & \hskip1cm  + \left(b_i\, -3\, l_{imn}\, v^m v^n + \beta \, l_i \, \right)\, x^{ij} \, \left(b_j\, -3\, l_{jmn}\, v^m v^n + \beta \, l_j \, \right) \\
& & \hskip-1cm {\cal M}_{0i} = -\frac{\left(b_i + a_{im}\,v^m + 3\, l_{imn}\, v^m v^n + \beta\, l_i\right)\, \left(b_j\, v^j - l_{jj'k'}\, v^j v^{j'} v^{k'} - \beta \, l_j \, v^j \right)}{x_0}\nonumber\\
& & \hskip0.5cm  -x_{ij}\,v^j + \left(6\, l_{imm'} v^{m'} + a_{im}\right)\, x^{mn} \,  \left(b_n\, -3\, l_{nm'n'}\, v^{m'} v^{n'} + \beta \, l_n \, \right) \equiv {\cal M}_{i0} \, \nonumber\\
& & \hskip-1cm {\cal M}_{ij} =   - \frac{\left(b_i + a_{im}\,v^m + 3\, l_{imn}\, v^m v^n + \beta\, l_i\right)\, \left(b_j + a_{jm'}\,v^{m'} + 3\, l_{jm'n'}\, v^{m'} v^{n'} + \beta\, l_j \right)}{x_0}\, \nonumber\\
& & \hskip2cm  +\, x_{ij}\, +\left(6\, l_{imm'} v^{m'} + a_{im}\right)\, x^{mn} \, \left(6\, l_{jnn'} v^{n'} + a_{jn}\right)\, ,\nonumber
\eea
where $\beta = \left(\frac{9 \gamma}{8\, l - \gamma}\right)$ have been used.

\subsubsection*{Explicit expressions for ${\cal M}_1$-matrices defined as ${\cal M}_1=({\cal M}+{\cal L})$}
For proving the symplectic identities in eqn. (\ref{eq:symp10}), let us simplify the various components of a new matrix ${\cal M}_1 = {\cal M}+ {\cal L}$. Using sets of eqns. (\ref{eq:explicitLmatrix1})-(\ref{eq:explicitLmatrix4}) and eqns. (\ref{eq:explicitMmatrix1})-(\ref{eq:explicitMmatrix4}), the same are given as under,
\bea
\label{eq:explicitM1matrix1}
& & \hskip-1.5cm {\bf (i). \quad \left({\cal M}_1\right)^{\Lambda\Sigma}: }\qquad \nonumber\\
& & \hskip-1.3cm \left({\cal M}_1\right)^{00} = \frac{2}{(4 l+\gamma)}, \quad \left({\cal M}_1\right)^{0i} =\frac{2 v^i}{(4 l + \gamma)} \equiv \left({\cal M}_1\right)^{i0}, \quad \left({\cal M}_1\right)^{ij}  = \frac{2\left(v^i v^j + u^i u^j\right)}{(4 l+\gamma)} 
\eea
\bea
\label{eq:explicitM1matrix2}
& & \hskip-0.5cm {\bf (ii). \quad \left({\cal M}_1\right)_{\Lambda}{}^{\Sigma}: }\qquad \nonumber\\
& & \hskip0.5cm \left({\cal M}_1\right)_{0}{}^{0} = \frac{2 \left(b_i \, v^i -l_{ijk}\, v^i \, v^j\, v^k +3 \, l_i\, v^i\right)}{(4\, l+\gamma)}, \\
& & \hskip0.5cm \left({\cal M}_1\right)_{i}{}^{0} = \frac{2 \left(b_i +a_{ij}\, v^j- 3\, l_i+3\, l_{ijk} \, v^j\, v^k \right)}{(4\, l+\gamma)}  , \nonumber\\
& & \hskip0.5cm \left({\cal M}_1\right)_{0}{}^{i} = \frac{2 \bigl[v^i \, \left(b_j\,v^j - l_{i'j'k'}\, v^{i'} \, v^{j'}\, v^{k'} + 3 l_j\, v^j\right)+ u^i \left(l + \gamma + b_j\,u^j - 3 l_{mn}v^m v^n\right)\bigr]}{(4\, l+\gamma)},  \nonumber\\
& & \hskip0.5cm \left({\cal M}_1\right)_{i}{}^{j} = \frac{2 \bigl[\left(b_i + a_{im}v^m -3\, l_{i} +3\,  l_{ij'k'}\,  v^{j'}\, v^{k'} \right) v^{j} \, +\left(6 \, l_{im}\, v^m + a_{im}\,u^m\right) u^j  \bigr]}{(4\, l+\gamma)}, \nonumber
\eea
\bea
\label{eq:explicitM1matrix3}
& & \hskip-7.7cm {\bf (iii). \quad \left({\cal M}_1\right)^{\Lambda}{}_{\Sigma}: }\, \, \, \, \quad \left({\cal M}_1\right)^{\Lambda}{}_{\Sigma} = - \left(\left({\cal M}_1\right)_{\Lambda}{}^{\Sigma}\right)^{T}\, 
\eea
\bea
\label{eq:explicitM1matrix4}
& & \hskip-0.4cm {\bf (iv). \quad \left({\cal M}_1\right)_{\Lambda\Sigma}: }\qquad \nonumber\\
& &  \hskip0.4cm \left({\cal M}_1\right)_{00} = -\frac{2}{(4\, l+\gamma)}  \biggl[\left(b_i\,v^i + 3\, l_i v^i -l_{ijk} v^i v^j v^k \right)^2 + \left(l + \gamma + b_j u^j - 3 l_{mn} v^m v^n\right)^2\, \biggr]\nonumber\\
& & \hskip0.4cm \left({\cal M}_1\right)_{0i} = -\frac{2}{(4\, l+\gamma)} \biggl[ \left(b_n\,v^n + 3\, l_n v^n -l_{njk} v^n v^j v^k \right)\left(b_i + a_{im}v^m -3\, l_i + 3\, l_{imn} v^m v^n\right) \nonumber\\
& & \hskip2.5cm + \left(a_{im}u^m + 6\, l_{im}v^m\right) \left(l + \gamma + b_j u^j - 3 l_{mn} v^m v^n\right) \biggr] \equiv \left({\cal M}_1\right)_{i0} \, \\
& & \hskip0.4cm \left({\cal M}_1\right)_{ij} = -\frac{2}{(4\, l+\gamma)} \biggl[\left(b_i + a_{im}v^m -3\, l_i + 3\, l_{imn} v^m v^n\right) \left(b_j + a_{jp}v^p -3\, l_j + 3\, l_{jpq} v^p v^q\right) \nonumber\\
& & \hskip2.5cm + \left(a_{im}u^m + 6\, l_{im}v^m\right) \left(a_{jn}u^n + 6\, l_{jn}v^n\right)\biggr]\,.\nonumber
\eea
\subsubsection*{Verifying the symplectic identities in eqn. (\ref{eq:symp10})}
Now using the fact that $e^{-K_{cs}} = -{2(4 \, l+\gamma)}$ and expanding the left hand sides of identities in eqn. (\ref{eq:symp10}), we find that the right hand sides are exactly what we expect from the explicit expressions of ${\cal M}_1$ matrices as presented in eqns. (\ref{eq:explicitM1matrix1})-(\ref{eq:explicitM1matrix4}). For example, if we consider the simplest case as being the first relation in eqn. (\ref{eq:symp10}), we have
\bea
& & \hskip-2cm Re(X^0 \,\ov X^0) = 1, \qquad Re(X^0 \,\ov X^i) = v^i, \qquad Re(X^i \,\ov X^j) = \left(v^i v^j + \, u^i u^j\right), \nonumber\\
%& & -\frac{1}{4}\, e^{-K_{cs}} \, \left({\cal M}_1\right)^{00} =- \frac{1}{4}\, \biggl[-{2(4 \, l+\gamma)}\biggr] \times \left(\frac{2}{(4 l+\gamma)}\right) = 1 \nonumber\\
& & \Longrightarrow \hskip1cm Re(X^\Lambda \ov X^\Delta) = -\frac{1}{4}\, e^{-K_{cs}} \, \left({\cal M}_1\right)^{\Lambda\Delta}. 
\eea
Similarly for the LHS of the second set of identities in eqn. (\ref{eq:symp10}), we find that,
\bea
\label{eq:ReFovX}
& & \hskip-2cm Re(F_0 \,\ov X^0)=Re(F_0) = \left(b_i \, v^i -l_{ijk}\, v^i \, v^j\, v^k +3 \, l_i\, v^i\right), \\
& & \hskip-2cm Re(F_i \,\ov X^0)=Re(F_i) = \left(b_i +a_{ij}\, v^j- 3\, l_i+3\, l_{ijk} \, v^j\, v^k \right), \nonumber\\
& & \hskip-2cm Re(F_0 \,\ov X^i) =v^i\, Re(F_0) + u^i Im(F_0) \nonumber\\
& & = v^i \, \left(b_j\,v^j - l_{i'j'k'}\, v^{i'} \, v^{j'}\, v^{k'} + 3 l_j\, v^j\right)+ u^i \left(l + \gamma + b_j\,u^j - 3 l_{mn}v^m v^n\right), \nonumber\\
& & \hskip-2cm Re(F_i \,\ov X^j) = Re(F_i)\, v^j + Im(F_i)\, u^j\nonumber\\
& & = \left(b_i + a_{im}v^m -3\, l_{i} +3\,  l_{ij'k'}\,  v^{j'}\, v^{k'} \right) v^{j} \, +\left(6 \, l_{im}\, v^m + a_{im}\,u^m\right) u^j, \nonumber
\eea
Considering eqn. (\ref{eq:ReFovX}) along with the explicit expressions of ${\cal M}_1$ matrices as presented in eqns. (\ref{eq:explicitM1matrix2}), it is straight to observe now, that the following relation holds,
\bea
%& & -\frac{1}{4}\, e^{-K_{cs}} \, \left({\cal M}_1\right)^{00} =- \frac{1}{4}\, \biggl[-{2(4 \, l+\gamma)}\biggr] \times \left(\frac{2}{(4 l+\gamma)}\right) = 1 \nonumber\\
& & Re(F_\Lambda \ov X^\Delta) = -\frac{1}{4}\, e^{-K_{cs}} \, \left({\cal M}_1\right)_{\Lambda}{}^{\Delta}. 
\eea
On the similar lines, the other two sets of symplectic identities of eqn. (\ref{eq:symp10}) can also be cross-checked component-by-component, and is found to hold generically. This completes the proof of symplectic identities in eqn. (\ref{eq:symp10}).

\subsubsection*{Verifying the symplectic identities in eqn. (\ref{eq:symp11})}
For proving the second class of relevant identities in eqn. (\ref{eq:symp11}) let us consider the various components of the factor $\left({\cal M}^{\Lambda}{}_{ \Sigma} \, {\cal L}^{\Sigma \Delta} + {\cal M}^{\Lambda \Sigma} \, {\cal L}_{\Sigma}{}^{ \Delta} \right)$, for which we have,
\bea
& & \hskip-0.5cm \left(\frac{1}{4} \, e^{-K_{cs}} \right)\, \left({\cal M}^{0}{}_{ \Sigma} \, {\cal L}^{\Sigma 0} + {\cal M}^{0 \Sigma} \, {\cal L}_{\Sigma}{}^{0} \right) = \frac{1}{4}\, \biggl[-{2(4 \, l+\gamma)}\biggr] \times 0 = 0, \\
& & \hskip-0.5cm \left(\frac{1}{4} \, e^{-K_{cs}} \right)\, \left({\cal M}^{0}{}_{ \Sigma} \, {\cal L}^{\Sigma i} + {\cal M}^{0 \Sigma} \, {\cal L}_{\Sigma}{}^{i} \right) = \frac{1}{4}\, \biggl[-{2(4 \, l+\gamma)}\biggr] \times \left(\frac{2\, u^i}{(4\, l+\gamma)}\right) = -u^i  \nonumber\\
& & \hskip-0.5cm \left(\frac{1}{4} \, e^{-K_{cs}} \right)\, \left({\cal M}^{i}{}_{ \Sigma} \, {\cal L}^{\Sigma j} + {\cal M}^{i \Sigma} \, {\cal L}_{\Sigma}{}^{j} \right) = \frac{1}{4}\, \biggl[-{2(4 \, l+\gamma)}\biggr] \times \left(\frac{2(v^i \, u^j - u^i\, v^j)}{4\,l+\gamma}\right) = u^i \, v^j - v^i \, u^j  \nonumber
\eea
Now it is straight forward to observe that these three relations verify the first identity of eqn. (\ref{eq:symp11}) which is,
\bea
& & Im({\cal X}^\Lambda \, \ov{\cal X}^\Delta) =\frac{1}{4} \, e^{-K_{cs}} \, \biggl[\left({\cal M}^{\Lambda}{}_{ \Sigma} \, {\cal L}^{\Sigma \Delta} + {\cal M}^{\Lambda \Sigma} \, {\cal L}_{\Sigma}{}^{ \Delta} \right) \biggr] \nonumber
\eea
Similarly the other relations of (\ref{eq:symp11}) can also be explicitly checked by using the collection of various intermediate symplectic quantities we have earlier derived. 

\section{Expanded version of the explicit c.s. moduli dependent scalar potential}
\label{sec_csVExpanded}
Utilizing the compact expressions of ${\cal M}$ and  ${\cal S}$-matrices given in eqns. (\ref{eq:M-matrixCompact}) and $(\ref{eq:S-matrixCompact})$, the various pieces of the scalar potential (\ref{eq:Vsss12345}) can be rewritten as under,
\bea
\label{eq:main4}
& & \hskip-1cm V_{\mathbb F \mathbb F} = -\frac{1}{4\,s\,{\cal V}_E^2}\biggl[\frac{\left({\mathbb F}_0 + v^i\, {\mathbb F}_i + r_0 \, {\mathbb F}^0 + r_i \, {\mathbb F}^i \right)^2}{x_0}  - \left({\mathbb F}^i - v^i \,{ \mathbb F}^0 \right)\, x_{ij} \, \left({\mathbb F}^j - v^j \,{ \mathbb F}^0 \right) \\
& & - \left({\mathbb F}_i + \epsilon_i \, {\mathbb F}^0 + \eta_{im} \, {\mathbb F}^m \right) \, x^{ij} \left({\mathbb F}_j + \epsilon_j \, {\mathbb F}^0 + \eta_{jm} \, {\mathbb F}^m \right)+ x_0\, ({\mathbb F}^0)^2\biggr] \nonumber\\
& & \hskip-1cm V_{\mathbb H \mathbb H} = -\frac{s}{4\,{\cal V}_E^2}\biggl[\frac{\left({\mathbb H}_0 + v^i\, {\mathbb H}_i + r_0 \, {\mathbb H}^0 + r_i \, {\mathbb H}^i \right)^2}{x_0}  - \left({\mathbb H}^i - v^i \,{ \mathbb H}^0 \right)\, x_{ij} \, \left({\mathbb H}^j - v^j \,{ \mathbb H}^0 \right) \nonumber\\
& & - \left({\mathbb H}_i + \epsilon_i \, {\mathbb H}^0 + \eta_{im} \, {\mathbb H}^m \right) \, x^{ij} \left({\mathbb H}_j + \epsilon_j \, {\mathbb H}^0 + \eta_{jm} \, {\mathbb H}^m \right)+ x_0\, ({\mathbb H}^0)^2 \biggr] \nonumber\\
& & \hskip-1cm V_{\hat{\mathbb Q} \hat{\mathbb Q}} = -\frac{1}{4\,s\,{\cal V}_E^2}\biggl[\biggl\{\frac{\left(\hat{\mathbb Q}_0 + v^i\, \hat{\mathbb Q}_i + r_0 \, \hat{\mathbb Q}^0 + r_i \, \hat{\mathbb Q}^i \right)^2}{x_0}  - \left(\hat{\mathbb Q}^i - v^i \,\hat{ \mathbb Q}^0 \right)\, x_{ij} \, \left(\hat{\mathbb Q}^j - v^j \,\hat{\mathbb Q}^0 \right) \nonumber\\
& & - \left(\hat{\mathbb Q}_i + \epsilon_i \, \hat{\mathbb Q}^0 + \eta_{im} \, \hat{\mathbb Q}^m \right) \, x^{ij} \left(\hat{\mathbb Q}_j + \epsilon_j \, \hat{\mathbb Q}^0 + \eta_{jm} \, \hat{\mathbb Q}^m \right)+ x_0\, (\hat{\mathbb Q}^0)^2 \biggr\} + \left(\frac{1}{4\,l+\gamma}\right)\nonumber\\
& & \hskip0cm  \times\left(\frac{4\, k_0^2}{9} \tilde{\cal G}_{\alpha \beta}- 4 \sigma_\alpha\, \sigma_\beta \right)\,\biggl\{\left(u^i\, \hat{\mathbb Q}_i{}^\alpha + \psi_0 \, \hat{\mathbb Q}^{\alpha 0} + \psi_i \, \hat{\mathbb Q}^{\alpha i} \right)  \left(u^i\, \hat{\mathbb Q}_i{}^\beta + \psi_0 \, \hat{\mathbb Q}^{\beta 0} + \psi_i \, \hat{\mathbb Q}^{\beta i} \right)\nonumber\\
& & +\left(\hat{\mathbb Q}_0{}^\alpha + v^i\, \hat{\mathbb Q}_i{}^\alpha + \phi_0 \, \hat{\mathbb Q}^{\alpha 0} + \phi_i \, \hat{\mathbb Q}^{\alpha i} \right)  \left(\hat{\mathbb Q}_0{}^\beta + v^i\, \hat{\mathbb Q}_i{}^\beta + \phi_0 \, \hat{\mathbb Q}^{\beta 0} + \phi_i \, \hat{\mathbb Q}^{\beta i} \right) \biggr\}\biggr]  \nonumber\\
& & \hskip-1cm V_{\hat{\mathbb P} \hat{\mathbb P}} = -\frac{s}{4\,{\cal V}_E^2}\biggl[\biggl\{\frac{\left(\hat{\mathbb P}_0 + v^i\, \hat{\mathbb P}_i + r_0 \, \hat{\mathbb P}^0 + r_i \, \hat{\mathbb P}^i \right)^2}{x_0}  - \left(\hat{\mathbb P}^i - v^i \,\hat{\mathbb P}^0 \right)\, x_{ij} \, \left(\hat{\mathbb P}^j - v^j \,\hat{\mathbb P}^0 \right) \nonumber\\
& & - \left(\hat{\mathbb P}_i + \epsilon_i \, \hat{\mathbb P}^0 + \eta_{im} \, \hat{\mathbb P}^m \right) \, x^{ij} \left(\hat{\mathbb P}_j + \epsilon_j \, \hat{\mathbb P}^0 + \eta_{jm} \, \hat{\mathbb P}^m \right)+ x_0\, (\hat{\mathbb P}^0)^2 \biggr\}+ \left(\frac{1}{4\,l+\gamma}\right)\nonumber\\
& & \hskip0cm  \times\left(\frac{4\, k_0^2}{9} \tilde{\cal G}_{\alpha \beta}- 4 \sigma_\alpha\, \sigma_\beta \right)\,\biggl\{\left(u^i\, \hat{\mathbb P}_i{}^\alpha + \psi_0 \, \hat{\mathbb P}^{\alpha 0} + \psi_i \, \hat{\mathbb P}^{\alpha i} \right)  \left(u^i\, \hat{\mathbb P}_i{}^\beta + \psi_0 \, \hat{\mathbb P}^{\beta 0} + \psi_i \, \hat{\mathbb P}^{\beta i} \right)\nonumber\\
& & +\left(\hat{\mathbb P}_0{}^\alpha + v^i\, \hat{\mathbb P}_i{}^\alpha + \phi_0 \, \hat{\mathbb P}^{\alpha 0} + \phi_i \, \hat{\mathbb P}^{\alpha i} \right)  \left(\hat{\mathbb P}_0{}^\beta + v^i\, \hat{\mathbb P}_i{}^\beta + \phi_0 \, \hat{\mathbb P}^{\beta 0} + \phi_i \, \hat{\mathbb P}^{\beta i} \right) \biggr\}\biggr]  \nonumber\\ 
&  & \hskip-1cm V_{\mathbb \mho \mathbb \mho} = -\frac{1}{4\,{\cal V}_E^2}\biggl[\left(\frac{1}{4\,l+\gamma}\right)\, {\cal G}^{ab}\,\biggl\{\left(u^i\, {\mathbb \mho}_{ai} + \psi_0 \, {\mathbb \mho}_{a}{}^{0} + \psi_i \, {\mathbb \mho}_a{}^i \right)  \left(u^i\, {\mathbb \mho}_{bi} + \psi_0 \, {\mathbb \mho}_{b}{}^{0} + \psi_i \, {\mathbb \mho}_b{}^i \right)\nonumber\\
& & +\left({\mathbb \mho}_{a0} + v^i\, {\mathbb \mho}_{ai} + \phi_0 \, {\mathbb \mho}_{a}{}^{0} + \phi_i \, {\mathbb \mho}_{a}{}^i \right)  \left({\mathbb \mho}_{b0} + v^i\, {\mathbb \mho}_{bi} + \phi_0 \, {\mathbb \mho}_{b}{}^{0} + \phi_i \, {\mathbb \mho}_{b}{}^i \right) \biggr\}\biggr]  \nonumber\\
& & \hskip-1cm V_{\mathbb H \hat{\mathbb Q}} = +\frac{1}{2\,{\cal V}_E^2}\biggl[\biggl\{\frac{\left({\mathbb H}_0 + v^i\, {\mathbb H}_i + r_0 \, {\mathbb H}^0 + r_i \, {\mathbb H}^i \right)\left(\hat{\mathbb Q}_0 + v^i\, \hat{\mathbb Q}_i + r_0 \, \hat{\mathbb Q}^0 + r_i \, \hat{\mathbb Q}^i \right)}{x_0}  + x_0\, {\mathbb H}^0 \hat{\mathbb Q}^0  \nonumber\\
& & - \left({\mathbb H}^i - v^i \,{ \mathbb H}^0 \right)\, x_{ij} \, \left(\hat{\mathbb Q}^j - v^j \,\hat{\mathbb Q}^0 \right) - \left({\mathbb H}_i + \epsilon_i \, {\mathbb H}^0 + \eta_{im} \, {\mathbb H}^m \right) \, x^{ij} \left(\hat{\mathbb Q}_j + \epsilon_j \, \hat{\mathbb Q}^0 + \eta_{jm} \, \hat{\mathbb Q}^m \right) \biggr\} \nonumber\\
& & \hskip0cm - \left(\frac{4}{4\,l+\gamma}\right)\,\biggl\{\left(u^i\, {\mathbb H}_i + \psi_0 \, {\mathbb H}^{0} + \psi_i \, {\mathbb H}^{i} \right)  \left(u^i\, \hat{\mathbb Q}_i + \psi_0 \, \hat{\mathbb Q}^{0} + \psi_i \, \hat{\mathbb Q}^{i} \right)\nonumber\\
& & +\left({\mathbb H}_0 + v^i\, {\mathbb H}_i + \phi_0 \, {\mathbb H}^{0} + \phi_i \, {\mathbb H}^{i} \right)  \left(\hat{\mathbb Q}_0 + v^i\, \hat{\mathbb Q}_i + \phi_0 \, \hat{\mathbb Q}^{0} + \phi_i \, \hat{\mathbb Q}^{i} \right) \biggr\}\biggr]\nonumber\\
& & \hskip-1cm V_{\mathbb F \hat{\mathbb P}} = -\frac{1}{2\,{\cal V}_E^2}\biggl[\biggl\{\frac{\left({\mathbb F}_0 + v^i\, {\mathbb F}_i + r_0 \, {\mathbb F}^0 + r_i \, {\mathbb F}^i \right)\left(\hat{\mathbb P}_0 + v^i\, \hat{\mathbb P}_i + r_0 \, \hat{\mathbb P}^0 + r_i \, \hat{\mathbb P}^i \right)}{x_0} + x_0\, {\mathbb F}^0 \hat{\mathbb P}^0  \nonumber\\
& & - \left({\mathbb F}^i - v^i \,{ \mathbb F}^0 \right)\, x_{ij} \, \left(\hat{\mathbb P}^j - v^j \, \hat{\mathbb P}^0 \right)- \left({\mathbb F}_i + \epsilon_i \, {\mathbb F}^0 + \eta_{im} \, {\mathbb F}^m \right) \, x^{ij} \left(\hat{\mathbb P}_j + \epsilon_j \, \hat{\mathbb P}^0 + \eta_{jm} \, \hat{\mathbb P}^m \right) \biggr\} \nonumber\\
& & \hskip0cm - \left(\frac{4}{4\,l+\gamma}\right)\,\biggl\{\left(u^i\, {\mathbb F}_i + \psi_0 \, {\mathbb F}^{0} + \psi_i \, {\mathbb F}^{i} \right)  \left(u^i\, \hat{\mathbb P}_i + \psi_0 \, \hat{\mathbb P}^{0} + \psi_i \, \hat{\mathbb P}^{i} \right)\nonumber\\
& & +\left({\mathbb F}_0 + v^i\, {\mathbb F}_i + \phi_0 \, {\mathbb F}^{0} + \phi_i \, {\mathbb F}^{i} \right)  \left(\hat{\mathbb P}_0 + v^i\, \hat{\mathbb P}_i + \phi_0 \, \hat{\mathbb P}^{0} + \phi_i \, \hat{\mathbb P}^{i} \right) \biggr\}\biggr]  \nonumber\\
& & \hskip-1cm V_{\hat{\mathbb P} \hat{\mathbb Q}} =\frac{1}{2 \, {\cal V}_E^2} \biggl[%\left(\hat{\mathbb P}_0 \, \hat{\mathbb Q}^{0} +\hat{\mathbb P}_i \, \hat{\mathbb Q}^i  - \hat{\mathbb P}^0 \,\hat{\mathbb Q}_0 - \hat{\mathbb P}^i \,\hat{\mathbb Q}_i \right)+
\frac{1}{2} \, \left(\frac{4\, k_0^2}{9} \tilde{\cal G}_{\alpha \beta}\,   - 4\, \sigma_\alpha \sigma_\beta \right)\biggl\{\hat{\mathbb P}^\alpha{}_0 \,  \tilde{\cal Q}^{\beta 0} + \hat{\mathbb Q}^{\alpha 0} \,  \, \tilde{\cal P}^{\beta}{}_{0}\nonumber\\
& & \hskip2cm  - \hat{\mathbb Q}^\alpha{}_0 \,  \tilde{\cal P}^{\beta 0} - \hat{\mathbb P}^{\alpha 0} \,  \, \tilde{\cal Q}^{\beta}{}_{0}+ \hat{\mathbb P}^\alpha{}_i \,  \tilde{\cal Q}^{\beta i} + \hat{\mathbb Q}^{\alpha i} \,  \, \tilde{\cal P}^{\beta}{}_{i} - \hat{\mathbb Q}^\alpha{}_i \,  \tilde{\cal P}^{\beta i} - \hat{\mathbb P}^{\alpha i} \,  \, \tilde{\cal Q}^{\beta}{}_{i} \biggr\} \biggr] \nonumber %\\
%& & \hskip-1cm V_{{\mathbb F} {\mathbb H}} =  -\frac{1}{4\, \, {\cal V}_E^2}  \, \biggl[(+2)\,\, \,\left({\mathbb F}_0 \, {\mathbb H}^0 + {\mathbb F}_i \, {\mathbb H}^i - {\mathbb F}^0 \, {\mathbb H}_0 - {\mathbb F}^i \, {\mathbb H}_i \right) \biggr]\nonumber\\
%& & \hskip-1cm V_{{\mathbb F} \hat{\mathbb Q}} = -\frac{1}{4\, s \, {\cal V}_E^2}  \, \biggl[(-2)\, \left({\mathbb F}_0 \, \hat{\mathbb Q}^0 + {\mathbb F}_i \, \hat{\mathbb Q}^i - {\mathbb F}^0 \,\hat{\mathbb Q}_0 - {\mathbb F}^i \,\hat{\mathbb Q}_i \right) \biggr] \nonumber\\
%& & \hskip-1cm V_{{\mathbb H} {\mathbb P}} = -\frac{s}{4\, \, {\cal V}_E^2}  \, \biggl[(-2)\, \left({\mathbb H}_0 \, \hat{\mathbb P}^0+{\mathbb H}_i \, \hat{\mathbb P}^i - {\mathbb H}^0 \,\hat{\mathbb P}_0 - {\mathbb H}^i \,\hat{\mathbb P}_i \right) \biggr] \nonumber
\eea
Here, the remaining collection of pieces which is independent of the complex structure saxions/axions are collected as $V_{tad}$ to be given as under,
\bea
\label{eq:tadpoleT}
& & \hskip-1cm V_{tad} = -\frac{1}{2\, s\, {\cal V}_E^2} \, \biggl[s\, \,\left({\mathbb F}_0 \, {\mathbb H}^0 + {\mathbb F}_i \, {\mathbb H}^i - {\mathbb F}^0 \, {\mathbb H}_0 - {\mathbb F}^i \, {\mathbb H}_i \right) - s\, \,\left(\hat{\mathbb P}_0 \, \hat{\mathbb Q}^{0} +\hat{\mathbb P}_i \, \hat{\mathbb Q}^i  - \hat{\mathbb P}^0 \,\hat{\mathbb Q}_0 - \hat{\mathbb P}^i \,\hat{\mathbb Q}_i \right) \nonumber\\
& & \hskip0cm -\, \left({\mathbb F}_0 \, \hat{\mathbb Q}^0 + {\mathbb F}_i \, \hat{\mathbb Q}^i - {\mathbb F}^0 \,\hat{\mathbb Q}_0 - {\mathbb F}^i \,\hat{\mathbb Q}_i \right)  - \,s^2\, \left({\mathbb H}_0 \, \hat{\mathbb P}^0+{\mathbb H}_i \, \hat{\mathbb P}^i - {\mathbb H}^0 \,\hat{\mathbb P}_0 - {\mathbb H}^i \,\hat{\mathbb P}_i \right) \biggr]\,.
\eea
Given that $V_{tad}$ corresponds to the generalized tadpoles to be canceled by the local sources on top of satisfying a set of Bianchi identities, we will ignore this piece in our discussion from now onwards, and will focus only on the remaining pieces in (\ref{eq:main4}).

Now let us observe here that at the level of scalar potential given in eqn. (\ref{eq:main4}), the sole effect of the real quantities $b_i$ and $a_{ij}$ appearing in the prepotential given in eqn. (\ref{eq:prepotential}) can be absorbed by some redefinitions of the flux parameters. For the illustration, let us consider the following redefinitions for the flux parameters,
\bea
& & \hskip-0.5cm \ov{\mathbb Y}_0 = {\mathbb Y}_0 + b_i \, {\mathbb Y}^i, \qquad \ov {\mathbb Y}_i = {\mathbb Y}_i + b_i \, {\mathbb Y}^0 + a_{ij} {\mathbb Y}^j \, , \qquad {\mathbb Y} \in \{ {\mathbb F}, {\mathbb H}, {\mathbb \mho}_a, \hat{\mathbb Q}^\alpha, \hat{\mathbb P}^\alpha \}
\eea
along with a redefined version of the subset of other parameters given in eqn. (\ref{eq:parameters4}),
\bea
& & \hskip-0.9cm \ov r_0 =  - l_{ijk} v^i v^j v^k -\beta \, l_i\, v^i , \quad  \ov r_i =  3 \, l_{ijk}\, v^j v^k+\beta \, l_i , \quad \ov\epsilon_i = \beta\, l_i\,- 3\, l_{ijk} v^j v^k,  \quad \ov\eta_{ij} = 6 \,l_{ijk}\, v^k, \nonumber\\
& & \hskip-0.9cm \ov \phi_0 =  - l_{ijk} v^i v^j v^k +3\, l_i\, v^i, \quad \ov \phi_i = 3 \, l_{ijk}\, v^j v^k  - 3 \, l_i, \quad \ov\psi_0 = l + \gamma - 3\, l_{ij} v^i v^j, \quad \ov\psi_i = 6 \,l_{ij}\, v^j\, , \nonumber
\eea
where the flux parameters ${\mathbb Y} \in \{ {\mathbb F}, {\mathbb H}, {\mathbb \mho}_a, \hat{\mathbb Q}^\alpha, \hat{\mathbb P}^\alpha \}$ have appropriate upper and lower $h^{21}_-(CY)$-indices. Now the total scalar potential (\ref{eq:main4}) is invariant under the following transformation,
\bea
& & \hskip1cm {\mathbb Y}_0 \to \ov{\mathbb Y}_0, \quad {\mathbb Y}_i \to \ov{\mathbb Y}_i, \quad r_0 \to \ov r_0, \quad r_i \to \ov r_i, \\
& & \hskip-1cm \phi_0 \to \ov \phi_0, \quad \phi_i \to \ov \phi_i, \quad \psi_0 \to \ov \psi_0, \quad \psi_i \to \ov \psi_i, \quad \epsilon_i \to \ov \epsilon_i, \quad \eta_{ij} \to \ov\eta_{ij}\, . \nonumber
\eea
This can be easily checked as the only types of flux combinations involving $b_i$ and $a_{ij}$ (rational) parameters sitting inside the various pieces of the scalar potential (\ref{eq:main4}) are as given under,
\bea 
\label{eq:newCSfluxOrbits}
& & \left({\mathbb Y}_0 + v^i\, {\mathbb Y}_i + r_0 \, {\mathbb Y}^0 + r_i \, {\mathbb Y}^i \right) = \left(\ov{\mathbb Y}_0 + v^i\, \ov{\mathbb Y}_i + \ov r_0 \, {\mathbb Y}^0 + \ov r_i \, {\mathbb Y}^i \right), \\
& & \left({\mathbb Y}_i + \epsilon_i \, {\mathbb Y}^0 + \eta_{ij} \, {\mathbb Y}^j \right) = \left(\ov{\mathbb Y}_i + \ov\epsilon_i \, {\mathbb Y}^0 + \ov\eta_{ij} \, {\mathbb Y}^j \right), \nonumber\\
& & \left(u^i\, {\mathbb Y}_i + \psi_0 \, {\mathbb Y}^{0} + \psi_i \, {\mathbb Y}^{i} \right) = \left(u^i\, \ov{\mathbb Y}_i + \ov\psi_0 \, {\mathbb Y}^{0} + \ov\psi_i \, {\mathbb Y}^{i} \right)\nonumber\\
& & \left({\mathbb Y}_0 + v^i\, {\mathbb Y}_i + \phi_0 \, {\mathbb Y}^{0} + \phi_i \, {\mathbb Y}^{i} \right) = \left(\ov{\mathbb Y}_0 + v^i\, \ov{\mathbb Y}_i + \ov\phi_0 \, {\mathbb Y}^{0} + \ov\phi_i \, {\mathbb Y}^{i} \right) \nonumber
\eea
Such observations of getting rid of the explicit dependences on the $a_{ij}$ and $b_i$ parameters, via a redefinition of fluxes appearing in the scalar potential, have also been also made in \cite{Blumenhagen:2014nba} with the presence of only standard fluxes, $H_3$ and $F_3$, and now we find the same to be true with the inclusion of (non-)geometric fluxes as well. However note that effects of $\gamma$, which comes from the perturbative $\alpha^\prime$-corrections on the mirror side, cannot be absorbed in such a manner. %One reason of this distinction among the real parameters $(a_{ij}, b_i)$ and $\gamma$  could be the fact that the former parameters do not appear in the simplified K\"ahler potential (\ref{eq:KcsSimp}) while the later does. 
Nevertheless the corrections involving $\gamma$ parameter can also be ignored (along with the non-perturbative effects) in the large complex structure limit.

Note that the new generalized flux orbits which have been defined in eqns. (\ref{eq:orbitsB1})-(\ref{eq:orbits11B}) have already taken care of a proper reshuffling of volume moduli, their respective axions as well as the odd axions, while from the new explicit collection of scalar potential in eqn. (\ref{eq:main4}),  we observe that the saxions/axions of the complex structure moduli are also incorporating some mixings of fluxes and moduli/axions to form a new set of peculiar flux combinations relevant for expressing the scalar potential.  For example, the following flux combinations do appear in the scalar potential collection (\ref{eq:main4}),
\bea 
& & \hskip-2cm \Theta^0[{\mathbb Y}] = {\mathbb Y}^0, \hskip5.3cm \Theta_0[{\mathbb Y}] =  {\mathbb Y}_0 + v^i \,  {\mathbb Y}_i + r_0  {\mathbb Y}^0 + r_i  {\mathbb Y}^i\, , \nonumber\\
& & \hskip-2cm \Theta^i[{\mathbb Y}] = {\mathbb Y}^i - v^i \, {\mathbb Y}^0, \hskip4.0cm \Theta_i[{\mathbb Y}] =  {\mathbb Y}_i + \epsilon_i\,  {\mathbb Y}^0 + \eta_{ij} \,  {\mathbb Y}^j\, ,\\
& & \hskip-1.9cm \Phi[{\mathbb Y}] = {\mathbb Y}_0 +v^i \, {\mathbb Y}_i +\phi_0\,  {\mathbb Y}^0+\phi_i \, {\mathbb Y}^i, \quad \qquad \Psi[{\mathbb Y}] =u^i \, {\mathbb Y}_i +\psi_0\,  {\mathbb Y}^0+\psi_i \, {\mathbb Y}^i\,, \nonumber
\eea
where the flux parameters ${\mathbb Y} \in \{ {\mathbb F}, {\mathbb H}, {\mathbb \mho}_a, \hat{\mathbb Q}^\alpha, \hat{\mathbb P}^\alpha \}$ have appropriate upper and lower $h^{21}_-(CY)$-indices. Moreover, here we have denoted $\Theta[\mathbb Y]$ etc. in such a way so that one could distinguish among various flux orbits; for example, $\Theta^0[{\mathbb H}] = {\mathbb H}^0$ and $\Theta_0[{\mathbb F}] =  {\mathbb F}_0 + v^i \,  {\mathbb F}_i + r_0  {\mathbb F}^0 + r_i  {\mathbb F}^i$ etc.

\newpage

%\bibliography{references}
\bibliographystyle{utphys}
\bibliography{reference}

\end{document}